\documentclass[twocolumn,times,tighten,twocolappendix]{aastex631} 
\usepackage{graphicx}
\usepackage{xcolor}
\usepackage{url}
\usepackage{mathtools}
\usepackage{enumitem}
\usepackage{rotating}
\usepackage{textcase}

\accepted{November 8, 2024}
\submitjournal{ApJ}
\shortauthors{L. L. Lee et al.}
\journalinfo{}

\definecolor{green}{RGB}{0,155,85}

\newcommand{\Hi}{\ion{H}{1}}
\newcommand{\Ha}{H${\alpha}$}

\newcommand{\cii}{[\ion{C}{2}]}
\newcommand{\co}{\ion{CO}{0}}

\newcommand{\mstar}{$M_{\star}$}

\newcommand{\vcirc}{$v_{\rm c}$}
\newcommand{\vrot}{$V_{\rm rot}$}
\newcommand{\sigo}{$\sigma_{0}$}
\newcommand{\vsig}{$v_{\rm rot}/\sigma_{0}$}
\newcommand{\fdmre}{$f_{\rm DM}(<$$R_{\rm e})$}
\newcommand{\bb}{\texttt{$^{\tt 3D}$Barolo}}
\newcommand{\gp}{\texttt{GalPak$^{\tt 3D}$}}
\newcommand{\dy}{\texttt{DysmalPy}}

\newcommand{\mvir}{$M_{\rm vir}$}

\newcommand{\fgas}{$f_{\rm gas}$}

\DeclareMathOperator{\sech}{sech}
\newcommand{\msun}{$\rm M_{\odot}$}
\newcommand{\reff}{$R_{\rm e}$}
\newcommand{\incl}{$i$}
\newcommand{\pa}{PA}

\newcommand{\znoon}{$z$\,$\sim$\,1$-$3}
\newcommand{\ReRbeam}{$R_{\rm e}$/beam}
\newcommand{\vrotrecRe}{$V_{\rm model}(R_{\rm e})/V_{\rm intrinsic}(R_{\rm e})$}
\newcommand{\sigrecRe}{$\sigma_{\rm model}(R_{\rm e})/\sigma_{\rm intrinsic}(R_{\rm e})$}

\begin{document}

\title{Disk kinematics at high redshift: \dy's extension to 3D modeling and comparison with different approaches}

\correspondingauthor{Lilian L. Lee}
\email{lilian@mpe.mpg.de; mail@lilianlylee.com}

\author[0000-0001-7457-4371]{Lilian ~L. Lee}
\affiliation{Max-Planck-Institut f\"ur Extraterrestrische Physik (MPE), Giessenbachstr. 1, D-85748 Garching, Germany}

\author[0000-0003-4264-3381]{Natascha ~M. F\"{o}rster Schreiber}
\affiliation{Max-Planck-Institut f\"ur Extraterrestrische Physik (MPE), Giessenbachstr. 1, D-85748 Garching, Germany}

\author[0000-0002-0108-4176]{Sedona ~H. Price}
\affiliation{Department of Physics and Astronomy and PITT PACC, University of Pittsburgh, Pittsburgh, PA 15260, USA}

\author[0000-0001-9773-7479]{Daizhong Liu}
\affiliation{Max-Planck-Institut f\"ur Extraterrestrische Physik (MPE), Giessenbachstr. 1, D-85748 Garching, Germany}
\affiliation{Purple Mountain Observatory, Chinese Academy of Sciences, 10 Yuanhua Road, Nanjing 210023, China}

\author[0000-0002-2767-9653]{Reinhard Genzel}
\affiliation{Max-Planck-Institut f\"ur Extraterrestrische Physik (MPE), Giessenbachstr. 1, D-85748 Garching, Germany}
\affiliation{Departments of Physics and Astronomy, University of California, Berkeley, CA 94720, USA}

\author[0000-0003-4949-7217]{Ric Davies}
\affiliation{Max-Planck-Institut f\"ur Extraterrestrische Physik (MPE), Giessenbachstr. 1, D-85748 Garching, Germany}

\author[0000-0002-1485-9401]{Linda ~J. Tacconi}
\affiliation{Max-Planck-Institut f\"ur Extraterrestrische Physik (MPE), Giessenbachstr. 1, D-85748 Garching, Germany}

\author[0000-0002-2125-4670]{Taro ~T. Shimizu}
\affiliation{Max-Planck-Institut f\"ur Extraterrestrische Physik (MPE), Giessenbachstr. 1, D-85748 Garching, Germany}

\author[0000-0003-1785-1357]{Amit Nestor Shachar}
\affiliation{School of Physics and Astronomy, Tel Aviv University, Tel Aviv 69978, Israel}

\author[0000-0001-6703-4676]{Juan ~M. Espejo Salcedo}
\affiliation{Max-Planck-Institut f\"ur Extraterrestrische Physik (MPE), Giessenbachstr. 1, D-85748 Garching, Germany}

\author[0009-0009-0472-6080]{Stavros Pastras}
\affiliation{Max-Planck-Institut f\"ur Extraterrestrische Physik (MPE), Giessenbachstr. 1, D-85748 Garching, Germany}

\author[0000-0003-3735-1931]{Stijn Wuyts}
\affiliation{Department of Physics, University of Bath, Claverton Down, Bath, BA2 7AY, UK}

\author[0000-0003-0291-9582]{Dieter Lutz}
\affiliation{Max-Planck-Institut f\"ur Extraterrestrische Physik (MPE), Giessenbachstr. 1, D-85748 Garching, Germany}

\author[0000-0002-7093-7355]{Alvio Renzini}
\affiliation{Osservatorio Astronomico di Padova, Vicolo dell’Osservatorio 5, Padova, I-35122, Italy}

\author[0000-0003-4891-0794]{Hannah \"Ubler}
\affiliation{Kavli Institute for Cosmology, University of Cambridge, Madingley Road, Cambridge, CB3 0HA, UK}
\affiliation{Cavendish Laboratory, University of Cambridge, 19 JJ Thomson Avenue, Cambridge CB3 0HE, UK}
\affiliation{Max-Planck-Institut f\"ur Extraterrestrische Physik (MPE), Giessenbachstr. 1, D-85748 Garching, Germany}

\author[0000-0002-2775-0595]{Rodrigo Herrera-Camus}
\affiliation{Departamento de Astronom\'{\i}a, Universidad de Concepción, Barrio Universitario, Concepción, Chile}

\author[0000-0001-5065-9530]{Amiel Sternberg}
\affiliation{School of Physics and Astronomy, Tel Aviv University, Tel Aviv 69978, Israel}
\affiliation{Max-Planck-Institut f\"ur Extraterrestrische Physik (MPE), Giessenbachstr. 1, D-85748 Garching, Germany}
\affiliation{Center for Computational Astrophysics, Flatiron Institute, 162 5th Avenue, New York, NY 10010, USA}

\begin{abstract}
Spatially-resolved emission line kinematics are invaluable to investigating fundamental galaxy properties and have become increasingly accessible for galaxies at $z$\,$\ga$\,0.5 through sensitive near-infrared imaging spectroscopy and millimeter interferometry. Kinematic modeling is at the core of the analysis and interpretation of such data sets,  which at high-$z$ present challenges due to lower signal-to-noise ratio (S/N) and resolution compared to data of local galaxies. 
We present and test the 3D fitting functionality of \dy, examining how well it recovers intrinsic disk rotation velocity and velocity dispersion, using a large suite of axisymmetric models, covering a range of galaxy properties and observational parameters typical of \znoon\ star-forming galaxies.
We also compare \dy's recovery performance to that of
two other commonly used codes, \gp\ and \bb, which we use in turn to create additional sets of models to benchmark \dy.
Over the ranges of S/N, resolution, mass, and velocity dispersion explored, the rotation velocity is accurately recovered by all tools.
The velocity dispersion is recovered well at high S/N, but the impact of methodology differences is more apparent. In particular,
template differences for parametric tools and S/N sensitivity for the non-parametric tool can lead to
differences up to a factor of 2. Our tests highlight and the importance of deep, high-resolution data and the need for careful consideration of: (1) the choice of priors (parametric approaches), (2) the masking (all approaches) and, more generally, evaluating the suitability of each approach to the specific data at hand.
This paper accompanies the public release of \dy.
\end{abstract}


\section{Introduction\label{sec:intro}}
Spatially-resolved kinematics provide fundamental insights into the nature, dynamical state, and mass assembly history of galaxies. Thanks to ever more powerful near-infrared/optical integral field unit (IFU) spectrographs on
4--8\,m-class telescopes and millimeter interferometers \citep{Bacon1995,Weitzel1996,Eisenhauer2003,Larkin2006,Sharples2013}, kinematics have become a widespread tool in galaxy evolution studies at redshifts $z$\,$>$\,0.5 \citep{Glazebrook2013,fs2020}. IFU studies, mainly targeting the \Ha\ rest-optical line, comprise the most comprehensive census of resolved
kinematics of massive star-forming galaxies (SFGs) at $z$\,$\sim$1$-$3.
These surveys cover well the ``main sequence'' (MS) of SFGs in stellar mass
\mstar\ vs.\ star formation rate SFR, which dominates the population and cosmic
star formation \citep[e.g.,][]{Rodighiero2011,Sargent2012,Madau2014}.
Resolved millimeter interferometric observations focused primarily on CO lines at $z$\,$\la$\,$4$ and on the bright \cii\,$\lambda$158\micron\ far-infrared line at $z$\,$\ga$\,$4$ in modest samples are now available for dynamical studies of the cold neutral gas.
With its NIRSpec instrument in IFU mode, the \textit{James Webb Space Telescope} (\textit{JWST}), has recently opened up the way to resolved \Ha\ kinematics mapping at $z$\,$\ga$\,$3$.

Such observations provide the most direct approach to probe the potential well and the physical processes that shape galaxies over time, including gas accretion, non-circular motions, galaxy interactions, mass and angular momentum transfer, and feedback from massive stars and active galactic nuclei (AGN). 
In this context, the gas velocity dispersion \sigo\ and rotational-to-dispersion support \vsig\ constitute important measures of disk structure and settling across cosmic time.  Despite increasing observational efforts, results remain mixed on the amount, evolution, and origin of gas dispersion and dynamical support of distant disks.

Part of the differences may be attributed to sample selection and tracer choices. A number of studies over the past couple of decades found increasing disk dispersions and decreasing \vsig\ towards higher redshift for samples probing mainly massive (stellar masses \mstar\,$\ga$\,$10^{10}$\msun) MS SFGs out to $z$\,$\sim$\,3, trends that have been interpreted in the framework of marginally unstable gas-rich disks given the increasing molecular gas fraction at earlier time \citep[see reviews by][and references therein]{Tacconi2020,fs2020}. Other studies reported dynamically fairly cold, regular disk rotation including among infrared-luminous dusty SFGs and at $z$\,$\ga$\,4, either unlensed or strongly gravitationally lensed \citep[e.g.,][]{Hodge2012,Sharda2019,Fraternali2021,Lelli2021,Rizzo2021,Rizzo2022,Tsukui2021}. At any given redshift, the scatter in \sigo\ and \vsig\ is substantial, even among the best data sets of MS SFGs \citep[e.g.,][]{Uebler2019}. Although part of the spread in \vsig\ at fixed redshift appears to be explained by a trend with galaxy stellar mass, it is less clear for \sigo\ \citep[e.g.][]{Kassin2012,Wisnioski2015,Simons2018,Johnson2018,Parlanti2023,Rowland2024}. Furthermore, there is a potentially inherent difference between the gas kinematics traced by warm ionized gas and colder atomic and molecular gas. Larger samples with high-resolution, high S/N data of multiple tracers for the same galaxies will be essential to establish whether, and how much, kinematic properties depend on the interstellar medium (ISM) phase \citep[contrast][with \citealt{Uebler2018,Genzel2023}]{Liu2023,Parlanti2023,Parlanti2024,Rizzo2023}.

Compounding the sample and tracer
differences described above, the observational picture is further complicated by the use of various modeling approaches among studies. In some cases this has even led to discrepant results from the same data sets (e.g., compare \citealt{Tadaki2018,Tadaki2020} to \citealt{Sharda2019,Roman-Oliveira2023}; see also \citealt{Lelli2023}). Because kinematics modeling serves as the foundation for linking observables to physical properties, tools should ideally rely on full spatial and 
velocity forward modeling that accounts for the appreciable beam-smearing
in high-redshift data and the limited S/N due to cosmological surface
brightness dimming. Several packages have been developed to this aim, including the parametric modeling codes \texttt{Dysmal\/}/\dy\ \citep[e.g.,][]{Davies2004a,Davies2004b,Davies2011,Cresci2009,Wuyts2016,Lang2017,Price2021}, \gp\ \citep[][]{Bouche2015,Bouche2022}, and the non-parametric modeling code \bb\ \citep[][]{DiTeodoro2015}\footnote{Other 3D-fitting tools exist, such as \texttt{TiRiFiC} \citep{Jozsa2007,Kamphuis2015}, \texttt{KinMS\/} \citep[][]{Davis2013,Davis2017}, GPU-accelerated \texttt{GBKFIT} \citep{Bekiaris2016}, \texttt{BLOBBY3D\/} \citep[][]{Varidel2019}, and \texttt{qubefit\/} \citep[][]{Neeleman2021}.}.
Parametric models are built around analytic descriptions of the mass distribution and/or the rotation and dispersion velocities \citep[e.g.,][]{Courteau1997},
making them inherently less sensitive to S/N than non-parametric models.
Moreover, parametric modeling that starts from mass models enables an immediate characterization of the mass profile and quantifies the amount of gas involved in out- or in-flowing motions.
On the other hand, non-parametric models have the advantage of more flexibility in a system's
description, which may deviate from common functional forms due to non-axisymmetric
features and kinematic perturbations 
\citep[e.g.,][]{Rogstad1974,Begeman1989,Sancisi2004}.
Each code has its strengths and has been internally benchmarked against observations of local galaxies, mock analytical models, and/or numerical simulations. 

All these tools, however, share a common advantage: they operate in full 3D space to generate models, thereby accounting for beam smearing and projection effects in the best possible way.

In this work, we expand on \citet{Price2021} by testing the \dy's performance in
fitting in
3D, accompanied by the first public release of the code\footnote{Available at \url{https://www.mpe.mpg.de/resources/IR/DYSMALPY/index.html}}.
We additionally benchmarked \dy\ against two widely-used modeling tools: the parametric \gp\ and the non-parametric \bb.
For this purpose, 
we employ an extensive set of synthetic galaxies with known input properties subjected to various realistic observational conditions. 
The mock galaxies consist of intentionally
simple axisymmetric disk models but with an empirically motivated range of properties.
The systematic inter-comparison expands on the validation tests for the individual codes presented by \citet[][see also \citealt{Davies2011}]{Price2021} for \dy, \citet[][]{Bouche2015} for \gp, and \citet{DiTeodoro2015} for \bb. We use modeling setups as consistently as possible between the tools
and discuss the impact of our adopted choices vs.\ recommended or widely used ones when they differ. 
We focus on fits performed in 3D for IFU and interferometric observations. 
\gp\ was designed to only fit in 3D, while \bb\ could also fit a 2D velocity field through its \texttt{2DFIT} task. \dy\ has mainly been applied to fit data in 1D or 2D (e.g., major axis kinematic profiles, moment maps), but by its 3D model construction was easily adapted to fit data cubes \citep[][]{Price2021}.

The paper is organized as follows. In Sec.~\ref{sec:methods}, we summarize the relevant features of the three fitting codes and describe the main setups employed in our analysis. In Sec.~\ref{sec:sample}, we describe the construction of the baseline mock data cubes used in this study. In Sec.~\ref{sec:results}, we compare the fitting results between the codes with the baseline setups and mock data sets and explore more deeply the impact of parametrization and treatment of S/N. 
In Sec.~\ref{sec:discussion}, we discuss the potential implications in studies of high-$z$ kinematics. 
In Sec.~\ref{sec:conclusion}, we summarize our findings.
Throughout, we assume a $\Lambda$-dominated cosmology with $H_{0} = 70~{\rm km\,s^{-1}\,Mpc^{-1}}$, $\Omega_{\rm m} = 0.3$, and $\Omega_{\Lambda} = 0.7$. For this cosmology, 1\arcsec\ corresponds to 8.37\,kpc at $z$\,$=$\,2.

\begin{deluxetable*}{llllll}
\centering
\tablecaption{Comparisons between the three software used in this study.}
\tablewidth{0pc}
\tablehead{
\colhead{Software} &  \colhead{Parametric} & \colhead{$V_{\rm rot}$} & \colhead{$\sigma$} &\colhead{Optimizer} &\colhead{References}}
\startdata 
\dy\ & Yes & axisymmetric mass model(s) & turbulence & \texttt{MPFIT}/MCMC & \cite{Price2021} and references therein \\
\gp\ & Yes & analytic functions & turbulence+thick disk+mixing & MCMC & \cite{Bouche2015} \\
\bb\ & No &free-form&  free-form&  Nelder-Mead & \cite{DiTeodoro2015}
\enddata
\tablecomments{\bb\ parameterizes the scale-height as Gaussian, $\sech^2$ or constant}
\label{tab:methods_compare}
\end{deluxetable*}

\section{Kinematic Modeling Codes and Setups}\label{sec:methods}
Complete details of \dy, \gp, and \bb\ can be found in the references below. For the comparisons, we focus on the recovery of the main kinematic properties: the intrinsic rotation velocity \vrot\ and local disk velocity dispersion $\sigma$. These properties are the ones that are most directly comparable between the codes.
As we are primarily interested in exploring the impact of model parametrization and parametric vs.\ non-parametric approaches, we maximize consistency as follows: (i) we perform the fitting in 3D space, (ii) within the architecture of the codes, we use prior on parameters, minimization algorithms, masking and weighting schemes that are as similar as possible,
and (iii) we keep the number of free parameters to a minimum.

For high-$z$ observations, beam smearing is important, S/N is modest, and the morphology of the emission line tracing kinematics can be prominently irregular due to spatial variations in dust extinction, stellar population properties, and gas distributions. 
To mitigate these challenges, geometric parameters are typically fixed or tightly constrained through narrow priors and the
morphology is not always directly used in the fitting (especially for parametric codes that typically assume smooth axisymmetric models).
This also helps to reduce well-known degeneracies (e.g., between mass and inclination), which are compounded by low resolution. 
Therefore, in running all codes, we fixed the center, size parameter(s), inclination (\incl) along the line-of-sight, and position angle (\pa) of the line of nodes on the sky plane. 

This intentional choice of a few degrees of freedom is an ideal case; if other parameters are fitted simultaneously (with or without priors), 
the outcome of all codes will be less accurate. 
We have verified this by additionally allowing $i$ and $R_{\rm e,d}$ to vary freely (for \dy\ and \gp), within ranges of $\pm15^\circ$ and $\pm20\%$, respectively, with initial guesses derived from the line intensity map. We find that there is no substantial improvement in fitting accuracy, and in some cases, there are more systematic offsets from the intrinsic values, while also resulting in reduced precision. More extensive exploration will be beneficial, but we consider it as a future study.


Table~\ref{tab:methods_compare} summarizes the key differences between the three packages, and Table~\ref{tab:prior_table} in Appendix~\ref{app:mocksetup} lists
 the setups and parameters employed for each of them. For \dy\ and \bb, we initially follow their built-in masking routine, with specific settings listed in Table~\ref{tab:prior_table}. Later in Sec.~\ref{subsec:masking_discussion}, we swap the masks between \dy\ and \bb, allowing us to explore the effects of using a common masking approach and to assess the impact of masking on the performance of each individual code. 

\subsection{DysmalPy}\label{subsec:dpy}

\subsubsection{Main Features}
The \texttt{Python}-based \dy, or its parent \texttt{IDL} version \texttt{Dysmal}, is a versatile forward-modeling tool based on multi-component mass models with a long history of development. It has been employed in near-IR/optical IFU and millimeter interferometric studies of disk galaxies at high-$z$ \citep[e.g.,][]{Genzel2006,Genzel2011,Genzel2014b,Genzel2017,Genzel2020,Genzel2023,Cresci2009,Wuyts2016,Burkert2016,Lang2017,Tadaki2017,Uebler2018,Uebler2019,Uebler2021,uebler2024b,Price2021,Herrera-Camus2022,Nestor2023}, and of local disks \citep[e.g.,][]{Davies2009,Davies2014,Sani2012,MuellerSanchez2013,Lin2016}. 
\texttt{Dysmal} has been tested, especially regarding the recovery of velocity dispersion by \citet{Davies2011}. \citet{Wuyts2016}, \citet{Burkert2016}, and \citet{Lang2017}  significantly expanded it to incorporate an improved treatment of the effects of disk finite thickness and pressure gradients and to add DM halos to the family of possible mass components. 
\citet{Uebler2018} adopted the \texttt{Python} version, introducing Markov chain Monte Carlo (MCMC) posterior sampling in addition to the original least-squares minimization.
\citet{Price2021} presented a substantial upgrade, including a wide set of DM halo parametrizations, radial flow motions (to represent, for instance, bar-induced inflows or feedback-driven outflows), and the ability to tie model component parameters and fit in 3D space.

A detailed description of \dy's model construction and optimization is given by \citet{Price2021}; we focus here on the aspects that are relevant to the present paper.
\dy\ is based on a mass distribution from which the kinematics are 
computed. Mass components are defined by azimuthally symmetric parametric functions,
with flexibility in terms of their number and mass-to-light ratio (M/L), as well as a
common center and inclination. Baryonic components are set up in the disk framework,
whereas the DM halo is spherically symmetric
(and contributes no light by definition).
The total circular velocity \vcirc\ in the mid-plane is obtained from those of
the individual mass components (summed in quadrature), computed in the spherical
approximation ($v_{\rm circ}^{2}(r) = GM(<r)/r$, where $G$ is the gravitational
constant and $M(<r)$ is the enclosed mass), with options to account for disk
geometry with finite thickness (via the prescriptions of \citealt{Noordermeer2008}
for oblate flattened spheroids) as well as the effects of pressure gradients and
DM halo adiabatic contraction \citep[following the formulations of][]{Burkert2010}.
The 3D model is cylindrical, with the mid-plane radial kinematics and structure
assigned to all vertical layers, with a Gaussian light distribution of standard deviation $h_{z}$.

The velocity dispersion is assumed to be locally isotropic and radially uniform, representing a dominant turbulence term \sigo.  
This choice is motivated by the lack of evidence for clear trends with inclination and radius in high resolution, high S/N IFU observations of extended star-forming disks at \znoon. Typical inferred values are significantly in excess of predicted values at large radii for (exponential) disks in hydrostatic equilibrium with constant vertical scale height $h_{z}$, $\sigma_{\rm d}(R) = h_{z} V_{\rm rot}(R) / R$ (e.g., \citealt{Genzel2011,Uebler2019,Liu2023}; but see also, e.g., \citealt{Rizzo2020,Rizzo2021} and \citealt{Lelli2021} for contrasting views).

The equilibrium assumption of \dy\ is admittedly simple and cannot account for merger perturbations. However, the merger fraction is $\lesssim$20\% at cosmic noon \citep{Madau2014}. 
On the other hand, disk accretion and violent disk instability can indeed perturb the system \citep{Dekel2022}, potentially impacting kinematic measurements. But 
(i) dynamics from stars and gas at cosmic noon \citep{uebler2024} are in good agreement, suggesting that in general the impact of such perturbations is not drastic; (ii) signatures of large dynamical perturbations would be evident in residuals (along with  observed velocity fields deviating substantially from spider diagrams), providing sanity  checks and ways to evaluate whether an object is well described by the equilibrium  assumption or not.

With the ingredients above,  \dy\ generates the intrinsic composite model as a 4D hypercube, summing up the components accounting for projection according to the inclination, \pa, and relative flux weighting. Each cell of the hypercube contains the total model flux in the ``sky'' coordinates ($x_{\rm sky}$, $y_{\rm sky}$, $z_{\rm sky}$) and its full line-of-sight velocity distribution, which is then collapsed along $z_{\rm sky}$ and convolved with a 3D kernel folding in the spatial point spread function (PSF) and spectral line spread function (LSF). This procedure accounts for beam smearing, velocity resolution, and broadening of the line-of-sight velocity distribution due to projection effects. The convolved data cube is the basis for fitting the observations directly in 3D or 1D/2D by applying identical profile and map extraction methods between the model and data to ensure full consistency in their comparisons. 
The fitting can be performed through least-squares minimization via the Levenberg-Markwardt iterative search technique (using the routine \texttt{MPFIT}; \citealt{Markwardt2009}), or in a Bayesian framework through affine invariant MCMC parameter space exploration (using the \texttt{emcee} implementation of \citet{Foreman-Mackey2016})
\footnote{\texttt{dynesty} \citep{koposov2023}, a dynamic nested sampling algorithm as an alternative Bayesian inference method to \texttt{emcee}, is now also incorporated in \dy\ version 2.0.0.}.
Masking and weighting schemes can be applied to exclude bad pixels, low S/N data, or any region as needed by a specific application and treat the impact of oversampling if relevant. In 2D or 1D, \dy\ can constrain the free parameters using either the observed flux, velocity, and velocity dispersion or solely the kinematics.

\subsubsection{Adopted Setups}
In this work, we use version 1.8.2 of \dy\footnote{The publicly available version is 2.0.0, but the main functionalities are the same, and the minor differences and improvements have no impact on our results.}. 
We fit models consisting of a baryonic disk, bulge, and DM halo. While this choice is motivated by the set of mock galaxies created for our tests (described in Sec.~\ref{sec:sample}), it also allows us to cover a wide enough range of realistic rotation curve shapes. To match as closely as possible the workings of \gp\ and \bb, we keep only a minimum of free parameters in the \dy\ fitting: the total baryonic mass $M_{\rm bar}$, the (turbulent) disk velocity dispersion \sigo, and the DM mass fraction within the disk's effective radius \fdmre.

The disk component is parametrized as a S\'ersic profile of index $n_{\rm d}$\,$=$\,$1$ (exponential disk) adopted for all fits, with flattening $q_{\rm d}$ and effective radius $R_{\rm e,d}$ (henceforth \reff) fixed to the particular values of the fitted mock model. Similarly, the bulge component is set with a S\'ersic profile of
$n_{\rm b}$\,$=$\,$4$  (de Vaucouleurs) and flattening $q_{\rm b}$\,$=$\,$1$ identical for all fits,
and $R_{\rm e,b}$ fixed to those of the mock galaxy that is modeled.

For the baryonic components, the mass is treated as a flattened 3D deprojected S\'ersic profile characterized by \reff\ and $n_d$.
Their total mass is left free while their relative masses are tied through
the bulge-to-total ratio (B/T) fixed to the mock model value.
The two-parameter NFW \citep*{Navarro1996} DM halo profile option is used, with
concentration parameter $c$ set to that of the input mock model, and virial
mass $M_{\rm vir}$ tied to the variable \fdmre; we do not apply adiabatic contraction in our fits.
The inclination is also fixed to the value of the mock model. 

We start from the \citet{Noordermeer2008} parametrization with thick disk geometry and apply the \citet{Burkert2010} corrections
to \vcirc\ to account for pressure support.
We use the option for a self-gravitating exponential disk with
constant velocity dispersion, such that
\begin{equation}
    V_{\rm rot}^2(R) = V_{\rm circ}^2(R) - 3.36\sigma_{0}^2(R/R_{\rm e}).
    \label{eqn:dy_pressure_support}
\end{equation}

The vertical Gaussian light weighting of the disk or bulge is controlled by the scale height $h_z$ in {\dy} and is through the inverse of $q$:
\begin{equation}
    h_z \equiv R_{\rm e}\cdot q/\sqrt{2\ln2}.   
    \label{eqn:scale_height}
\end{equation}
The denominator accounts for the conversion of full width at half maximum (FWHM) to dispersion for a Gaussian thickness profile. $h_z$ is fixed to the true value of the mock model. 
The rotation curve is always extracted from the midplane for all radial and vertical ($z$-direction) positions, with the vertical Gaussian light weighting.

Given the simple axisymmetric light distribution of the mock models, we fit in 3D to account simultaneously for the morphology and
kinematics. We use the least-squares minimization (\texttt{MPFIT}) option. This choice is adequate given the small number of free parameters ($N_{\rm free\,params}=3$).
\citet{Price2021} showed a very good agreement between results obtained with
least-squares minimization and the Bayesian MCMC approach in {\dy}.
We do not apply weighting but use masking to exclude low S/N data in the fits.
Specifically, we mask out entire spaxels with insufficient integrated
line flux S/N (S/N\,$<$\,3) to avoid fitting overly masked line profiles.

\subsection{\gp}\label{subsec:gp}

\subsubsection{Main Features}
\gp\ \citep{Bouche2015} is a \texttt{Python}-based parametric forward-modeling
tool in a Bayesian framework. It was the first such public tool
available\footnote{\url{https://galpak3d.univ-lyon1.fr/index.html}}
that was designed specifically for fitting disk models directly to 3D data cubes of high-redshift galaxies and has been applied in a variety of studies from optical to mm wavelengths
\citep[e.g.,][]{Peroux2013,Bacon2015,Contini2016,Mason2017,Girard2018,Tadaki2018,Sharon2019,Zabl2019,Zabl2020,Zabl2021,Bouche2022,Hogan2021,Huang2023,Puglisi2023}.
The original conceptual approach differs from \dy\ in that the models are defined
by the light distribution and kinematics, returning best-fit parameters and rotation
curves that can be used for subsequent mass decomposition modeling outside of \gp.
The code was developed with an emphasis on applications to distant low-mass
galaxies and extensively tested by \citet{Bouche2015,Bouche2021} using parametric
axisymmetric  models and numerical hydrodynamical simulations.
\citet{Bouche2021,Bouche2022} upgraded the code notably to expand the set of rotation curve parametrizations and fitting algorithms and augment its capabilities to disk-halo mass decomposition in which a multi-component mass model is first generated and sets the kinematics (more similarly to \dy).
We focus below on \gp's baseline framework where kinematics define the models.

\gp\ first creates a 2D model light distribution following a choice
of S\'ersic radial profile parametrizations in the disk plane and vertical Gaussian,
exponential, or $\sech^2$ profiles with thickness following Eqn.~\ref{eqn:scale_height}.  
Circular velocity cubes are generated that contain the velocity components in
the disk plane from parametric rotation curves and propagating them to the vertical
layers. Several functional forms are implemented, including notably an $\arctan$,
inverted exponential, or $\tanh$ profile, motivated by typical shapes of
local disk rotation curves, and with turnover radius $R_{\rm t}$ and maximum intrinsic
rotation velocity $V_{\rm max}$ as parameters\footnote{For completeness, the publicly available version also has an option to calculate the circular velocity in the spherical 
approximation using the light distribution as a proxy for mass.}.
The model is then rotated according to \incl\ and \pa\ to create a cube in
projected sky plane coordinates and in wavelength, using intermediate 2D projected flux, flux-weighted mean velocity along the line-of-sight, and total line-of-sight
velocity dispersion $\sigma_{\rm tot}$ maps.
The latter combines (i) the contribution from disk self-gravity
$\sigma_{\rm d}$\,$=$\,$h_{z}\,V(r)/r$,
(ii) a broadening term $\sigma_{\rm s}$ due to mixing of velocities along the
line of sight for a thick disk and computed as the flux-weighted variance of
the projected circular velocities along each sightline,
(iii) and an isotropic and spatially constant term \sigo\ to capture additional
turbulence.
The three terms are added in quadrature, 
\begin{equation}
    \sigma_{\rm tot}^2(R)  = \sigma_{\rm d}^{2}(R) + \sigma_{\rm s}^{2}(R) + \sigma_{0}^2.
    \label{eqn:gp_disprof}
\end{equation}
Throughout, for consistent comparison with other codes, we adopt the $\sigma$ value at \reff\ of the above total dispersion profile for \gp.
The rotated data cube is then convolved with the PSF and LSF.
Fitting in 3D is performed in a Bayesian framework, with a choice of several MCMC samplers. No masking nor weighting schemes are performed by \gp\ \citep{Bouche2021}.

Key differences between \gp\ and \dy\ model construction that we exploit for our
analysis lie in the explicit parametrization of rotation curves in \gp\ and the
treatment of velocity dispersion. The former implies different families of
intrinsic rotation curve shapes. We note that the options implemented implicitly 
account for the effects of a DM halo, as captured by the asymptotically flat 
behavior of the functions to large radii.

For the velocity dispersion, \dy\ explicitly considers the full velocity distribution of each spatial
coordinate from 4D space, encoding both circular velocity and (turbulent) dispersion
before projection of the data in 3D space. \gp\ approximates line broadening due
to projection solely from the circular velocities, adds self-gravity and turbulence
terms assuming isotropic local dispersion, and then generates the line-of-sight
velocity distribution. This procedure speeds up the code but assumes Gaussianity.
\dy's approach preserves higher-order moments. In practice, at modest to low
S/N and for reasonably regular morphologies and kinematics, higher-order moments
are difficult to discern, and the mixing term due to projection is usually small
compared to the other terms, so these aspects have little impact on our analysis.
More importantly, in the regime of low turbulence, the model dispersion profile
in \gp\ exhibits an appreciable radial dependence from $\sigma_{\rm d}$ that is
absent from current \dy\ models.

\subsubsection{Adopted Setups}
We use the publicly available version 1.32.0 of \gp.
We employ an exponential disk profile (S\'ersic model with
$n$\,$=$\,$1$) with Gaussian vertical distribution of thickness tied to the disk size by Eqn.~\ref{eqn:scale_height}, identical to the disk component assumed for the \dy\ modeling.
We use an $\arctan$ rotation curve shape with turnover radius also tied to the galaxy size via $R_{\rm t}$\,$=$\,$0.25\,R_{\rm e}$. The free parameters in our fits are the maximum rotation velocity $V_{\rm max}$ and 
the intrinsic turbulence term \sigo, assuming flat bounded priors. Other \gp\ input parameters are the center, \reff, \incl, and \pa\ fixed to the values of each mock galaxy modeled. We adopt the default MCMC method in {\gp}, which uses a Metropolis--Hasting (MH) algorithm and a Cauchy (or Lorentzian) proposal distribution that converges faster than a Gaussian distribution thanks to its broader wings. The maximum iteration is set to 3000, which is sufficient to pass the burn-in phase; tests with a subset of our mock sample show that increasing the number of iterations does not significantly change the results.

\subsection{\bb}\label{subsec:bb}

\subsubsection{Main Features}

\bb\footnote{Available at \url{https://editeodoro.github.io/Bbarolo/}}
\citep{DiTeodoro2015} is a non-parametric modeling tool that extends the ``tilted-ring''
approach from its classical 2D applications in modeling high-resolution velocity fields 
\citep[e.g.,][]{Rogstad1974,vanAlbada1985,Begeman1989,vdH1992} to fitting full 3D
observations of disk-like systems. It was developed for a wide range of applications to emission-line data cubes, with a special emphasis on lower-resolution data. It has been widely used in studies of galaxy kinematics at high-$z$
\citep[e.g.,][]{DiTeodoro2016,Fan2019,Loiacono2019,Bischetti2021,Fraternali2021,Fujimoto2021,Fujimoto2024,Jones2021,Sharma2021,Sharma2022,Sharma2023,Hogan2022,Lelli2023,Pope2023,Posses2023,Rizzo2023,Roman-Oliveira2023}
and low--$z$ \citep[e.g.,][]{Iorio2017,MancenaPina2019,BewketuBelete2021,Deg2022,Perna2022,Su2022,Biswas2023,Cao2023}.
Continuous developments and testings have been made since the original tool release, notably to add the \texttt{pyBBarolo} \texttt{Python} wrapper running the \texttt{C++} core code, to incorporate the option of accounting for pressure support on rotational velocities \citep[][available since version 1.3]{Iorio2017}, and to improve estimates of geometric parameters \citep[via the \texttt{CANNUBI}\footnote{\url{https://www.filippofraternali.com/cannubi}} \texttt{Python} script;][]{Roman-Oliveira2023}.

\bb\ constructs a 3D disk model as a series
of concentric rings characterized by their radius and width, spatial center and
systemic velocity $V_{\rm sys}$, inclination and \pa, rotational velocity $V_{\rm rot}$ and
velocity dispersion $\sigma$, face-on gas surface density, and scale-height
$z_{0}$.  All details and extensive testing of the impact of spatial and spectral
resolution, inclination, and S/N with data of local galaxies and mock models are
presented by \citet{DiTeodoro2015}. Complementary tests using zoom-in numerical
cosmological simulations are presented by \citet{Rizzo2022}. Our analysis extends
these tests mainly by expanding the explored space to regimes of higher disk
velocity dispersions.

In brief, the model generation in \bb\ derives from the \texttt{GALMOD} routine
\citep{Sicking1997} incorporated in the \texttt{GIPSY} software environment
\citep{vdH1992}. Each ring is randomly populated via a Monte Carlo procedure by
``clouds'' represented as Gaussian point sources, drawn from uniform distributions
in radius (within the ring width) and azimuth, and non-uniform vertical distributions
(Gaussian, $\sech^2$, exponential, Lorentzian, or top-hat). Each ring is rotated
according to its \incl\ and \pa.  The observed velocity distribution along the
line-of-sight is computed from the combination of systemic, rotational, and random
motions, splitting the clouds at each location into sub-clouds 
distributed around the average velocity according to the sum-squared of dispersions
accounting for intrinsic random motions and the LSF. The resulting model rings are
individually convolved with a 2D Gaussian PSF and normalized such that the full
model surface mass density matches the observed distribution (in column density,
or light as a proxy) either on a spaxel-by-spaxel basis or to the azimuthally-averaged
ring flux. 
Normalization can be disabled, allowing either a predefined functional form for the surface density distribution to be provided or leaving it free to be fitted along with the other parameters.
The effects of pressure support, if chosen to be accounted for,
are computed following the classical asymmetric drift formulation
\citep[e.g.,][]{Oh2015,Iorio2017}.

Fitting in 3D is performed ring-by-ring via the Nelder-Mead multidimensional
downhill simplex solver for non-analytic functions \citep{NelderMead1965}.
At each ring, the sum of the residuals $F$ over individual valid pixels
is passed to the minimization algorithm. Valid pixels in 3D are identified based on the source
finding results through the \texttt{DUCHAMP} algorithm \citep{Whiting2012}, and as those exceeding a flux threshold defined by the root-mean-square (rms) noise of the cube with or without prior smoothing.  

There is an option to let \bb\ automatically estimate initial guesses and to perform regularization to avoid unphysical discontinuities in the returned best-fit radial profiles when geometrical
parameters (\incl\ and \pa) are left free. Residuals between model $M$ and data $D$ values can be computed as a pseudo-$\chi^2$ $(M-D)^2/\sqrt{D}$, the absolute difference
$|M-D|$, or as $|M-D|/(M+D)$ to upweight fainter emission regions.  The minimized
quantity $F$ scales the residuals by $w(\theta) = |\cos(\theta)|^m$, where $m$ can be $0$ (unweighted sum of residuals), $1$ or $2$ (giving increasing weight to regions along
the kinematic major axis defined as $\theta = 0$).

By construction, the tilted-ring approach and the 3D implementation of \bb\ leave many degrees of freedom. This allows, for instance, capturing non-axisymmetric features
such as thin disk warping or other local irregularities in high-resolution rotation
curves. As stressed by \citet{DiTeodoro2015}, this flexibility must be used with
caution depending on the resolution, inclination, and S/N.
In applications to high-$z$ data, fixing global parameters (such as center and systemic velocity) and adopting radially constant inclination and \pa\ may be necessary.

\subsubsection{Adopted Setups}
We use version 1.6 of \bb\
and perform the 3D fitting through the \texttt{3DFIT} task.
To keep the setup as uniform as possible between the modeling tools considered in
this paper, and similar to high-$z$ studies using \bb, we fix the rings
to identical centers, $V_{\rm sys}$, $i$, \pa, and Gaussian vertical surface
density distribution as in Eqn.~\ref{eqn:scale_height}, according to the values of each modeled mock galaxy.
We set the radial bin width and separation to one-third of the beam FWHM size
(corresponding to a physical scale of $0.228$\,kpc on average for our mock galaxies),
and adopt local flux normalization (i.e., on a pixel-per-pixel basis).
We verified that varying the bin widths to one-half to full-size of the beam 
has no significant statistical effect on the results, consistent with \citet{Varidel2019}.
\bb\ assumes the velocity dispersion within each ring to be isotropic.
For the baseline runs, we choose a masking threshold of S/N $= 3$ from unsmoothed data,
and uniformly weighted pseudo-$\chi^2$ residuals as the closest analogs to the
procedures in \dy\ and \gp, and discuss the impact of these choices in detail
in Sec.~\ref{sec:results}.
The rings $V_{\rm rot}$ and dispersion $\sigma$ are left free to vary within bounded intervals ($[0,400]{\rm\,km\,s^{-1}}$ and $[0,150]{\rm\,km\,s^{-1}}$, respectively). 
We do not employ the asymmetric drift correction option because our focus 
 is on comparing the recovery of the rotation velocity $V_{\rm rot}$ (corrected for inclination and resolution), not the circular velocity $V_{\rm circ}$.

 \section{Mock galaxies set} \label{sec:sample}
\begin{deluxetable}{llc}
\centering
\tablecaption{Range of parameters of the mock galaxy models}
\tablewidth{0pc}
\tablehead{
\colhead{Parameter} & \colhead{[$5^{\rm th}$,$95^{\rm th}$] percentile}  &\colhead{Constrained by}}
\startdata 
$z$ & $[0.72, 2.43]$ & RC100 \\
$\log{( M_*/ M_\odot)}$ & $[9.9, 11.1]$ & RC100\\
$R_{\rm e,d}$ [kpc]& $[3.2, 9.2]$& \citet{vanderWel2014}\\
$R_{\rm e,b}$ [kpc]& $[0.5, 1.1]$& \citet{Lang2014}\\
B/T & $[0.08, 0.53]$& \citet{Lang2014}\\
$\sigma_0$ [${\rm km\,s^{-1}}$]& $[16.2, 88.8]$& RC100\\
SFR [$M_\odot {\,\rm yr^{-1}}$] & $[7, 121]$&\citet{Speagle2014}\\
$\log{( M_{\rm gas}/ M_\odot)}$ & $[10.2, 11.4]$ & RC100\\
$f_{\rm gas}$ & $[0.31, 0.63]$ & RC100\\
$i$ [deg] & $[28.6, 75.0]$& RC100 \\
$\log{( M_{\rm vir}/ M_\odot)}$ & $[11.5, 12.7]$& \citet{Moster2018}\\
$c$ & [3.8, 7.1]& \citet{Dutton2014}\\
$\langle S/N(<R_e)\rangle$ & $[2, 21]$& RC100\\
PSF FWHM [$\arcsec$] & [0.2, 1.2] & RC100\\
$R_{\rm e}$/beam$_{\rm HWHM}$ & $[1.3, 5.4]$& RC100
\enddata
\tablecomments{PA and LSF are kept constant at $90^\circ$ and $40{\,\rm km\,s^{-1}}$, respectively. Other parameters are also held fixed.}
\label{tab:param_table}
\end{deluxetable}

Our primary goal is to assess the 3D self-recovery performance of \dy. We then take it a step further by comparing \dy's performance against other popular 3D modelling tools, which are \gp\ (parametric) and \bb\ (non-parametric), to understand the factors that lead to any differences between the fitting results.  We thus employ a baseline set of analytical axisymmetric model disk galaxies, with exact knowledge of the intrinsic kinematic parameters of interest, \vrot\ and \sigo, created by \dy.
We also create variants of this suite to explore the effects of
irregularities in the light distribution and of different families of rotation and
dispersion profiles. 

In parametric modeling, a mismatch between the assumed model and reality is inevitable. Given that \gp\ and \dy\ employ different templates for $V_{\rm rot}(R)$ and $\sigma(R)$, we also create a subset of mock models using \gp\ in Sec.~\ref{sec:gpmock} 
to investigate the impact of template mismatch on \dy. 
In contrast, the non-parametric \bb\ should not be restricted by specific templates. Our comparison of 1000 face-on mock cubes generated by \texttt{GALMOD} (the core routine of \bb, see Sec.~\ref{subsec:bb}) and \dy, sharing the same $V_{\rm rot}(R)$ and $\sigma(R)$ profiles, 
reveals that the differences between the two are negligible within $1.5$\reff\ ($\lesssim5\%$), with discrepancies primarily attributed to numerical noise. Nevertheless, to complete the comparison, we repeat the same exercise with mock cubes generated by \bb\ in Sec.~\ref{sec:bbmock}.

In the following Sections, we outline the setup for generating the mock models using \dy, \gp\ and \bb. The number of mock galaxies generated exceeds the currently available observational data with comparable properties.
It is clear that using models ``blindly" on a large set of samples can be problematic. In reality, results should be examined critically on an individual basis.
Initially, we began with a limited dataset but soon discovered that it was insufficient to properly identify systematic behaviours. 
To address this limitation, we expanded our mock sets while maintaining a minimal number of fitted parameters. 
We prioritized fixing parameters that are known to be observationally uncertain and have a significant impact on the outcome, such as the dynamical center, \reff, PA, and inclination, as detailed in Sec.~\ref{sec:methods}.

\subsection{Baseline Mock Models with \dy}\label{sec:dymock}
To explore a realistic range of disk properties and observational parameters and to ensure a sufficiently large mock data set to identify statistical trends in the recovery analysis, we build a baseline set with 9000 mock galaxies guided by the
properties of 100 \znoon\ MS SFGs, the ``RC100'' sample discussed by \citet{Nestor2023}.  
This sample has high-quality 3D kinematics from deep observations
(median on-source integration time of $10.7$\,hr) with typical S/N of 10 per pixel in the brightest channel averaged within \reff, and FWHM
angular resolution from $0\farcs2$ up to $1\farcs2$ ($5^{\rm th}$ and $95^{\rm th}$ percentile).
RC100 is drawn from the large parent sample of the KMOS$^{\rm 3D}$\,$+$\,SINS/zC-SINF
near-infrared IFU surveys targeting \Ha\ emission and the PHIBSS$+$NOEMA$^{\rm 3D}$
millimeter interferometric surveys of CO emission, totaling $\sim$800 galaxies that
probe well the massive SFG population at 0.6\,$<$\,$z$\,$<$2.6 over nearly two orders
of magnitude in stellar mass and SFR
\citep[][]{nmfs2009,nmfs2018,Mancini2011,Tacconi2013,Tacconi2018,Wisnioski2015,Wisnioski2019,Freundlich2019}.

The defining properties of the mock galaxy population are the stellar mass ($M_*$) and
redshift ($z$), from which all other physical properties are obtained via scaling relations
and accounting for their scatter in drawing values at fixed \mstar\ and $z$.
We use \dy\ to create the model data cubes, constructing each galaxy as baryonic
thick disk$+$bulge with total mass $M_{\rm bar} = M_{\star} + M_{\rm gas}$ embedded
in a spherical \citet*{Navarro1996} DM halo.  
Table~\ref{tab:param_table} lists the relevant parameters for the mock galaxies' construction, and Appendix~\ref{app:mockparams}
illustrates the match to the RC100 distributions in the main parameters.

We randomly draw 9000 times from the \mstar\ and redshift distributions of RC100,
split equally between the redshift ranges $z$\,$=$\,$[0.6,1.1]$, $[1.15,1.8]$, and
$[1.9,2.6]$ (for which \Ha\ falls in the $YJ$, $H$, and $K$ near-IR atmospheric
bands).  We set the SFR and gas-to-baryonic mass fraction \fgas\ based on the relationships for SFR($M_{*},z$) from \citet{Speagle2014} and
$f_{\rm gas}(M_{*},z,{\rm SFR})$ from \citet{Tacconi2020}.
The DM halo virial mass and concentration are derived from the \mstar\,$-$\,\mvir\
and $c$(\mvir,$z$) relationships of \citet{Moster2018} and \citet{Dutton2014},
respectively.  The effective radius of the disk is taken from the stellar mass-size relation of \citet{vanderWel2014}.  
The (stellar) bulge mass is assigned following
the B/T ratios relation of \citet{Lang2014}. 
The bulge effective radius $R_{\rm e,b}$ is fixed at $1{\,\rm kpc}$ and it does not emit light.
Given that the observed disk velocity dispersion exhibits primarily a trend with redshift, 
with a large scatter and no clear dependence on physical galaxy properties
\citep[e.g.,][]{Johnson2018, Uebler2019}, for each mock model, we assign the \sigo\
of the galaxy in RC100 that most closely matches it in \mstar, SFR, and {\fgas}.
The inclination \incl\ is drawn randomly from the RC100 distribution.
Other parameters specifying the disk and bulge components are fixed or tied to
those mentioned above as described in Sec.~\ref{subsec:dpy}.

The model cubes are created on a grid with a spaxel size of
$0\farcs 125 \times 0\farcs 125$ over a FOV of $6\farcs 375 \times 6\farcs 375$,
and a velocity channel width of 10\,km\,s$^{-1}$ over the range
$\pm$1000\,km\,s$^{-1}$.
Since for the effects of beam smearing, the number of linear resolution elements across
the source is most relevant, we assign the PSF FWHM by drawing from the RC100
distribution of \reff/beam, where the beam is the PSF half-width at half-maximum (HWHM).
This results in a mock data set covering \ReRbeam\ from $0.93$ to $8.5$ ([$5^{\rm th}$,$95^{\rm th}$] percentile $=[1.3,5.4]$).
The velocity resolution is fixed and represented by a Gaussian dispersion
$\sigma_{\rm instrument}$$=$\,40\,km\,s$^{-1}$.  
The adopted velocity resolution is higher than the \sigo\ of
$37\%$ galaxies in our baseline sample.  
However, only $6.5\%$ of the galaxies have $\sigma_0$\,$<$\,$0.5\sigma_\mathrm{instrument}$ and the minimum
$\sigma_0/\sigma_\mathrm{instrument}$\,$\sim$\,$0.3$ affects $<1\%$ of the sample.
As demonstrated by \citet{Wisnioski2015}, the presence of galaxies with velocity
dispersion below the spectral resolution limit in IFU surveys is not uncommon and
will amount to $\sim$\,$30$--$60\%$ error (depending on the S/N) in the recovered
velocity dispersion when the resolution decreases from
$\sigma_{\rm intrinsic}$\,$\approx$\,$\sigma_\mathrm{instrument}$
to $\sigma_{\rm intrinsic}$\,$\approx$\,$0.3\sigma_\mathrm{instrument}$,
compared to $20\%$
error when $\sigma_{\rm intrinsic}$\,$>$\,$\sigma_\mathrm{instrument}$.
Finally, random Gaussian noise is added to the model cubes to match the RC100 S/N
distribution, using our adopted definition of the average flux to rms noise ratio
in pixels within \reff\ for the velocity channel with the brightest line emission.

\subsection{Clumpy Mock Models}\label{sec:clumpymock}

Disks at high redshift commonly exhibit prominently clumpy or irregular light 
distributions.  To explore how the different modeling tools respond to light-weighting
effects, we created an additional suite by adding two massless clumps into 400 of the
baseline mock galaxies that lie at the higher ranges of \fgas\ ($\geq$\,$50\%$) and $z$ ($\gtrsim$\,$1$).
This subset follows the SFR and $M_{\rm bar}$ distributions of the baseline sample.
The clump sizes and brightnesses are motivated by observations of (unlensed) massive 
\znoon\ SFGs
\citep[e.g.,][]{Elmegreen2005,Genzel2008,nmfs2011b,Wuyts2012,Wuyts2013,Guo2015},
and are consistent with the Toomre scales predicted for gravitational instabilities
in high-$z$ gas-rich turbulent disks \citep[e.g.,][]{Genzel2008,Dekel2009}.
The clumps are represented by circular Gaussian light distributions with random
contributions to the total light between $2\%$ and $6\%$, and effective
radii of $R_{\rm e,clump} = 0.5~{\rm kpc}$.
They are placed randomly at galactocentric radii $0.75R_{\rm e,d}$\,$<$\,$R$\,$<$\,$2R_{\rm e,d}$, azimuthal angles in the range $[0, 2\pi]$ on the plane of the disk, and with a minimum azimuthal angle separation of $10^{\circ}$.  
The clumps are purely light sources with no intrinsic mass or kinematics.
They co-rotate with the galaxy, and the velocity dispersion corresponds to that of the host galaxy at the same location.
The two clumps always differ in luminosity to ensure the final model is asymmetric in light distribution.  Examples are shown in Appendix~\ref{app:clumpygal} Fig.~\ref{fig:clumpgal}.

\subsection{Mock Models with \gp}\label{sec:gpmock}
To test the impact of the choice of analytical prescription in kinematic modeling, we also
consider a set of $500$ model cubes generated with \gp.  For simplicity, we randomly draw
this subset from the full baseline sample and use the best-fit parameters returned 
from the \gp\ fitting as the ``true values'' to compare with in the recovery analysis.
We generate the input model cube by introducing Gaussian noise to the noiseless best-fit model, which mirrors the baseline model with identical spatial and spectral sampling and already incorporates beam-smearing and spectral broadening.
We verified that the
resulting distributions in the galaxy and observational properties are similar to those of the full baseline mock sample in terms of the total baryonic mass, size, \reff/beam, \incl\ and S/N to avoid strong biases stemming from model properties in comparing the recovery performances.

\subsection{Mock Models with \bb}\label{sec:bbmock}
To complete the comparison, we evaluate the performance of all codes using mock galaxies generated by \bb. Similar to Sec.~\ref{sec:gpmock}, we utilize the 500 best-fit model cubes returned by \bb\ when fitting the baseline models. The resulting distributions of physical properties are similar to those of the baseline models. We select cubes corresponding to galaxies with original S/N $\gtrsim20$, as the best-fit profiles of $V_{\rm rot}(R)$ and $\sigma(R)$ returned by \bb\ are better behaved and will serve as the new intrinsic reference profiles. The new intrinsic $V_{\rm rot}(R)$ profiles capture a variety of shapes, ranging from rising to declining profiles. However, the new intrinsic $\sigma(R)$ profiles are no longer constant but show mild declining trends, due to the S/N sensitivity of \bb\ when modeling the baseline models, as will be discussed in Sections~\ref{sec:dispcomp} and \ref{subsec:radial_prof}. Following the same approach as in Sec.~\ref{sec:gpmock}, we reintroduce Gaussian noise into these noiseless models (projected, beam-smeared, and spectral-broadened) to achieve an S/N distribution similar to that shown in the last panel of Fig.~\ref{fig:hist_obs} in Appendix~\ref{app:mockparams}.

 \section{Model Comparisons} \label{sec:results}
\begin{figure*}
    \centering
    \includegraphics[width=0.7\textwidth]{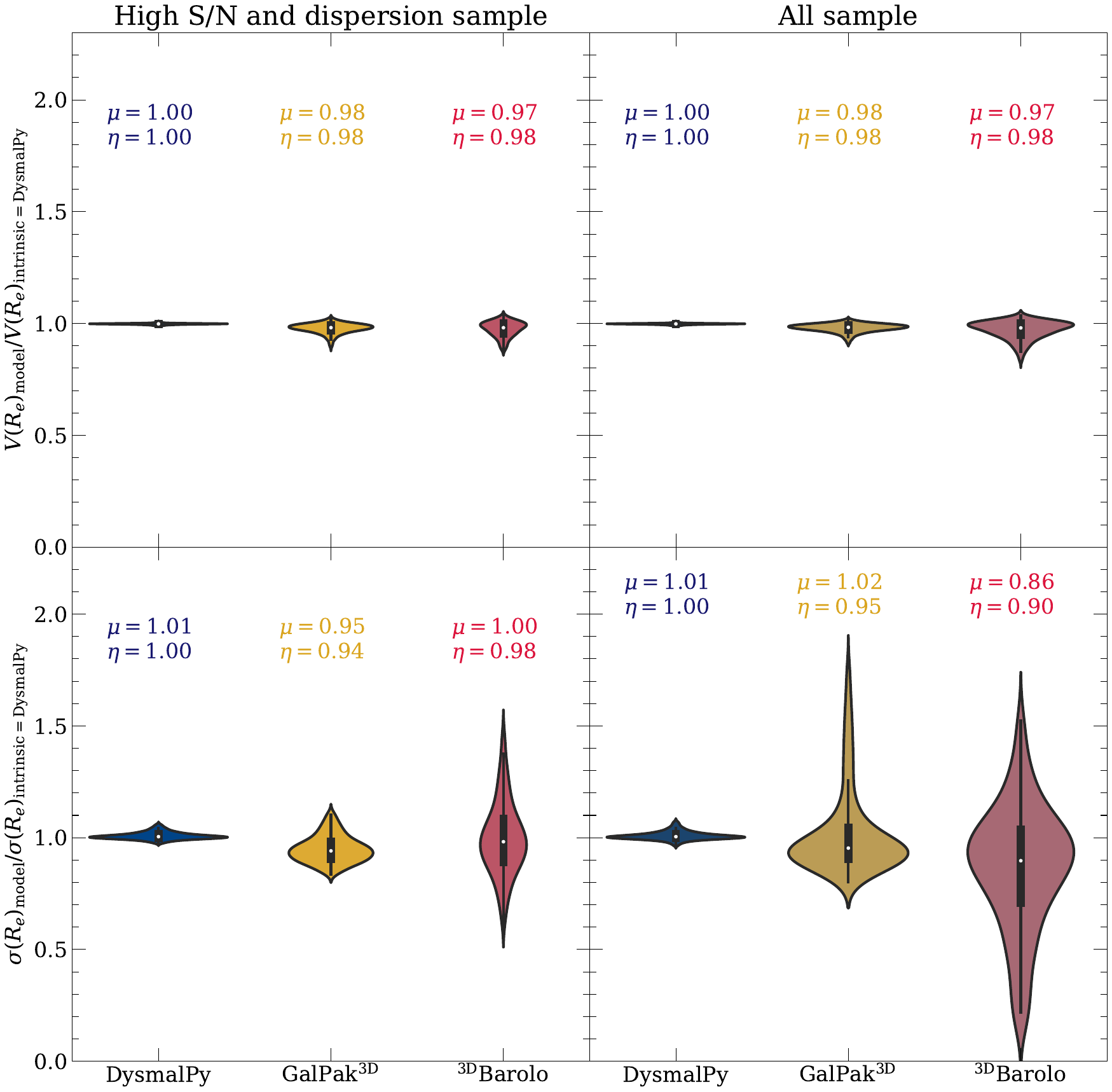}
    \caption{Comparison of the best-fit properties derived
    at the effective radius \reff\ from modeling the baseline mock data set with \dy,
    \gp, and \bb.  The results are illustrated via violin plots of the ratios of
    recovered to intrinsic rotation velocity \vrotrecRe\ (\textit{upper row}), 
    and recovered to intrinsic disk velocity dispersion \sigrecRe\ (\textit{lower
    row}).  
    The mean $\mu$ and median $\eta$ of
    each distribution is given in the plots.
    The \textit{first column} shows results from samples that are less susceptible 
    to template mismatch and signal-to-noise ratio (S/N) sensitivity issues specifically for \gp\ and \bb, in which all codes demonstrate good recovery performance
    with $\mu\approx\eta\approx1$.
    For the entire sample in the \textit{second column}, the rotation velocity is still very well recovered by all three packages.
    The largest differences are in velocity dispersion and mainly in the
    scatter and asymmetry of the distributions.  By construction, the properties
    of the baseline \dy-generated mock data cubes are best recovered by \dy.}
    \label{fig:violin_both}
\end{figure*}

In this section, we examine the performance of each code in
recovering the rotation velocity and disk velocity dispersion of the model galaxies.
Throughout, we quantify the goodness of recovery through the ratio of the best-fit value returned by the modeling tool to the known intrinsic model
value, i.e., $V_{\rm rot,model}/V_{\rm rot,intrinsic}$ and
$\sigma_{\rm model}/\sigma_{\rm intrinsic}$.
We exclude numerical catastrophic fits that do not converge in \dy\ (indicated by the \texttt{MPFIT} status or when model values hit the prior boundaries) and {\gp}. 
We also reject \bb\ fits if there are fewer than $3$--$4$ consecutive successfully modeled rings. 
The latter is a conservative choice but allows us to investigate recovered radial trends.
On average, all three tools achieve $\sim$\,$80\%$ ($\sim$\,$7000$ of the initial $9000$ baseline models) successful fits of the mock sample.
The S/N, \ReRbeam, $i$, and $\sigma_0$ distributions of the successfully modeled sample are shown in the side panels of
Fig.~\ref{fig:input_param_space} in Appendix~\ref{app:covar_plot}. 
All three tools share very similar distributions of these parameters. 
However, due to variations in the consideration of catastrophic fits among different codes, the final effective samples differ. 
Specifically, \bb\ has the fewest retained fits ($\sim$70\%), most notably at low S/N and \ReRbeam. 
Consequently, the effective samples of \dy\ and \gp\ include more of the low-S/N and poor-resolution mocks.

We stress that the \vrot\ refers to the intrinsic {\rm rotation} velocity,
corrected for beam-smearing and inclination but not for pressure support, in order to keep the comparison as simple and consistent as possible between
the modeling tools. \vrot\ is directly output by all three tools.
For the velocity dispersion, we adopt in all cases the total intrinsic $\sigma$
corrected for the effects of beam smearing, projection, and velocity resolution.
As described in Sec.~\ref{sec:methods}, the total intrinsic $\sigma$ at a given
radius $R$ corresponds to the global and radially constant \sigo\ for \dy,
to the sum of disk self-gravity, line-of-sight velocity mixing, and constant 
turbulence for \gp, and to the total velocity dispersion interpolated from the two closest rings
to $R$ for \bb.  These differences in implementation play a role in the results as discussed below, but beyond the fitting exercise. They also imply a different physical interpretation of the recovered dispersion that should be kept in mind.

\subsection{Overall Recovery of the Baseline Models}{\label{subsec:general_results}}
We begin by considering the rotation velocity and velocity
dispersion recovered at the disk effective radius (\reff) for the baseline set of mock data 
cubes.  
As velocity and dispersion are not described by parametric functions in {\bb}, comparing parametric modeling results from \dy\ and \gp\ is less straightforward.  For the rotation velocity, we use the returned intrinsic $V_{\rm rot}$ at \reff\ (linearly interpolated from the two nearest annuli).  
For the velocity dispersion, we measure the value at $R_e$ to ensure consistency across codes. 
As shown in the top panel of Fig.~\ref{fig:violin_both}, all three tools
overall perform very well for \vrot(\reff), which is recovered within $<$\,$5\%$ in
the mean $\mu$ and median $\eta$, and with small scatter of $<$\,$0.04$. 
In contrast, the distributions have larger scatter and are more asymmetric for
$\sigma(R_{\rm e})$.  The \gp\ results tend to underestimate the intrinsic values
by $5\%$ in the median, with a more pronounced tail extending to \sigrecRe\,$>$\,$1$
and $13\%$ of the sample lying above one standard deviation (SD) of
the mean.
The velocity dispersion recovered by \bb\ is $\sim$\,$90\%$ (mean and median)
of the input values, with a more extended tail towards lower values and
$16\%$ of the sample $1$ SD below the mean.
\dy\ performs best for both \vrot\ and $\sigma$, which is unsurprising given
the \dy-generated baseline mock data set and simply reflects the better match
in intrinsic and model parametrizations.  The tests carried out here use the
3D-space fitting functionality of \dy, and thus extend the validation tests 
performed in 1D and 2D presented by \citet{Davies2011} and \citet{Price2021}.

\subsection{Trends with Input Parameters}\label{subsec:biassource}
Next, we investigate the dominant source of scatter and asymmetry in
the recovery results for the baseline data set.  We searched for trends
in \vrotrecRe\ and \sigrecRe\ with observational parameters and galaxy
physical properties, based on the Spearman's rank correlation coefficient
\citep[$\rho$;][]{Spearman1904} as well as visual inspection.
However, as $\rho$ is only sensitive to monotonic trends between variables, it may not capture all possible relationships. 
To address this limitation, we also compare our results to the Maximal Information Coefficient (MIC) \citep{Reshef2011}, which is more adept at detecting non-single-valued functions.
The MIC scores mostly agree with $\rho$ in terms of identifying the stronger trends
in our results. Therefore, we will only report $\rho$ henceforth.
The S/N, \incl, angular resolution, and intrinsic velocity dispersion
have the largest impact on our results, consistent with previous findings
from validation tests \citep[e.g.,][]{Davies2011, Bouche2015, DiTeodoro2015}.
We thus focus on these four parameters.

\begin{figure*}
    \centering
    \includegraphics[width=\textwidth]{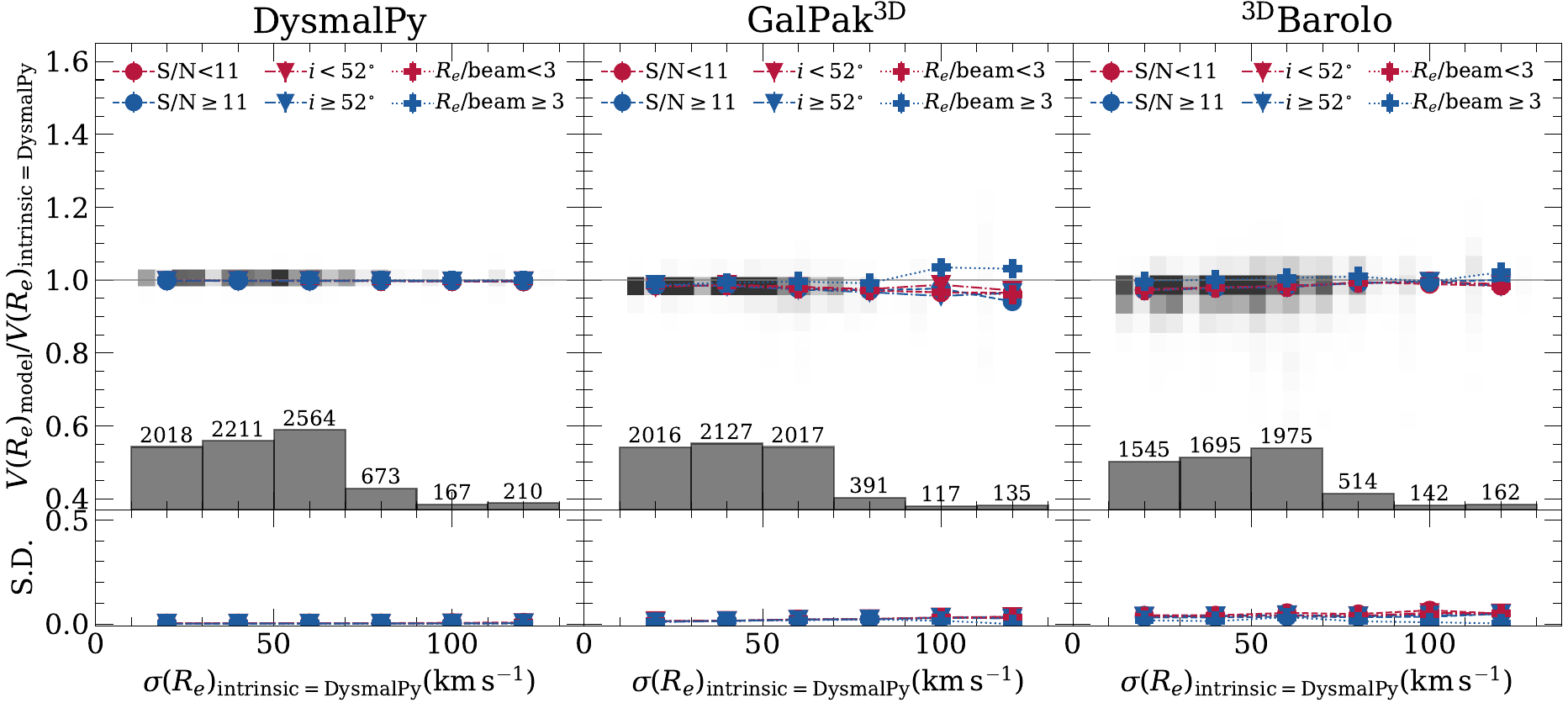}
    \caption{Comparison of best-fit to intrinsic rotation velocity derived at
    \reff\ from modeling the baseline mock models with \dy\ (\textit{left}),
    {\gp} (\textit{middle}) and {\bb} (\textit{right}).  The \vrotrecRe\ is
    plotted as a function of the \dy\ model intrinsic (and radially constant)
    velocity dispersion \sigo.
    The gray-scale background image illustrates the density distributions of the full
    set of models, and the overplotted curves correspond to running median trends
    for different subsets split by S/N, inclination, and \ReRbeam\ as labeled in
    each panel.  Error bars associated with the data points are the $68\%$ confidence
    interval for the medians derived from bootstrapping. 
    The gray histograms give the number of galaxies included in each $\sigma(R_e)$ bin.
    The panels in the \textit{bottom} row are the associated standard deviation (S.D.) values of
    each $\sigma(R_e)$ bin for each subset.
    Overall, all three
    modeling tools recover well the intrinsic \vrot(\reff), with no significant
    dependence on S/N, disk inclination, and angular resolution in the regimes
    tested by our models.
    }
    \label{fig:velratio_vel}
\end{figure*}

\begin{figure*}
    \centering
    \includegraphics[width=\textwidth]{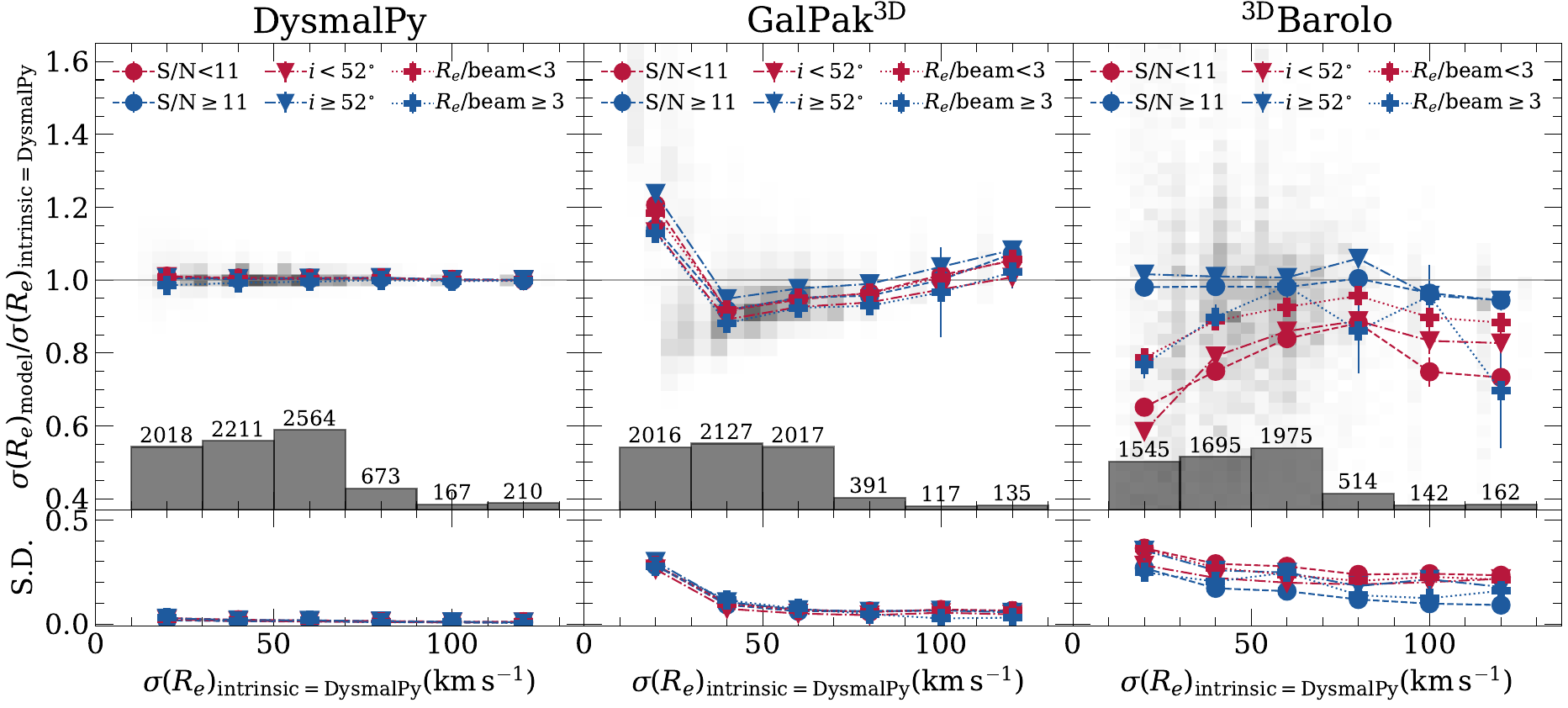}
    \caption{Similar to Fig.~\ref{fig:velratio_vel}, but for the ratio of
    \sigrecRe\ as a function of \dy-generated model intrinsic velocity
    dispersion \sigo.  The large typical overestimate by \gp\ at the lowest
    \sigo\ values is a direct result of the template mismatch between the
    constant and radially varying profiles adopted by {\dy} and {\gp}, 
    respectively.  This effect is much reduced as \sigo\ increases and
    there is little dependence on S/N, \incl, and \ReRbeam.  For \bb\, the
    curves show that the large scatter reflects, in part, a fairly strong
    dependence on S/N and \incl. 
    }
    \label{fig:dispratio_disp}
\end{figure*}

Figs.~\ref{fig:velratio_vel} and \ref{fig:dispratio_disp} show 2D histograms
of the distributions of recovered to intrinsic \vrot\ and $\sigma$ (at 1\reff) as a function
of intrinsic velocity dispersion for the full sample (excluding catastrophic fits).
Different curves are overplotted to illustrate the running median trends of subsets
split in terms of (i) S/N, (ii) inclination \incl, and (iii) \ReRbeam, with error bars showing
$68\%$ confidence intervals derived from bootstrapping.  The dividing values
correspond to the sample median values of S/N\,$=$\,$11$, \incl\,$=$\,$52^\circ$, and \ReRbeam\,$=$\,$3$.
The variations of standard deviations are also plotted. Fig.~\ref{fig:cm_spearman_dy} in Appendix~\ref{app:corr_matrix} report the Spearman's $\rho$ between the ratios \sigrecRe\ and the S/N, \incl, \ReRbeam, and intrinsic $\sigma$.
Fig.~\ref{fig:running_med_disp} is similar to Figs.~\ref{fig:velratio_vel} and \ref{fig:dispratio_disp}, 
but shows instead the distributions of recovered to intrinsic $\sigma$ vs.\  (i) S/N, (ii) inclination \incl, 
(iii) \ReRbeam\ and (iv) intrinsic $\sigma$ directly. 
For the baseline models under comparison, only the light green curves (labeled $\sigma(R_{\rm e})_\text{mock=DysmalPy}$) are pertinent. Other trends will be addressed in subsequent sections.

\subsubsection{Rotation Velocity}\label{sec:vel_results}
Fig.~\ref{fig:velratio_vel} indicates that there is overall a minor
impact of the parameters considered on the recovered $V_{\rm rot}(R_e)$.
\ReRbeam\ has the largest impact on the results,
causing the slight tail towards lower values in $V_{\rm rot}(R_e)$ for
\gp\ and \bb\ in Fig.~\ref{fig:violin_both}, but this is a very small
effect.  In our tests, the reliability of all three tools in recovering
\vrot(\reff) is fairly robust against varying S/N over the range explored.
Closer inspection shows that at lower S/N, the scatter becomes larger 
for {\bb} for which the standard deviation in \vrotrecRe\ increases
from ${\rm SD}$\,$=$\,$0.04$ at S/N\,$\ga$\,11 to 0.09 at S/N\,$<$\,11.

\subsubsection{Velocity Dispersion}\label{sec:dispcomp}
Fig.~\ref{fig:dispratio_disp} shows stronger
trends in median recovered velocity dispersion with different behavior for
\gp\ and \bb.  
For \dy, the weak or absent trends in median values and for different subsets are partly attributable to the match in parametrization between mock
models and fitted models.  Taken at face value, the \ReRbeam\ may play the
most important role, but the correlation is weak ($\rho$\,$=$\,$0.24$).

For \gp, the strongest sensitivity is to the intrinsic dispersion.
The most salient feature is the ``L-shaped'' trend with an upward tail at
$\sigma_{0}$\,$<$\,$30{\,\rm km\,s^{-1}}$ regardless of S/N, \incl, and \ReRbeam.
About a quarter ($\sim$\,$26\%$) of the baseline sample falls into this regime and 
is the main cause of the asymmetric distribution in Fig.~\ref{fig:violin_both}.
The overestimated dispersion at low $\sigma_{\rm intrinsic}$ is the direct consequence of the different velocity dispersion parametrizations between 
{\gp} and {\dy}.  To visualize this behavior, in Fig.~\ref{fig:dispprof_example} 
we compare the profiles (corrected for beam
smearing) of the best-fit \gp\ models for two baseline mock data sets
with high \sigo$\,=53{\rm\,km\,s^{-1}}$ and low \sigo$\,=13{\rm \,km\,s^{-1}}$.
At higher velocity dispersion, the radially-dependent $\sigma_{\rm d}$ term 
is sub-dominant, and \gp\ better matches the uniform dispersion through
its radially-constant turbulent term.  On the other hand, when \sigo\ is low \gp\
has more difficulty recovering the value around \reff\ because $\sigma_{\rm d}$
more strongly dominates out to larger radii.

\begin{figure*}
    \centering
    \includegraphics[width=\textwidth]{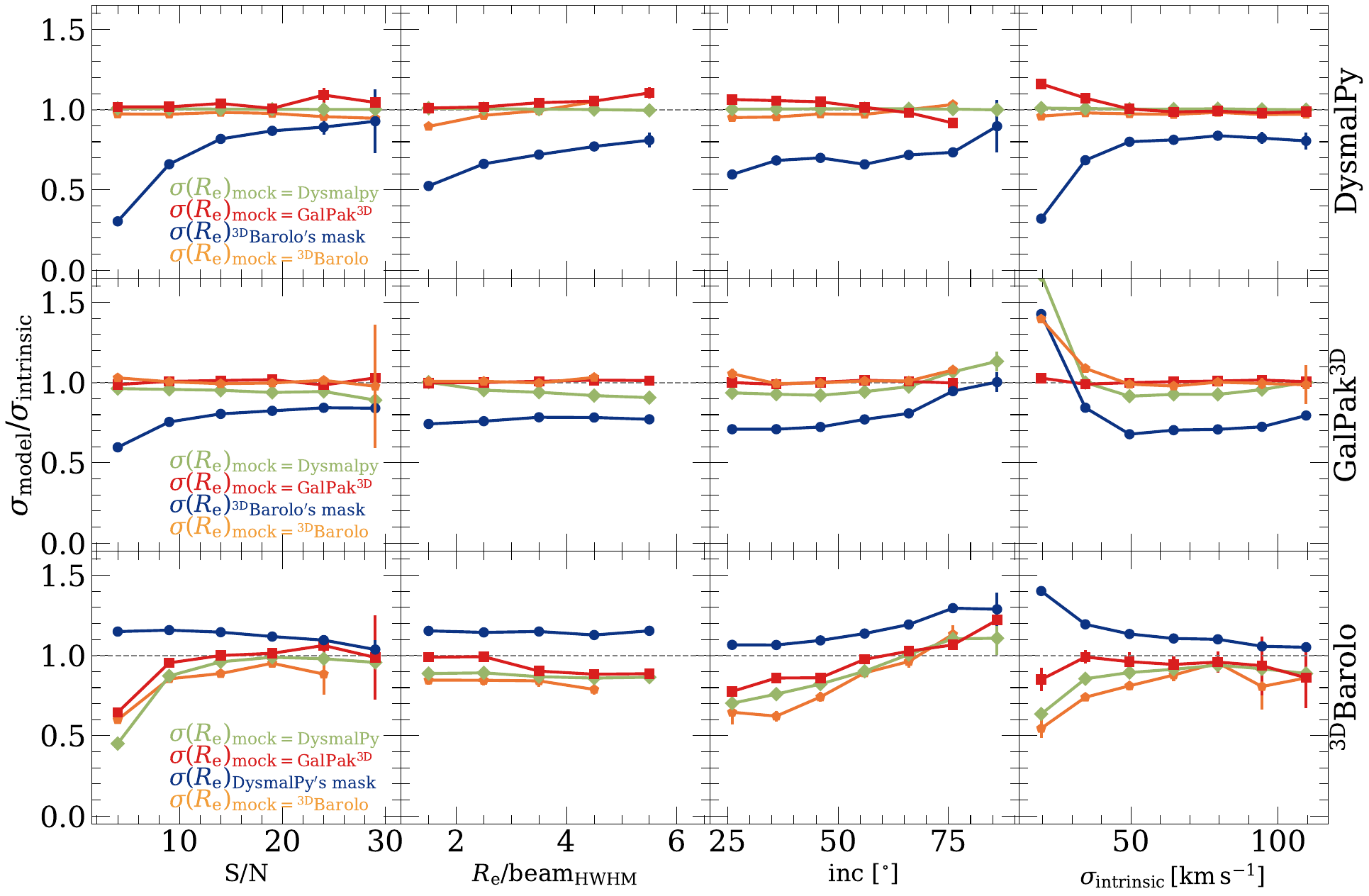}
    \caption{An expansion of Figure~\ref{fig:dispratio_disp} to show the ratios of recovered ($\sigma_{\rm model}$) and intrinsic ($\sigma_{\rm instrinsic}$) velocity dispersions against 4 different parameters, signal-to-noise ratio (S/N) (\textit{first column}), 
    ratio between the effective radius and beam in half-width-half-maximum (HWHM) (\textit{second column}), inclination ($i$) (\textit{third column}) and intrinsic dispersion ($\sigma_{\rm instrinsic}$) (\textit{fourth column}). 
    Intrinsic values correspond to baseline models unless specified by ``mock={\gp}" or ``mock={\bb}". In such case, $\sigma_{\rm intrinsic}$ would be the value taken at $R_e$ of the total intrinsic dispersion profile of {\gp} or {\bb}, respectively. The colored curves are the running median of their corresponding distributions as labeled, 
    with errors representing $68\%$ confidence interval derived by bootstrapping. 
    {\bb}-recovered $\sigma_{\rm model}$ shows positive dependence on $i$ and S/N, with the latter converging to the intrinsic values when S/N $\gtrsim8$. 
    For comparison, we show the same plot with only S/N $\geq 11$ mocks in Figure~\ref{fig:running_med_disp_highSN} in Appendix~\ref{app:highsn_recover}.
    Trends persist across different $\sigma_{\rm model}$ definitions, even with {\gp}- or {\bb-}generated  models. 
    The source of such S/N dependence is due to \bb\ spectral overmasking, and if the same masking is applied to \dy\ and \gp\ when modeling, a similar asymptotic trend (light green curves) is also recovered. S/N dependency in \bb\ vanishes (although now with systematic overestimate) 
    when {\dy}'s masking is adopted. Dependence on $i$ persists nevertheless.  
    \gp\ and \dy\ show negligible S/N and \reff/HWHM dependencies. Dependency on $\sigma_{\rm intrinsic}$ is attributed to template mismatch, as detailed in the main text.
    }
    \label{fig:running_med_disp}
\end{figure*}

\begin{figure}
    \centering
    \includegraphics[width=0.4\textwidth]{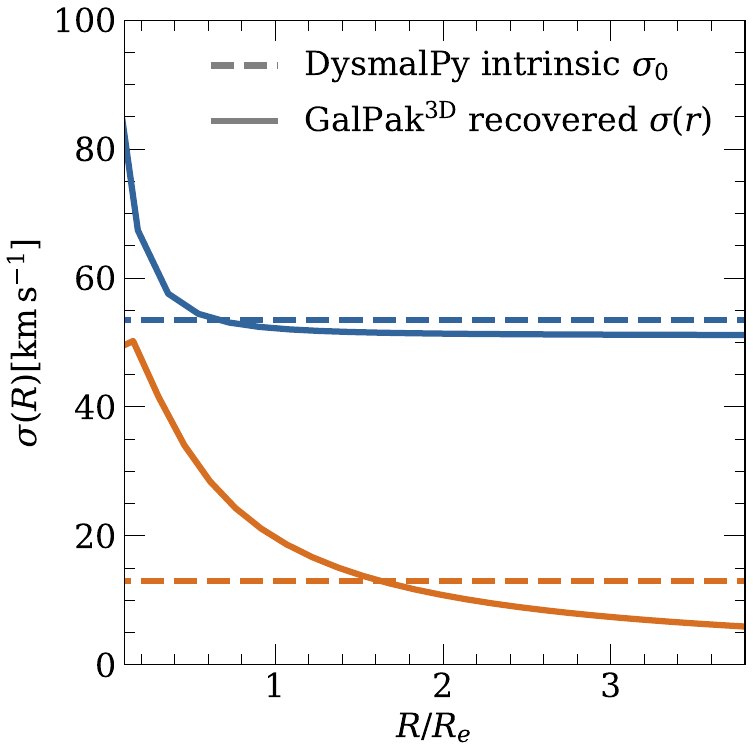}
    \caption{Examples of velocity dispersion profiles $\sigma(r)$ derived by {\gp} (\textit{solid}) (beam-smearing corrected) and the intrinsic profiles (\textit{dashed}) of two mock samples at high (\textit{orange}) and low (\textit{blue}) intrinsic dispersion constructed by {\dy}. In high intrinsic dispersion conditions, {\gp}'s dispersion profile coincides with the intrinsic dispersion value around $R_{\rm e}$, while in low intrinsic dispersion conditions, the radially dependent disk self-gravity term dominates and causes the intrinsic dispersion value at $R_e$ to be overestimated
    compared to a model that adopts a flat intrinsic dispersion profile.}
    \label{fig:dispprof_example}
\end{figure}

\begin{figure*}
    \centering
    \includegraphics[width=\textwidth]{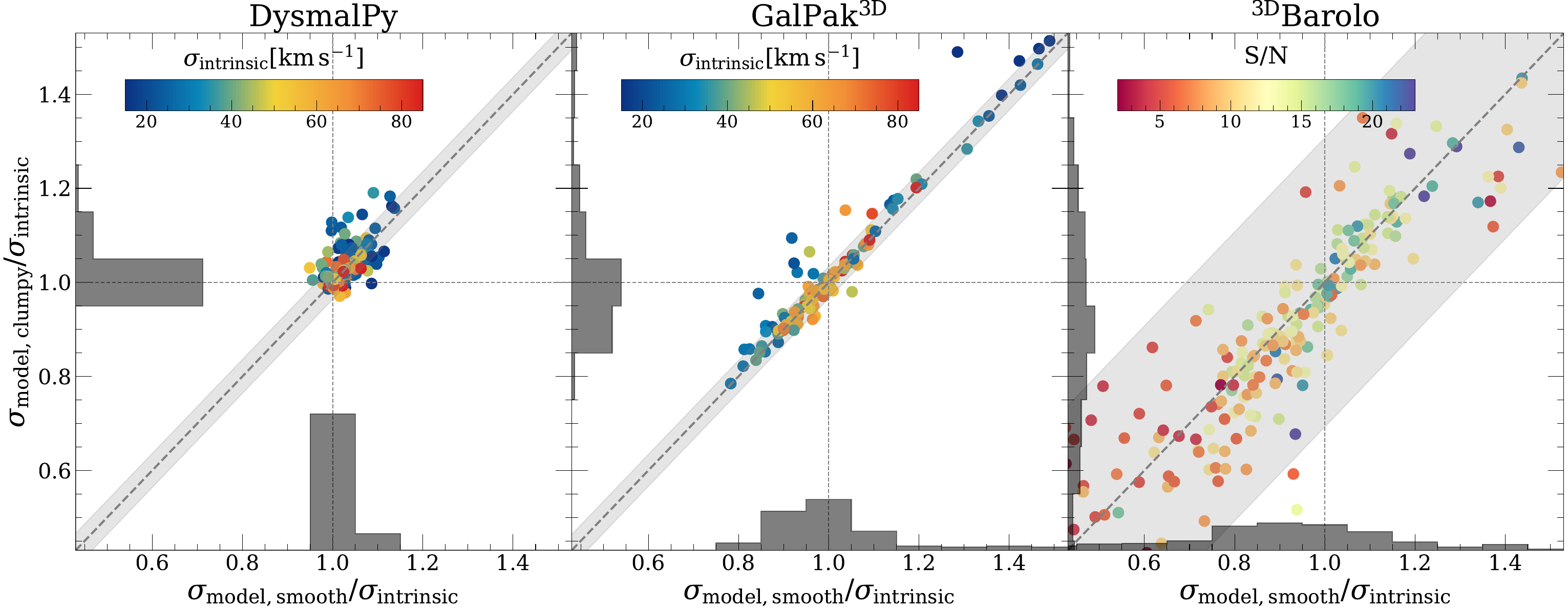}
    \caption{Comparison plots of velocity dispersions recovered from clumpy mock galaxies ($\sigma_{\rm model,clumpy}$) and those from smooth galaxies with otherwise identical properties ($\sigma_{\rm model,smooth}$) by {\dy} (\textit{left}), {\gp} (\textit{middle}) and {\bb} (\textit{right}). 
    The gray histograms along the x- and y-axis are the distributions of ratios between the recovered and intrinsic values for smooth and clumpy galaxies, respectively.
    The colors of the points represent the intrinsic velocity dispersion ($\sigma_{\rm intrinsic}$) in the \textit{left} and \textit{middle} panels, while the \textit{right} panel shows signal-to-noise (S/N). 
    The choice of colors for \gp\ and \bb\ is determined by the parameters that exhibit the strongest correlation with the recovered $\sigma$ in the smooth galaxies sample.
    The shading around the one-to-one line represents the typical fitting errors of the respective codes. 
    The points that lie along the one-to-one line are those not affected by the
    presence of clumps, and those lying outside the shading are more affected.
    For {\dy}, {\gp}, and {\bb}, roughly $30\%$, $12\%$, and $5\%$ of the points
    lie outside the shaded regions, respectively. In other words, {\em relative to
    their typical fitting uncertainties\/}, the parametric models are more affected
    by light clumps.  As for the smooth axisymmetric models, the scatter for the
    non-parametric modeling is larger towards lower S/N.
}
    \label{fig:clumps_comp}
\end{figure*}

\begin{figure*}
    \centering
    \includegraphics[width=\textwidth]{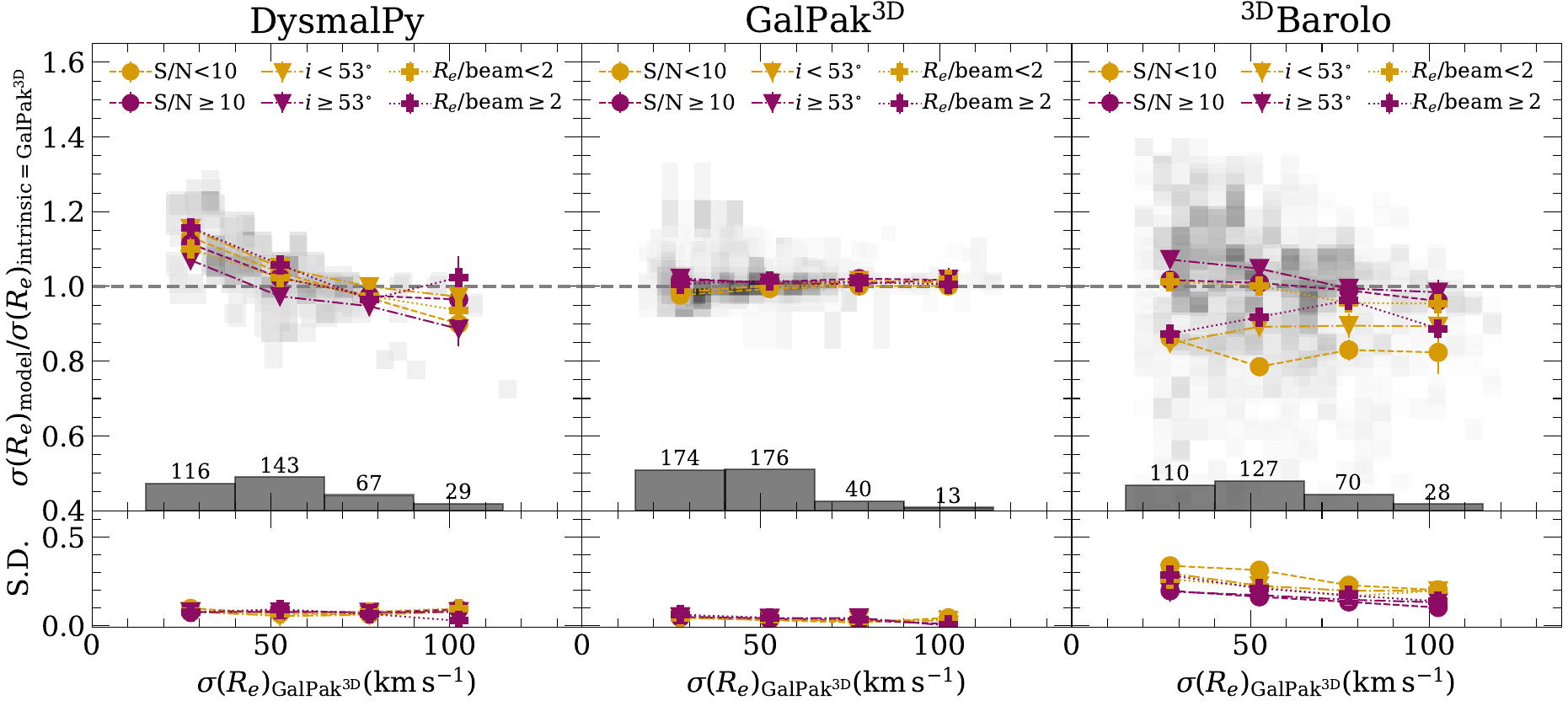}
    \caption{Similar to Fig.~\ref{fig:dispratio_disp}, but for {\gp} generated mocks. 
Due to the inherent template mismatch between {\dy} and {\gp} as explained in the main text, {\dy} in the \textit{left} panel tends to overestimate the dispersion values when the intrinsic values are low, whereas {\gp} could accurately recover the intrinsic values at all ranges. In the \textit{right} panel, {\bb} shows a similar behavior as before, with a comparable dependence on the signal-to-noise, but with an overall better recovery when compared to Figure~\ref{fig:dispratio_disp}.}
    \label{fig:gp_mocks_dispratio}
\end{figure*}

For {\bb}, the recovered $\sigma$(\reff) is comparably affected by S/N and \incl,
typically leading to an underestimate of the intrinsic velocity dispersion 
driving the asymmetric distribution shown in Fig.~\ref{fig:violin_both}.
Although globally, there is only a weak correlation with median intrinsic
$\sigma$, the results for the S/N\,$<$\,$11$ and $i$\,$<$\,$52^{\circ}$ subsets
exhibit a stronger dependence with more pronounced downturns at both low and
high dispersion ends.
That both S/N and \incl\ can affect \bb\ modeling results have been discussed previously
\citep[e.g.,][]{DiTeodoro2015,Bacchini2020,Deg2022}.  The sensitivity to intrinsic $\sigma$ could be related to these two factors.  At higher velocity dispersions, the line
flux is spread over more velocity channels, resulting in a lower S/N per pixel
in the brightest channel (our adopted definition of S/N).  At lower velocity
dispersions, the line emission gets narrower, especially at lower inclinations, 
potentially leading to the overmasking of line wings in velocity
from the \bb\ algorithm
and underestimating the line width.  Masking effects are discussed
in more detail in Sec.~\ref{subsec:masking_discussion} \citep[see also][]{Davies2011}.

For all three tools, we find little difference in the median trends as a function of angular resolution but note that this could be due to the limited range probed by our mock models: the \ReRbeam\ varies only from
$1.4$ to $7.2$, with a median of $3$.  For \bb, these results are consistent
with a very modest dependence on the angular resolution for similar ranges of 
\ReRbeam\ reported by \citet{DiTeodoro2015} and \citet{Rizzo2022}, based on different test models.

The tightness of the distributions is different between
the three tools, as evidenced by the lower panels in Fig.~\ref{fig:dispratio_disp}.
The scatter systematically decreases from 
low to high S/N and angular resolution regimes mainly for \dy\ and \bb. 
For \dy, the scatter (in standard deviation) ranges from $\sim$\,$0.06$ at S/N\,$<$\,$11$
and \ReRbeam\,$<$\,3 to $\sim$\,$0.03$ at S/N\,$>$\,$11$ and
\ReRbeam\,$>$\,$3$. For {\bb}, the corresponding drop is from about $0.36$ to $0.23$ as S/N increases, with comparable scatter from low to high \ReRbeam. There is no significant change in scatter for either parameter in {\gp}.

\subsection{Light-Weighting Effects Tested with Clumps}
  \label{subsec:lightclump}

One of the potential advantages of {\bb} over {\dy} and {\gp} is that its surface brightness distribution can take an arbitrary form, whereas {\dy}\footnote{\dy\ is capable of modeling clumps, but here we are interested in benchmarking its performance against \gp\, which offers smooth model only.}
and {\gp} assume a smoothly varying and axisymmetric analytic distribution.
Since accounting for the effects of beam smearing is driven by the smearing of the underlying flux profile, if the underlying flux distribution is clumpy, {\dy} and {\gp} may recover the main kinematic properties 
less accurately. 
We focus on the velocity dispersion, which is the property most sensitive to modeling approach and tool. We find similarly good performance in \vrot\ recovery among the three codes as in the case of the smooth mocks.

We present the recovery performance of the three codes in $\sigma_{\rm model, clumpy}/\sigma_{\rm intrinsic}$ for comparison with the smooth mocks $\sigma_{\rm model, smooth}/\sigma_{\rm intrinsic}$ in 1D histograms along the $y$- and $x$-axes of Fig.~\ref{fig:clumps_comp}, respectively. The systematic scatter between the two cases is comparable for \gp\ and \bb, while \dy\ exhibits slightly more overestimation in the case of clumpy galaxies compared to smooth galaxies. To isolate the pure effect of asymmetric light distribution introduced by light clumps versus axisymmetric effects, we consider light clumps to affect the model fitting if $\sigma_{\rm model, clumpy}/\sigma_{\rm intrinsic}$ differs from $\sigma_{\rm model, smooth}/\sigma_{\rm intrinsic}$ by more than the average fitting error of the same galaxy returned by the respective code. In Fig.~\ref{fig:clumps_comp}, those outside the shaded region around the one-to-one line would meet this criterion. Data points are color-coded based on the primary factor that most strongly correlates with the $\sigma$ recovery: $\sigma_{\rm intrinsic}$ for \gp, and S/N for \bb.

Of the three tools, {\dy} is the most affected by light clumps, with $\sim$\,$30\%$ of
the models differing by more than the fitting error. They tend to happen for
intrinsically low dispersion.  The fraction drops to $\sim$\,$12\%$ and $\sim$\,$5\%$
for \gp\ and \bb, respectively.  The scatter, however, is large for {\bb}.
Although \bb\ is less systematically affected by clumps relative to the fitting errors,
it is also less accurate in our recovery exercise.
In cases of clumpy galaxies, the Bayesian kinematic modeling tool 
\texttt{Blobby3D} \citep{Varidel2019,Varidel2023} could be a potentially
preferable choice, as it was more specifically designed to treat irregular
clumpy systems.  
Testing this code against others, as done here, would be valuable but is beyond the scope of this paper.

\begin{figure}
    \centering
\includegraphics[width=0.45\textwidth]{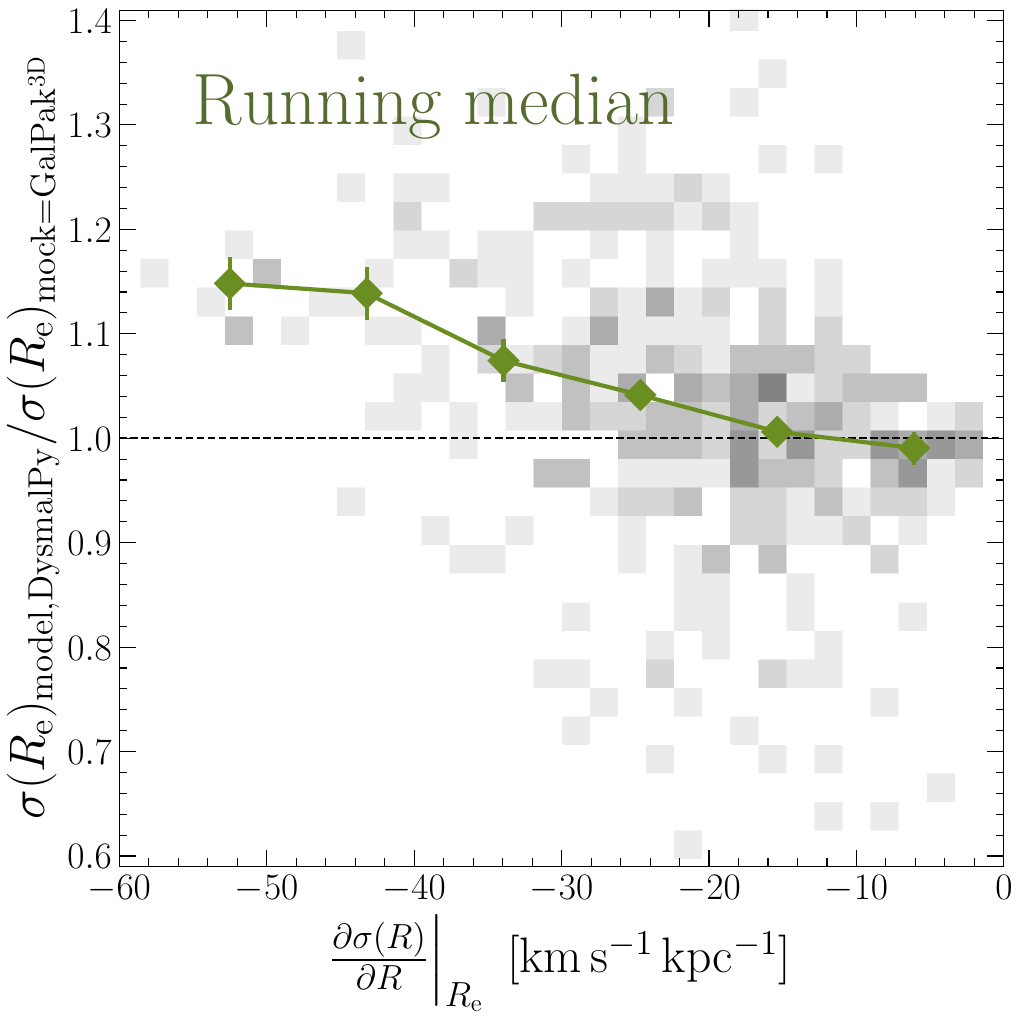}
    \caption{Similar to the leftmost panel of Figure~\ref{fig:gp_mocks_dispratio}, but with no distinction on signal-to-noise, inclination and resolution. The x-axis is replaced by the gradient of the intrinsic {\gp} dispersion profile at the effective radius \reff: $\left.\frac{\partial\sigma(R)}{\partial R} \right\vert_{R_{\rm e}}$.}
    \label{fig:gp_sigprof_gradient}
\end{figure}

\begin{figure*}
    \centering
    \includegraphics[width=0.95\textwidth]{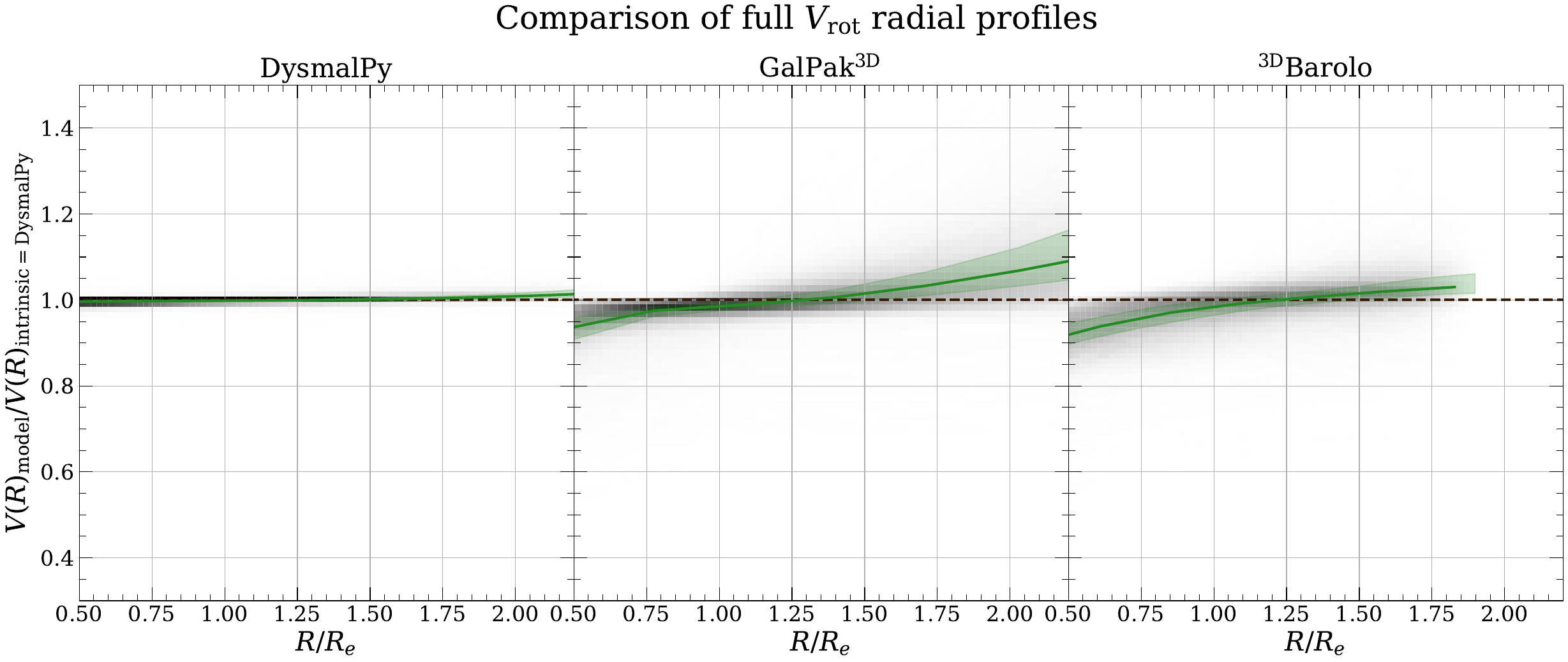}
    \caption{Running medians of the ratios between model and intrinsic rotation velocity ($V(R)_{\rm model}/V(R)_{\rm intrinsic}$) of {\dy} (\textit{left}), {\gp} (\textit{middle}), {\bb} (\textit{right}) in the range of  $[0.5,2.2]R_e$, where $R_e$ is the effective radius. The light green shading represents $1\sigma$ spread from the nominal median trend.
    }
    \label{fig:vel_profile_radial_dymock}
\end{figure*}

\begin{figure*}
    \centering
    \includegraphics[width=0.95\textwidth]{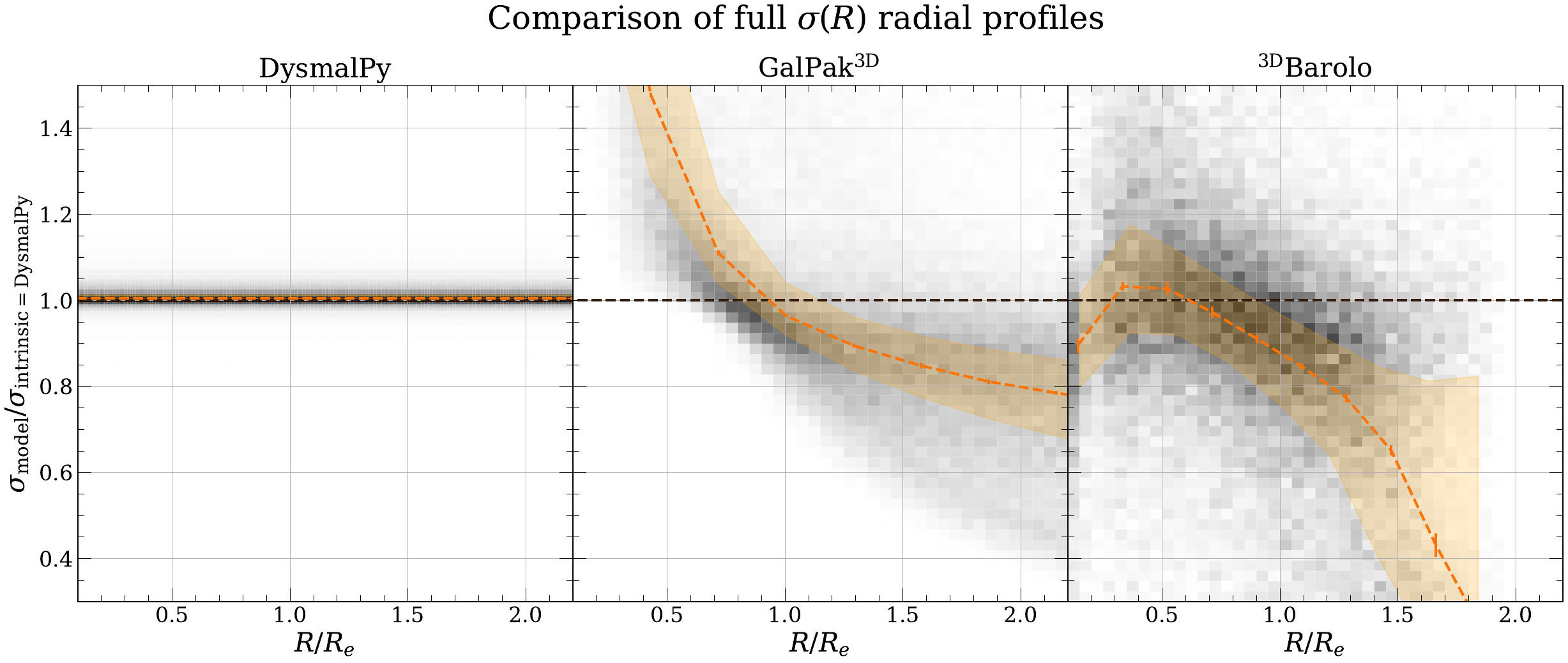}
    \caption{Similar to Figure\,\ref{fig:vel_profile_radial_dymock} but for velocity dispersions ($\sigma_{\rm model}/\sigma_{\rm intrinsic}$) of {\dy} (\textit{left}), {\gp} (\textit{middle}), {\bb} (\textit{right}) in the range of  $[0.1,2.2]R_e$, where $R_e$ is the effective radius. The errors of the running medians are the $95\%$ confidence interval derived by boot-strapping. The shading indicates $1$--$\sigma$ spread. In the \textit{left} panel, {\dy} shows an excellent recovery of the intrinsic values with minimal scatter,
    as template match is maximized. In the \textit{middle} panel, the decreasing trend is primarily due to the mismatch between {\dy}'s flat dispersion profile and the centrally peaked profile of {\gp}, as evidenced by the opposite trend in the \textit{left} panel in Figure\,\ref{fig:dispvel_profile_radial_gpmock} (see Section\,\ref{sec:dispcomp} for details). 
    In the \textit{right} panel, {\bb} underestimates the intrinsic values at radii $\gtrsim$\,$R_e$. As discussed in Section\,\ref{sec:dispcomp}, given the strong dependence of {\bb} performance on signal-to-noise (S/N), the observed trends are most likely caused by decreasing S/N at increasing radii from the centers.
    }
    \label{fig:dispvel_profile_radial_dymock}
\end{figure*}

\subsection{Impact of Parametrization}\label{subsec:impact_parametrisation}
We now examine the results when using \gp-generated mock data sets.  We focus
on the recovery of the velocity dispersion around \reff\
to compare the performance of the codes when galaxies have a radially-dependent dispersion.
The results are plotted in Fig.~\ref{fig:gp_mocks_dispratio}, where
the reference model intrinsic dispersion $\sigma(R_e)_{\rm gp}$ now corresponds
to the total dispersion at \reff\ including the hydrostatic equilibrium, line-of-sight
velocity mixing, and turbulence terms (Eqn.~\ref{eqn:gp_disprof}) but excluding any broadening by beam smearing.
For convenience, we denote the results from modeling the \gp\ mock models
with \dy, \gp, and \bb\ below as dy(gp), gp(gp), and bb(gp), and those from modeling
the \dy-generated baseline mock models as dy(dy), gp(dy), and bb(dy).

Because of the different analytical prescriptions, now the \gp\
fits perform better than those with \dy, as expected.  In both mock sets, there is no significant
correlation in the median \sigrecRe\ trend with S/N, \incl, \ReRbeam, 
The pronounced tail of systematically overestimated $\sigma$(\reff) at low
intrinsic dispersion observed for gp(dy) is now absent in the gp(gp) fits.

For the fits with \dy, the median \sigrecRe\ now shows a systematic behavior
reflecting again the template mismatch as seen in Fig.~\ref{fig:dispprof_example}.
At intrinsic velocity dispersions $\la$\,50\,km\,s$^{-1}$, \dy\ typically 
overestimates the dispersion at \reff\ by up to $\sim$\,$10$--$15\%$
in the median (a lesser effect than the sharp and steep tail in the case of
gp(dy) discussed in Sec.~\ref{sec:dispcomp}).
The effect reverses to a typical underestimate at
$\sigma_{\rm intrinsic}$\,$\ga$\,50\,km\,s$^{-1}$ by a few up to $\sim$\,$10\%$.
This behavior can be explained by the relative contribution to the
overall profile of the radially-dependent $\sigma_{\rm d}$ term in \gp\
(Eqn.~\ref{eqn:gp_disprof}), which contributes more importantly and over
a wider radial range at lower $\sigma(R_e)_{\rm gp}$.  

To illustrate the sensitivity of \dy\ to the slope of the intrinsic dispersion
profile, Fig.~\ref{fig:gp_sigprof_gradient} plots the \sigrecRe\ for dy(gp)
as a function of the intrinsic gradient at \reff\ of the model,
$\frac{\partial \sigma(R)}{\partial R}\Big|_{R_{\rm e}}$.
The Figure shows that more generally, by construction \dy\ will have a tendency
to overestimate the dispersion
in the case of radially declining intrinsic 
dispersion when the slope around \reff\ is
$\frac{\partial \sigma(R)}{\partial R}\Big|_{R_{\rm e}}$\,$\la$\,$-$30\,km\,s$^{-1}$\,kpc$^{-1}$, 
which is a fairly steep slope compared to local CO studies \citep[e.g.,][]{Wilson2011}.

\bb\ recovered values at \reff\ are slightly less underestimated than bb(dy) by $5\%$ in median of $\sigma(R_e)_{\rm model}/\sigma(R_e)_{\rm intrinsic}$.
The trends and scatter in Fig.~\ref{fig:gp_mocks_dispratio} show no significant difference compared to Fig.~\ref{fig:dispratio_disp} when S/N $\geq$\,$10$ and $\sigma_{\rm intrinsic}$\,$\gtrsim$\,$40{\,\rm km\,s^{-1}}$. 
The smaller scatter and overall slightly better recovery of \bb\ stem from its improved performance when $\sigma_{\rm intrinsic}$\,$\lesssim$\,$40{\rm km\,s^{-1}}$, as demonstrated in Fig.~\ref{fig:running_med_disp}: 
when $\sigma_{\rm intrinsic}\lesssim$\,$40{\rm km\,s^{-1}}$, it was underestimated by $\sim$\,$25\%$ when baseline (\dy-generated) models were used, 
and now this has improved to $\sim$\,$5\%$. There is an overall slightly milder dependence with respect to S/N, as also reflected in Fig.~\ref{fig:running_med_disp}.

Finally, a comparison with mocks generated using \bb\ reveals consistent trends for all codes, similar to those observed in the case of baseline models (orange lines in Fig.~\ref{fig:running_med_disp}). \bb\ displays similar trends with S/N, \reff/beam, $i$, and $\sigma_{\rm intrinsic}$. The persistent S/N trend of \bb\ is unsurprising, as it is not limited by any template assumption, and S/N sensitivity is universal regardless of intrinsic profiles. In other words, as long as the S/N is insufficient, \bb\ would tend to underestimate the $\sigma$.
In contrast, {\gp} still suffers from the template mismatch problem, as seen in the baseline models recovery, characterized by the same ``L-shaped" tail at the low dispersion end, albeit to a lesser extent. This is because most intrinsic profiles have declining slopes, but are much shallower than the profile assumed in \gp\ (Eqn.~\ref{eqn:gp_disprof}). Meanwhile, {\dy} performs similarly to the baseline models, but with a slightly stronger dependence on \reff/beam and $i$. This additional exercise highlights once again the limitations of parametric modeling when the assumed template deviates substantially from the truth. In such cases, non-parametric modeling may be more effective, although it requires a higher S/N.

\begin{figure*}
    \centering
    \includegraphics[width=0.95\textwidth]{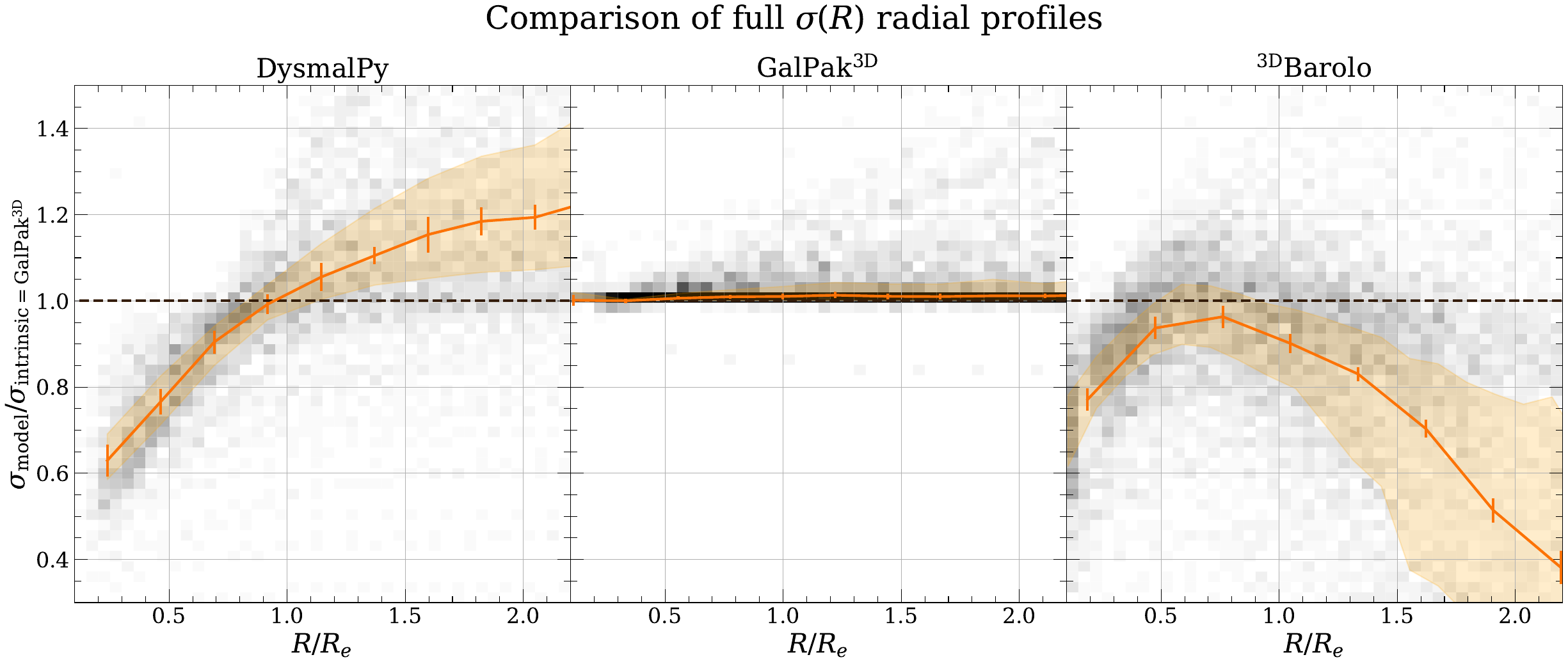}
    \caption{Similar to Figure~\ref{fig:dispvel_profile_radial_dymock}, but the mock galaxies are generated by {\gp}. The increasing trend in the \textit{left} panel, opposite to the decreasing trend in the \textit{middle} panel of Figure~\ref{fig:dispvel_profile_radial_dymock}, is due to the template mismatch that is described in detail in the main text (see also Figure~\ref{fig:dispprof_example} for the difference in profile shapes). {\gp} recovers itself excellently with the running median close to unity up to large radii, albeit with a larger scatter. \bb\ exhibits a similar radial trend as in the same panel in Figure~\ref{fig:dispvel_profile_radial_dymock}.
    }
    \label{fig:dispvel_profile_radial_gpmock}
\end{figure*}

\begin{figure*}
    \centering
    \includegraphics[width=0.95\textwidth]{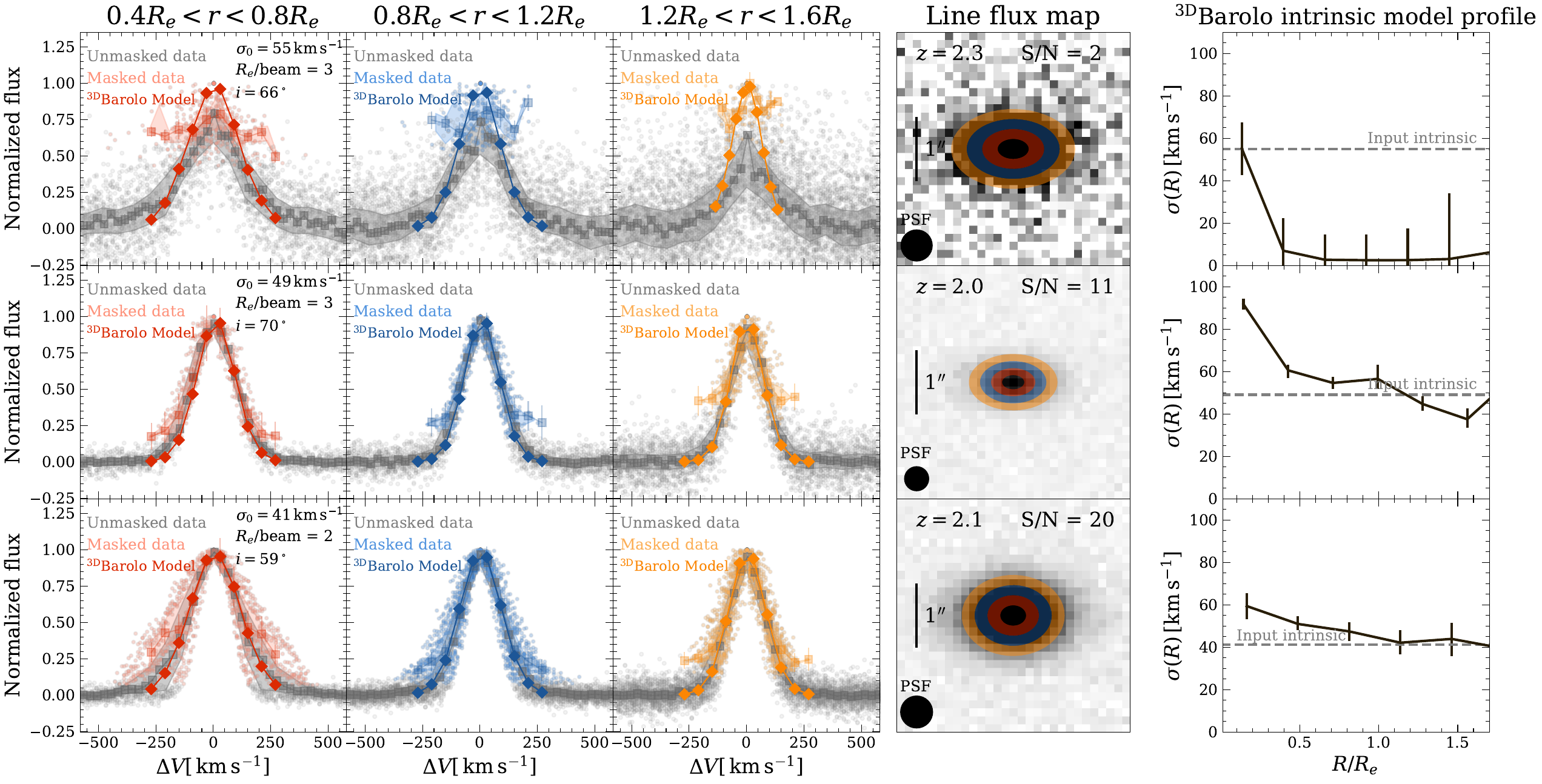}
    \caption{Velocity-shifted spectra of three {\dy}-generated mock galaxies, extracted from spaxels in 3 radial elliptical bins annotated on top of the \textit{first three} columns. 
    The spectra of the original unmasked mock data cube are represented by individual gray points, while the \bb-masked spectra (hereafter, masked data) are shown in pale red, blue, and orange points for each radial bin, respectively, across 3 different signal-to-noise (S/N) levels (S/N=$[2,11,20]$) in each row. 
    The axis ratio of the bins is determined by 2D S\'ersic fitting of the line flux map displayed in the \textit{fourth} column, overlaid with the color-coded radial bins. 
    The curves with squares and shading overlay on the spectra are the median trends, and the central $68\%$ 
    of the full distributions for both the unmasked and masked data. 
    The solid curves with diamond markers represent the velocity-shifted spectra extracted from the \bb's model. 
    The \textit{fifth} column shows the model intrinsic (accounted for observational effects) dispersion profile of the respective galaxies, with radii normalized to the effective radius $R_e$.
    At large radii, the line profile can be severely under-sampled because of marginal S/N, 
    where in the extreme case, only the brightest few pixels remain to determine the line width. This causes a systematic underestimation of the line width at large radii.
    Consequently, across all three galaxies, a clear radially declining trend is evident in the recovered dispersion profiles, with the steepest drop observed in the lowest S/N model. 
    The position-velocity diagrams and channel maps of these three galaxies are shown in Figures~\ref{fig:channelmaps} and \ref{fig:pv_diagrams} in Appendix \ref{app:pv_chan}, respectively.}
    \label{fig:bbmasked_spectrum}
\end{figure*}

\subsection{Further Insights from Full Radial Profiles}\label{subsec:radial_prof}

We have discussed so far the recovery of $V_{\rm rot}$ and $\sigma$ by the three codes at $1R_e$. Here we explore the full radial profiles of $V_{\rm rot}$ and $\sigma$ of the three codes. The full $V_{\rm rot}$ and $\sigma$ profiles would be essential in mass decomposition, one of the key applications of kinematic modeling. 

Fig.~\ref{fig:vel_profile_radial_dymock} shows the running medians and 1-$\sigma$ spread of $V_{\rm rot,model}/V_{\rm rot,intrinsic}$ over the radial range $[0,2.2]R_e$ for each of the packages for the baseline models. 
Because the same parametric models are used, \dy\ recovers itself well. {\gp} shows monotonically increasing median values and scatter from small to larger radii. The template mismatch between \gp\ and \dy\ becomes more apparent when $R \gtrsim$\,$1.25R_e$, as the choice of $\arctan$ in \gp\ only accommodates rising to flat shapes, in comparison with \dy\ multicomponent mass models, 
which can take on a variety of RC shapes, from rising to flat to declining. 
For {\bb}, the median increases with radius, albeit more mildly and with a smaller scatter. Too few galaxies are modeled by {\bb} at $\gtrsim$\,$2 R_e$, so values are not plotted in this range. 
Overall, \bb\ recovers the intrinsic $V_{\rm rot}(R)$
within $\lesssim 10\%$ accuracy, thanks to the flexibility of the tilted-ring approach that can adjust to any shape of the $V_{\rm rot}(R)$ rotation curve. 
The mild increasing trend is likely attributed to two factors  (i) known limitation of \bb\ when applied to thick disk (see Sec.~7.1 in \citet{Iorio2017}) (ii) residual beam smearing correction in \bb\ as noted in \citet{Varidel2019}. 
Factor (i) is in general true for tilted-ring modeling 
as also discussed in Sec.~4.5 in \citet{Jozsa2007} for \texttt{TiRiFiC}.
Both of these effects would underestimate the inner velocity gradient while overestimating it at the outer disk.

Similarly, we compare the radial variation of $\sigma$ in Fig.~\ref{fig:dispvel_profile_radial_dymock} for the recovery of the baseline models and Fig.~\ref{fig:dispvel_profile_radial_gpmock} for the \gp\ generated mocks. As expected, \dy\ and \gp\ show very good self-recovery when modeling their respective mocks. The expected template mismatch signatures are apparent when \gp\ models the baseline mocks, 
and vice versa, as reflected by the declining trend in the 
middle panel of Fig.~\ref{fig:dispvel_profile_radial_dymock} and rising trend in the left panel of Fig.~\ref{fig:dispvel_profile_radial_gpmock}, 
both with a typical transition radius at $\sim$\,$R_e$. Comparing $\sigma$ at $R_e$ as we did in the previous discussion should generally reduce the effect of template mismatch, 
albeit the scatter is still significant. 

The non-parametric \bb\ shows consistent radially declining behavior regardless of the mock model suite.
\bb\ best recovers the input $\sigma$ at $0.75R_{\rm e}$.
At $1R_e$, $\sigma$ is typically underestimated by $\sim$\,$10$--$15\%$ and worse at larger radii, accompanied by larger scatter. 
The S/N rapidly decreases with radius in the models, and \bb's sensitivity to S/N becomes increasingly apparent at larger radii. We discuss the implications of this trend on estimates of intrinsic velocity dispersion based on the full-fitted dispersion profile in Sec.~\ref{sec:discussion}.
Towards smaller radii, the systematic overestimate in $\sigma$ and 
the small but systematic underestimate in $V_{\rm rot}$ is the
result of its tilted-ring approach. The approach constrains parameters for each ring independently 
and is unable to account for disk thickness accurately. This limitation is discussed in detail in, for example, Sec.~5.1.2 of \citet{Roper2023} and \citet{Iorio2017}.

\subsection{Effect of masking in the recovery of $\sigma$}\label{subsec:masking_discussion}
We next explore an alternative masking routine in {\bb}. We did not apply smoothing before generating the mask to boost consistency across codes (by specifying \texttt{SEARCH} in the \texttt{MASK} option), which is also the default option in {\bb}. Nonetheless, \texttt{SEARCH} is commonly used in the literature \citep[e.g.,][]{Sharma2021,Hogan2022,Rizzo2023,Pope2023,Roman-Oliveira2023}. 
Another popular choice is \texttt{SMOOTH\&SEARCH} \citep[e.g.,][]{ManceraPina2020,Fraternali2021,Lelli2023}, where \bb\ first smooths the data cube before calling the source finding algorithm. 
Repeating our recovery tests instead using \texttt{SMOOTH\&SEARCH} masking (with default option \texttt{FACTOR=2}, which doubles the beam size) shows little difference in best-fit results, with similar trends obtained without smoothing
illustrated in Fig.~\ref{fig:running_med_disp}.
A more detailed inspection of {\bb} indicates that both masking routines 
often lead to over-masking in velocity, with only a few of the brightest channels being passed to the fitting algorithm. We verified that this behavior remains with different choices of 
\texttt{MINCHANNELS} in the \texttt{SEARCH} routine.

To illustrate the impact of spectral masking on low S/N data more clearly, we present spectra extracted from individual pixels from inner to outer regions (in elliptical annuli with axis ratio of the bins set by 2D S\'ersic fitting of the line flux map) of three selected galaxies in Fig.~\ref{fig:bbmasked_spectrum}. These galaxies share similar \ReRbeam, inclination ($\sim$\,$60^\circ$--$70^\circ$), and redshifts ($z$\,$\sim$\,$2$) but differ in S/N, from low S/N (S/N = 2) to high S/N (S/N = 20).
The spectra are shifted in velocity space to align with a common normalized emission line peak, and we refer to them as ``velocity-shifted spectra" to distinguish them 
from more typical integrated spectra extracted in circular aperture.
The gray points and the solid line with square markers represent the individual velocity-shifted spectrum and the running median trends of this ensemble of the spectrum. The data left after masking is applied are shown in colored points with the running median trend overlaid. 
The solid-colored curves are the median trends of the velocity-shifted spectra extracted from the \bb\ model cube.
The recovered $\sigma(R)$ profiles by \bb\ are shown in the last column of Fig.~\ref{fig:bbmasked_spectrum}. 
Additionally, we provide the position-velocity (PV) diagrams extracted using the \texttt{PVSLICE} task in \bb, as well as channel maps at every 4 channels in Figs.~\ref{fig:channelmaps} and \ref{fig:pv_diagrams}, respectively, in Appendix \ref{app:pv_chan}. We also show those from \dy\ and \gp in addition to \bb\ results. These diagnostics are supplementary to Fig.~\ref{fig:bbmasked_spectrum} to demonstrate the quality of the fitted model, particularly for $V_{\rm rot}$, but Fig.~\ref{fig:bbmasked_spectrum} is more informative of the effects of masking on velocity dispersion.

At higher S/N, the impact of spectral masking is minimal, and the recovered dispersion profile at large radii closely resembles the intrinsic constant profile. 
The upward trends (second and third panels in the last column) towards the inner region could be due to residual correction of beam smearing, consistent with the behavior in Fig.~8 in \citet{DiTeodoro2015} for similar inclinations.

At lower global S/N or towards larger radii within the same galaxy, 
the aggressive clipping of high- and low-velocity wings below 
the same specified S/N inevitably leads to 
underestimating the true velocity dispersion.
Within the same galaxy, this could manifest as an apparent radially declining profile, as illustrated in the first and second panels in the last column.
Indeed, as noted already in \citet{DiTeodoro2015}, a satisfactory fit at low S/N typically requires a channel count ranging from 8 to 12, depending on the spatial resolution, and the source should be detected over multiple channels with S/N $\gtrsim$\,$3$. This highlights the importance of validating model results with direct data-based measurements, particularly at large radii, similar to the velocity-shifted spectra presented here.

To further investigate the effects of spectral masking, we exchanged the masks between \dy\ and {\bb}. 
Given that \dy\ determines the mask based on the integrated S/N along a spaxel, it does not apply spectral masking, 
ensuring the same number of channels are modeled for every included spaxel, following the recommendation of \citet{Davies2011}.
We multiplied the mock cube with {\dy}'s mask and set \texttt{MASK} to \texttt{NONE} in {\bb}. 
To complement the test, we repeat the same exercise for \gp\, by multiplying the input mock cubes by the {\bb}'s masks.
When such a mask is used in \bb\ and S/N approaches $\sim$\,$30$, {\bb} and {\dy} generated masks are in good agreement. If the mask generated by \bb\ is applied to \dy\ and {\gp}, we recover a similar trend with S/N as in {\bb} when its mask is used. 
Overall, the S/N dependence of {\bb}'s recovery of $\sigma$ vanishes, as evidenced by the flat trend with S/N in Fig.~\ref{fig:running_med_disp} (leading to a very low Spearman's and MIC score).
Contrary to the previous underestimation trends, 
the systematic overestimation observed now is likely due to {\bb}'s sensitivity to the available valid pixels for modeling when S/N is limited.

In summary, we tested (1) \texttt{SMOOTH\&SEARCH} masking, (2) vary the \texttt{MINCHANNELS} parameter, and
(3) \dy\ masking (entire spaxel masking using integrated S/N).
We found no significant difference in recovery trends and scatter for (1) and (2),
and the dependence on S/N was appreciably reduced when (3) was adopted.

This simple experiment underscores the critical role of masking in $\sigma$ recovery across all codes, with non-parametric methods exhibiting particular sensitivity. 
The general impact of masking has been extensively discussed in a similar context, notably by \citet{Davies2011}, \citet{Deg2022}, and Sec.~7.3 (Fig.~14) in \citet{deBlok2024}. 
It is unsurprising that fitting is affected by the data to which the fitting is applied.
Whether and how (e.g., \dy\ and {\bb}) or not (e.g. {\gp}) masking is applied, this aspect should be taken with extra care when analyzing and interpreting fitting results.

Since all codes yield good agreement toward very high S/N, regardless of the masking routine, deeper integration data is ideal for reducing systematic differences across different codes.

\subsection{Effect of alternative settings in the recovery of $\sigma$ for \bb}\label{subsec:alt_settings}

To maintain consistency in comparing the best-fit results of the three kinematic modeling tools
discussed in this paper, we adopted settings for \bb\ that come closest to those implemented
in \dy\ and {\gp}. However, these are not necessarily those most commonly employed in the literature.
This pertains in particular to the metric employed for the goodness of fit and the masking procedure.
First, we experiment with a different residual function for minimization in {\bb}. While there are three options listed in Sec.~\ref{subsec:bb}, we opted for the pseudo-$\chi^2$: $(M-D)^2/\sqrt{D}$.
The default option in {\bb}, the absolute difference $|M-D|$, is also a common choice in the literature \citep[e.g.,][]{Lelli2023,Posses2023}. 
Similar to our findings using other minimization settings, our results reveal a similar trend with S/N and other parameters regardless of the residual function used.

 \section{Implications For The Physical Properties of High-\lowercase{$z$} Galaxies}\label{sec:discussion}
The recovery tests presented in this paper were performed with sets of 3D mock models
covering a realistic range of galaxy properties for massive MS SFGs at \znoon, and
representative observational parameters with emphasis on higher S/N than many current observations.
Intentionally, we used very large suites of simple axisymmetric models and a minimum of free parameters to robustly assess trends and scatter in the relative performance of the tools and to reduce the impact of known degeneracies affecting all modeling approaches (e.g., between mass and inclination).

Tests based on observations would be very informative, but for high-$z$ galaxies
the intrinsic parameters would themselves be derived from modeling and thus would
require the highest resolution and S/N possible.  Very few such data sets currently
exist as they rely on very deep observations (difficult to obtain) or very bright
targets (rare).  Data of local disk galaxies at very high S/N and resolution enable
a better characterization of intrinsic kinematics, but because of the different 
conditions prevailing at higher redshift (higher accretion rates, cold gas fractions,
and star formation rates; smaller disk sizes), local disks may not provide the most
realistic templates at high-$z$. 
Numerical cosmological simulations of galaxy evolution may better capture the
conditions and complexities of real high-$z$ disks, although their use is not
straightforward.  Complications include the mismatch in how
properties are derived between simulations and observations, and the reliance
on sub-grid recipes that are typically tuned to reproduce the final stage of
present-day galaxy properties
(see discussions by, e.g., \citealt{Wellons2020} and \citealt{Uebler2021}).

Nevertheless, our experiments highlight 
a key aspect that should be kept
in mind when interpreting high-$z$ kinematics data, especially in the context of two
key applications: 
challenges determining the disk velocity dispersion and the resulting implications for mass decomposition. 
Our tests suggest that it remains difficult to pin down the disk velocity dispersion
to better than $\sim$\,20$-$50\% depending on the tool considered, unless there is
a close match between the true and the assumed model profile and the S/N per spatial
pixel at the line peak velocity is above 10 over a sufficiently large and well
resolved radial range.  
In \dy, the intrinsic velocity dispersion (\sigo) is assumed to be isotropic and
spatially constant across disks, representing a dominant turbulence term.  
This assumption was empirically motivated by high S/N, adaptive optics (AO) assisted IFU
observations of \Ha\ resolved on $\sim$\,$1$ kpc scales of large
$z$\,$\sim$\,2 MS SFGs from the SINS/zC-SINF survey, after accounting
for instrumental and beam smearing effects through modeling \citep{Genzel2017,Uebler2019}.
In those galaxies, no significant spatial variation was observed in derived intrinsic 
velocity dispersion beyond the innermost radii (where residual beam smearing could still
play a role) out to a few $R_e$.  Similarly, no evidence for appreciable radial variations
in intrinsic dispersion to $\sim$\,10\,kpc was found from modeling of a much larger sample
of 240 \znoon\ MS SFGs with deep integrations in excellent near-IR seeing of
$\approx 0\farcs 5$ from the KMOS$^{\rm 3D}$ survey \citep{Wuyts2016}.

As the dispersion values
 derived for these galaxies are large ($>$\,30\,km\,s$^{-1}$),
in line with expectations in the framework of gas-rich marginally (un)stable disks
with the typically high \fgas\ at high redshift, the contribution from self-gravity
for radially decreasing mass densities is sub-dominant (see Fig.~\ref{fig:dispprof_example})
and would be difficult to discern as it would be small and apparent only in the 
innermost regions.  
If the full velocity dispersion is intrinsically declining, as considered by
\citet{Rizzo2020,Rizzo2021}, any significant radial gradient would have left
a measurable systematic trend in 
the observed dispersion profiles and residuals in the best-resolved
galaxies examined by \cite{Genzel2017} and \cite{Uebler2019}, which was not observed.

Ideally, non-parametric modeling would be best suited to examine galaxies' dispersion profiles.
Some past studies reported radially declining velocity dispersion using \bb\ \citep[e.g.,][]{Lelli2021,Rizzo2023,Roman-Oliveira2023}, with profiles similar to those of local \Hi\ and \co studies \citep[e.g.,][]{Boomsma2008,Tamburro2009,Wilson2011,Mogotsi2016}.
If so, this highlights the challenge of distinguishing radial variations due to the intrinsic profile versus the potential impact of S/N and beam smearing.
In this context, we note that our recovery tests compared $\sigma$ at a fixed radius
between the tools, taken as \reff\ in Sections \ref{subsec:general_results}--\ref{subsec:lightclump}, to maximize consistency.
The common convention for calculating $\sigma$ from \bb\ modeled profiles includes taking the mean or median value of all modeled rings or, less commonly, the average
of the two outermost rings
\citep[e.g.,][]{DiTeodoro2016,Iorio2017,Fraternali2021,Lelli2021,Rizzo2023,Sharma2023,Neeleman2023,Roman-Oliveira2023}.
Repeating our exercise using the former two definitions, 
we find
no significant difference in trends
identified when using the value around $1R_e$.

Determinations of the intrinsic velocity dispersion of high-$z$ disks are important
in the context of disentangling the drivers of disk gas turbulence (stellar feedback,
gas transport induced by internal gravitational/disk instabilities or by external
accretion), the relative contribution of which is thought to vary with redshift
\citep[e.g.,][]{Krumholz2018, Genzel2011,Hung2019,Ginzburg2022,Jimenez2023}.
It is also of interest in terms of the disk thickness, as more pressure support
implies geometrically thicker disks such that elevated dispersion in high-$z$ disks may be linked to the formation of today's thick disk components \citep[e.g.,][]{ElmegreenElmegreen2006,Bournaud2009}.
Moreover, if hydrostatic equilibrium holds, radially constant disk dispersion should
imply disk flaring at larger radii.  Ultimately, substantially higher sensitivity
and both angular and velocity resolution would be needed to accurately pin down 
disk velocity dispersions and their evolution at high redshift.  In the meantime,
one way forward could be to obtain deep data at (sub-)kpc resolution and
$R$\,$\ga$\,10000 (instrumental LSF of $\sigma$\,$\la$\,15\,km\,s$^{-1}$)
of low-inclination galaxies, reducing the contribution to emission line broadening
from projected rotation and even allowing model-independent estimates directly from
observed line widths.

Knowledge of the velocity dispersion and its profile is also important for dynamical 
mass estimates and mass modeling.
Since the rotation velocity curve \vrot($R$) is fairly well recovered by all
three tools, as long as the pressure support is small (i.e., \vrot\ approximates
$V_{\rm circ}$ as a tracer of the full potential well), results for $M_{\rm dyn}$ 
and mass decomposition derived from \dy, \gp, and \bb\ kinematic modeling should agree
very well for the same assumptions on the underlying mass distribution.
However, if dispersions are elevated, and especially relative to \vrot, results based on modeling with the different tools may lead to different conclusions. 
This could arise from possible significant misestimates in recovered
velocity dispersion depending on template mismatch (for parametric models like \dy\ and \gp), and on the S/N regime and inclination (most relevant for \bb).
The other important reason lies in the treatment of the pressure support, which can differ between studies \citep[e.g.,][]{Burkert2016,Iorio2017,Kretschmer2021,Price2021,Sharma2021}. Additional factors are obviously the specific choice for the mass model component(s).

Empirically-motivated pressure support corrections and well-constrained mass components from high-$z$ data of stars, warm and cold gas distributions, and kinematics would be ideal but may need to await future more powerful and efficient observational capabilities.
At the very least, deep observations probing as far out as possible in radius will help by providing better leverage for the relative contributions of DM and baryons, and sub-kpc resolution can tighten constraints on the inner core/bulge component.  
Independent observations of multiple ISM phases are valuable in
augmenting the constraints, giving better priors for the gas (which makes up an important fraction of the baryonic component at high-$z$), and complementing each
other in tracing radial coverage and mitigating optical depth effects.  These are
within reach of current facilities; time estimates for typical massive MS SFGs with
ERIS, ALMA and NOEMA imply on-source integration of $\sim$\,10$-$20\,hours,
and such deep observations already carried out demonstrate their potential \citep[e.g.,][]{Genzel2017,Genzel2020,Genzel2023,Uebler2018,Nestor2023,Puglisi2023}. Even stellar kinematics are feasible up to at least $z$\,$\sim$\,$1$ \citep[e.g.,][]{vanHoudt2021,Straatman2022,uebler2024} for very deep integrations.

 \section{Summary}\label{sec:conclusion}
We have assessed the 3D kinematic recovery performance of the kinematics modeling tool \dy, which is publicly released as part of this work. We also compared its performance with two other packages whose methodologies are based on different motivations but share the similarity of being 3D forward-modeling algorithms: \gp and \bb. We simulated a large number of mock disk galaxies
matching the galaxy parameter space and S/N and resolution distributions of
a sample of deep \znoon\ MS SFGs \citep[RC100;][]{Nestor2023},
which includes high-quality kinematics data
sets from near-IR IFU and mm interferometry.
Our experiment intentionally kept a minimum number of free parameters to help highlight the root causes of potential differences in different modeling approaches to mitigate them in applications to real data.
We focussed on evaluating the reliability in recovering
the important kinematics properties: $V_{\rm rot}$ and $\sigma$. We summarize the key results as follows:

\begin{enumerate}[label*=\arabic*.]
    \item Recovery of $V_{\rm rot}$ at $R_e$ is largely independent of the choice of modeling tool (e.g., Fig.~\ref{fig:violin_both} in Sec.~\ref{subsec:general_results}).
    In terms of recovery of the full $V_{\rm rot}$ profile, unsurprisingly, template mismatch can affect parametric modeling with \dy\ and \gp\, whereas
    \bb\ can accommodate different $V_{\rm rot}$ shapes more easily (Sec.~\ref{subsec:radial_prof}).

    \item  The recovery of $\sigma_{\rm model}/\sigma_{\rm intrinsic}$ from different tools can vary significantly depending on $\sigma_{\rm intrinsic}$, S/N, and $i$
    (Sec.~\ref{subsec:biassource}). The disagreement between {\gp} and {\dy} can be primarily explained by the inherent template mismatch (Sec.~\ref{subsec:impact_parametrisation}, Fig.~\ref{fig:gp_mocks_dispratio} and \ref{fig:gp_sigprof_gradient}). The flexibility of {\bb} is hampered by its stringent demand on S/N,
    which is challenging to fulfill for high-$z$ galaxies with typical allocated observing time. 
    Aspects to be cautious about include: 
\begin{itemize}
      \item the choice of parametric functions, for example, constant vs.\ radially varying velocity dispersion profile. The choice should be informed ideally from empirical evidence, which is still scarce at $z$\,$\gtrsim$\,$1 $;
      \item the flexibility of non-parametric methods comes 
      with higher S/N requirements for a robust recovery. If S/N is insufficient, tilted-ring modeling is more sensitive to the masking choice.
    \end{itemize}
\item The presence of light clumps (Sec.~\ref{subsec:lightclump}) 
    can affect the results from parametric models due to the inflexibility of the analytic light profile. Non-parametric modeling, on the other hand, is less systematically affected by asymmetric light distributions, although we find there is still a large scatter in the recovered-to-intrinsic values. The non-parametric model flexibility is still primarily hindered by poor sensitivity in low S/N samples.

\end{enumerate}

Based on our recovery exercise, we strongly recommend that, before applying any modeling to real data, it is crucial to assess the impact of
\begin{itemize}
\item  any prior assumptions on radial kinematics profiles, especially when these are parametrized (intrinsically or otherwise);
\item  masking using model-independent diagnostics, such as the example shown in Fig.~\ref{fig:bbmasked_spectrum};
\item  S/N on the recovered properties, especially towards the outer edges of the detected regions;

\end{itemize}
These steps additionally help to gauge whether the adopted template and the resulting modeled values deviate significantly from the data, thereby mitigating template mismatch and S/N sensitivity issues, as discussed above.

The mock models used in this paper are idealized in many respects: they are axisymmetric, and the center, PA, and $i$ are known (fixed in modeling). Those quantities are, however, difficult to recover in reality from low S/N and resolution data with highly irregular light distributions. 
In our study, we only tested one out of many possible scenarios of asymmetry by introducing a fixed number of light clumps. 
Real galaxies, however, can possess a range of different features, such as rings and bars.
Nevertheless, our exercise of using simple mock models here should shed light on the discrepancies of the measured kinematic quantities, especially the velocity dispersion across studies, when the same set of galaxies is analyzed.

In light of the significant difference in recovering the velocity dispersion between different modeling approaches, 
we need standardized metrics and modeling assumptions for accurate comparisons between different samples.
Only then can we assess robustly the evolution of disk velocity dispersion with redshift, and correlations with stellar mass, star formation activity, and other galaxy properties.

Obtaining a subset of galaxies with enhanced spectral resolution and higher S/N to larger radii is crucial for constraining whether velocity dispersion varies radially. This information is pivotal for selecting an appropriate template in parametric modeling. 
Deep observations with radio interferometers such as NOEMA and ALMA
are well-suited for that purpose. The
near-IR IFU ERIS on the VLT is also ideal, affording a spectral resolution capability of $R$\,$\sim$\,$11000$ 
and, combined with adaptive optics, a high spatial resolution with a high
Strehl ratio.  High S/N and resolution are as important as sample size for
characterizing the global velocity dispersion of high-$z$ disks, spatial
variations, and the origin of scatter among galaxies.

\begin{acknowledgments}
We thank the anonymous referee for the constructive comments.
We are very grateful for the useful discussions and insightful comments at various
stages of this work by Tim de Zeeuw, St\'ephane Courteau, Nathan Deg, Nicolas Bouch\'e, Emily Wisnioski, Jianhang Chen and Minju Lee.
N.M.F.S. and J.M.E.S. acknowledge financial support
from the European Research Council (ERC) Advanced Grant under the European Union's
Horizon Europe research and innovation programme (grant agreement AdG GALPHYS, No.\ 
101055023).
HÜ gratefully acknowledges support by the Isaac Newton Trust and by the Kavli Foundation through a Newton-Kavli Junior Fellowship.
\end{acknowledgments}
\vspace{5mm}
\software{DysmalPy \citep{Davies2004a,Davies2004b,Davies2011,Cresci2009,Wuyts2016,Lang2017,Price2021}, GalPak$^{\rm 3D}$ \citep{Bouche2015}, $^{\rm 3D}$Barolo \citep{DiTeodoro2015}, 
            Numpy \citep{harris2020}, MPFIT \citep{Markwardt2009},
            Scipy \citep{Virtanen2020}, Matplotlib \citep{Hunter2007}, 
            corner \citep{Foreman-Mackey2016}, Astropy \citep{Astropy2013},
            Imfit \citep{erwin2015},
            minepy \citep{Albanese2012}.
          }

\appendix

\section{Mock models setup}\label{app:mocksetup}
Table~\ref{tab:prior_table} lists the names and priors of the parameters in each code for the modeling as performed in this work. There are in total $3$, $2$ and $2$ free parameters in {\dy}, {\gp} and {\bb}, respectively.

\section{Distributions of mock galaxies' parameters}\label{app:mockparams}
\begin{figure}
    \centering
    \includegraphics[width=0.45\textwidth]{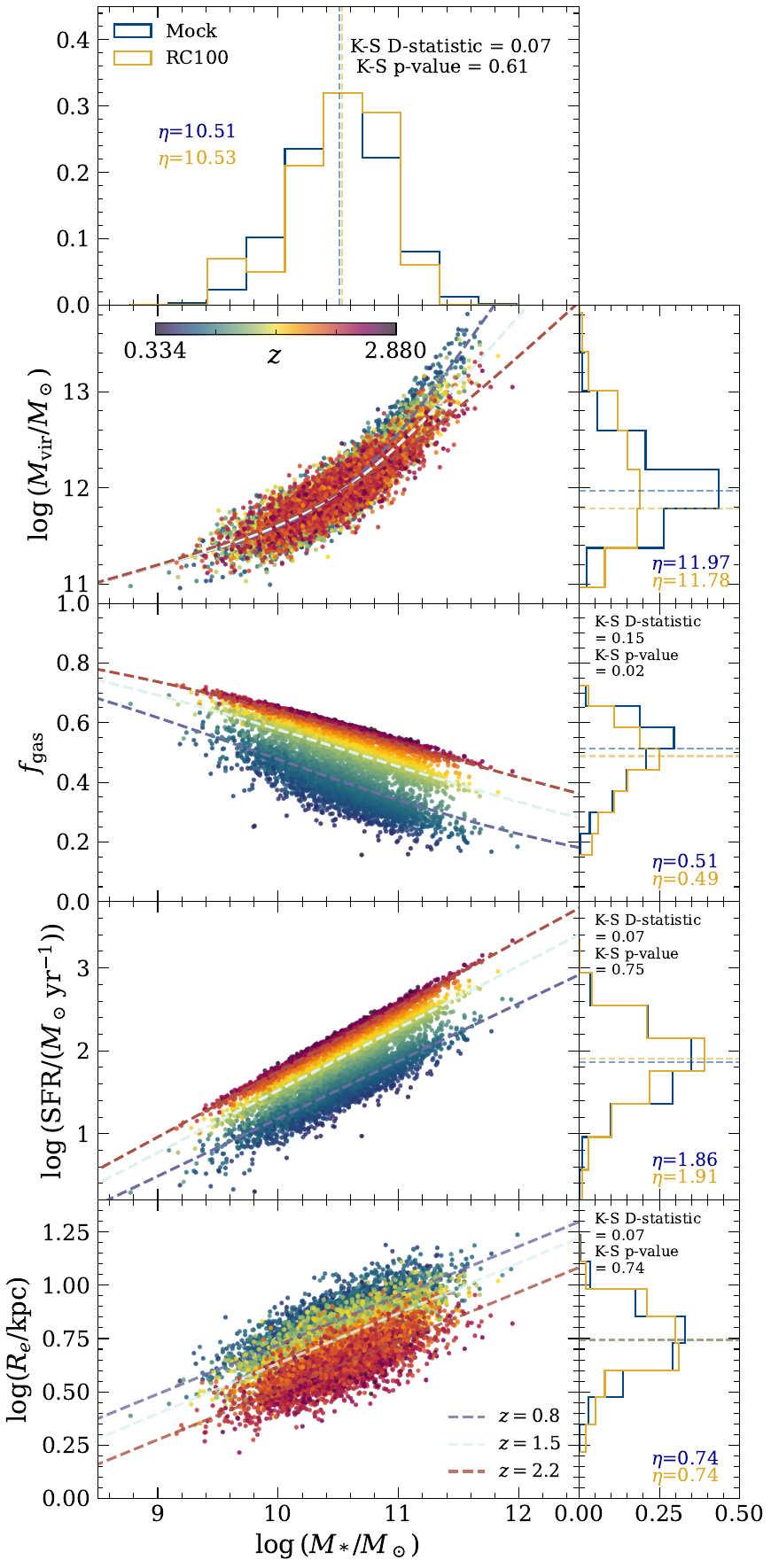}
    \caption{The distributions of stellar mass $M_*$, star formation rate (SFR), gas fraction ($f_{\rm gas}$) and virial mass $M_{\rm vir}$ of the baseline
    set of model galaxies generated with \dy.  The mock galaxy properties are
    guided by the parameter space coverage of the RC100 sample of 
    \citet{Nestor2023}, and derived from scaling relations as described in
    Sec.~\ref{sec:sample} and listed in Table~\ref{tab:param_table}.
    The empirical relations are plotted in dashed lines colored by redshifts. 
The mock sample (blue) is compared with the RC100 sample (yellow) in the histograms, annotated also by the Kolmogorov-Smirnov (K-S) statistic scores and the medians.}
    \label{fig:phy_distri}
\end{figure}

\begin{figure*}
    \centering
    \includegraphics[width=\textwidth]{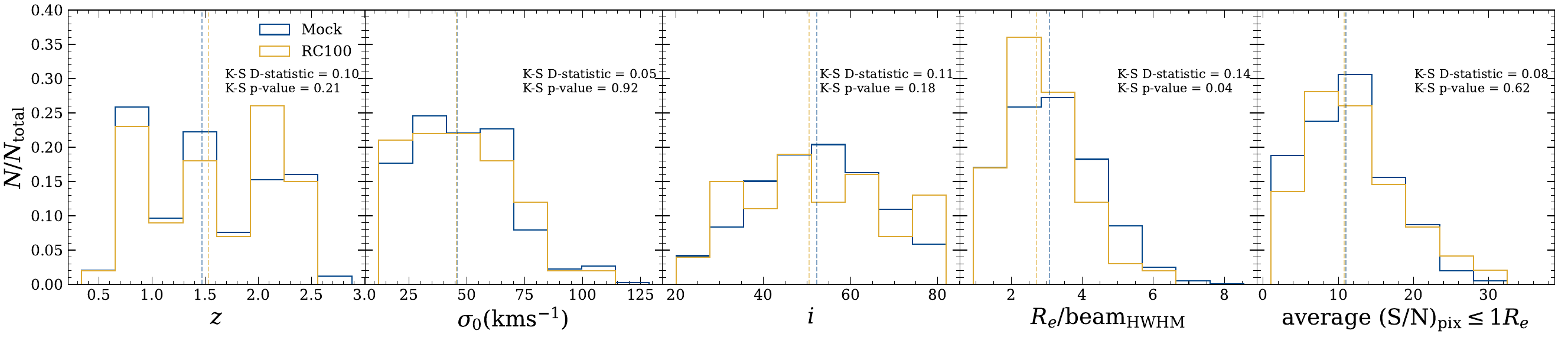}
    \caption{The distributions of redshift ($z$), velocity dispersion ($\sigma_0$), inclination ($i$), number of resolution elements within $R_e$ ($R_e/{\rm beam_{HWHM}}$) and signal-to-noise (S/N) of RC100 and the baseline mock sample. D-statistics and the associated p-value from the two-sample Kolmogorov-Smirnov (K-S) test are also shown to illustrate the resemblance of the resulting mock galaxies distribution and RC100.}
    \label{fig:hist_obs}
\end{figure*}

Fig.~\ref{fig:phy_distri} shows the distribution of the \dy-generated
baseline set of model galaxies in $M_*$, SFR, $f_{\rm gas}$, and $M_{\rm vir}$. Fig.~\ref{fig:hist_obs} plots the distribution of $z$, $\sigma_0$, $i$, $R_e$/beam$_{\rm HWHM}$, and S/N (in the brightest spectral channel and averaged over spaxels within \reff). In both figures, histograms compare the distributions of the baseline models with those of the RC100 disks, and the median values are indicated. The baseline sample's $M_*$, SFR, $z$, $R_e$/beam are statistically equivalent to the RC100 sample with K-S score $\lesssim 0.07$. In other parameters, although the K-S score is larger, the median values are in close agreement.

\subsection{Covariant distributions of parameters}\label{app:covar_plot}
Fig.~\ref{fig:input_param_space} shows the covariant distributions of $i$, \reff/beam, and S/N for the baseline set of model galaxies. The contours indicate $[1,2,3]\sigma$ of the distributions. Round-shaped contours imply the distributions are sufficiently randomized and not expected to introduce substantial biases in the analysis.  The histograms compare the original distribution of the full baseline set, as well as of the successfully modeled subsets by \dy, \gp, and \bb, indicating no important bias in the recovery analysis is introduced by the failed or excluded fits.

\subsection{Clumpy sub-sample}\label{app:clumpygal}
Fig.~\ref{fig:clumpgal} displays the integrated line intensity maps (0th-moment) of 12 randomly selected clumpy mocks generated by \dy.

\section{Recovery of $\sigma$ of high S/N mocks}\label{app:highsn_recover}
\begin{figure*}
    \centering
    \includegraphics[width=\textwidth]{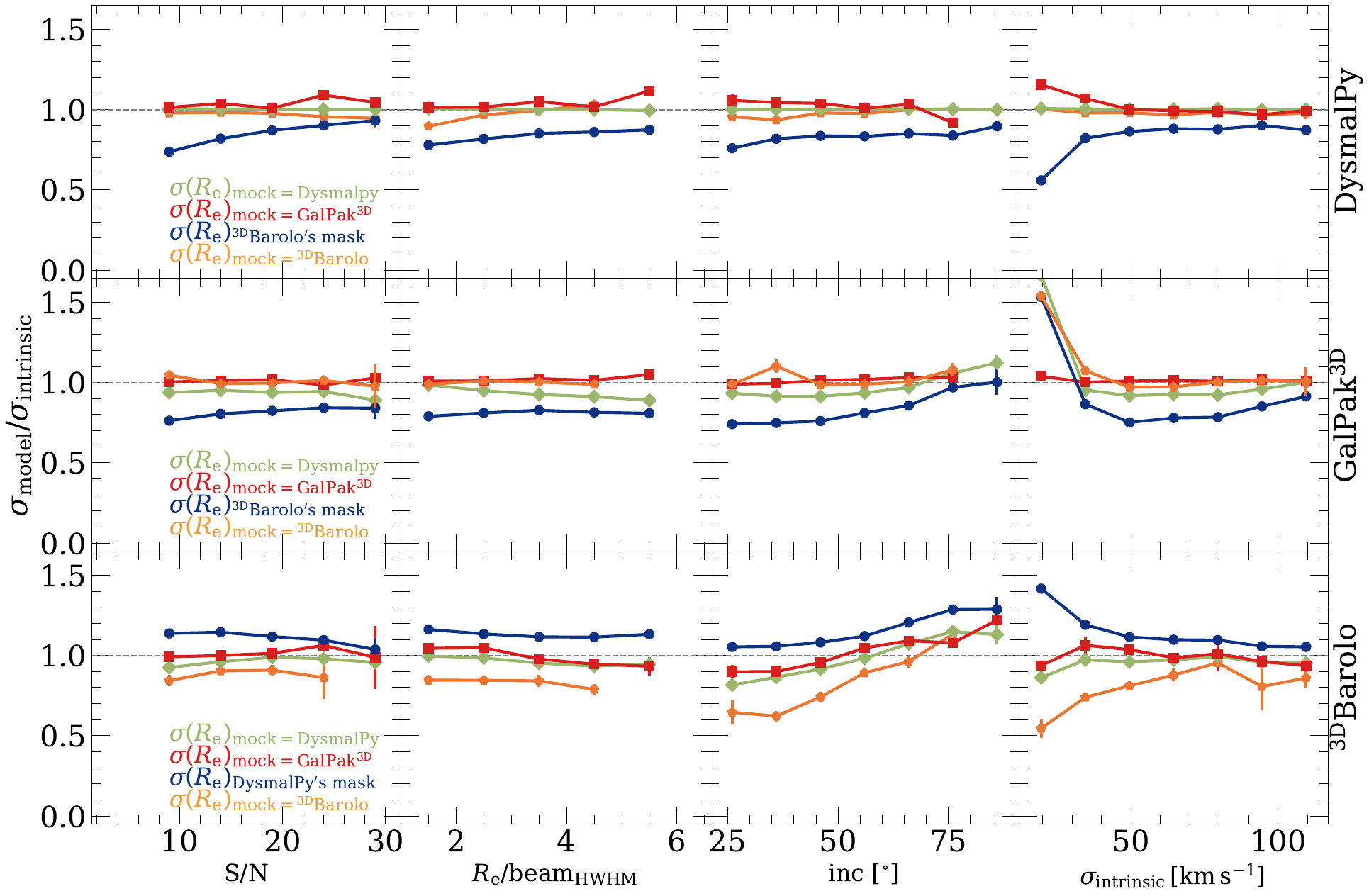}
    \caption{Similar to Figure~\ref{fig:running_med_disp} but only including mock models with signal-to-noise (S/N) $\geq11$.
    }
    \label{fig:running_med_disp_highSN}
\end{figure*}

\begin{figure*}
    \centering
    \includegraphics[width=0.8\textwidth]{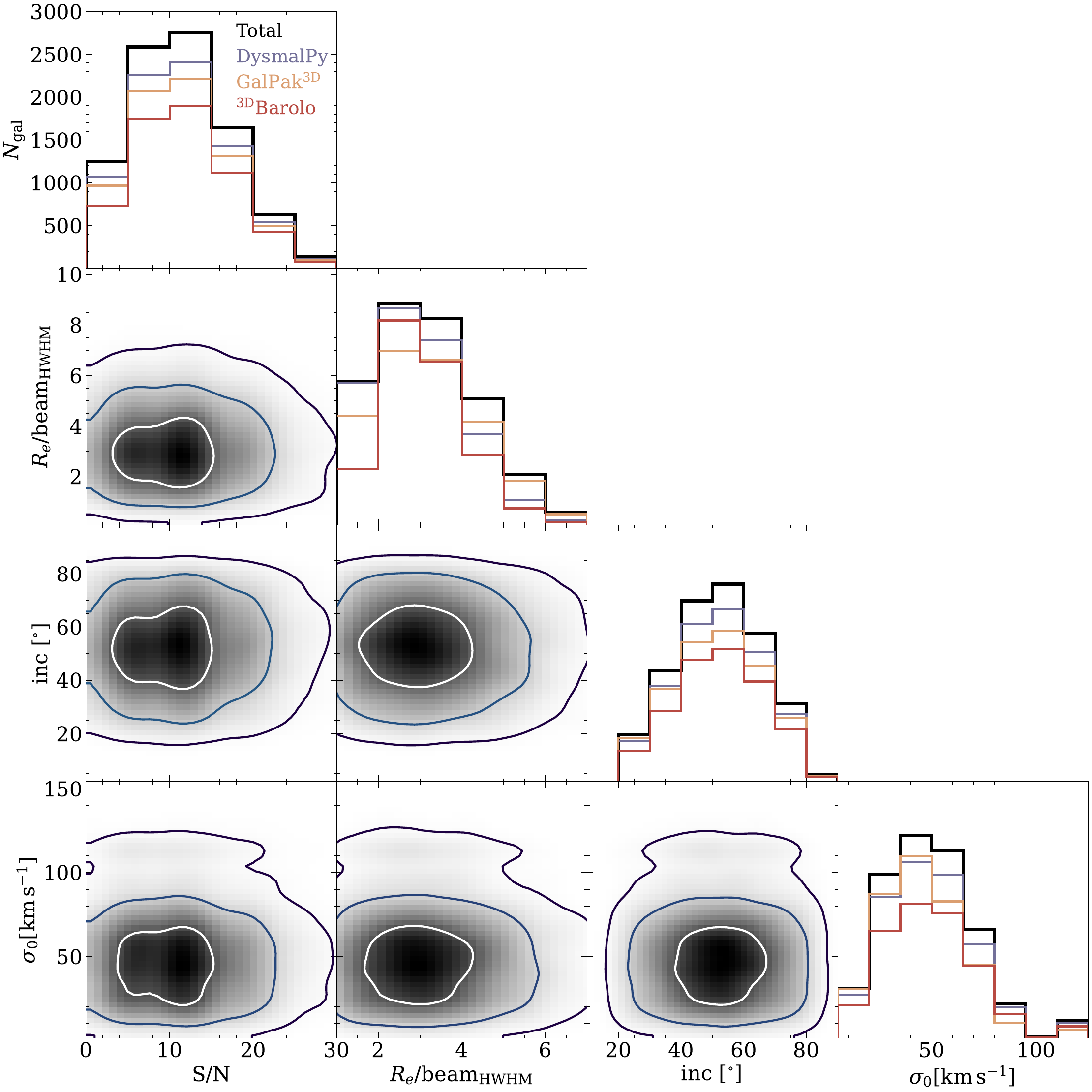}
    \caption{Corner plot showing the covariance distributions of velocity dispersion $\sigma_0$, inclination $i$, the number of resolution elements in one effective radius $R_e/{\rm beam}_{\rm HWHM}$ and signal-to-noise ratio (S/N) used for the baseline mock sample. 
    The contours correspond to $[1,2,3]\sigma$ of the distributions. 
    The histograms indicate the individual distributions of the original baseline samples and the successfully modeled samples by \dy, \gp, and \bb, respectively, in
    black, violet, orange, and red.
}
    \label{fig:input_param_space}
\end{figure*}

\begin{figure*}
    \centering
    \includegraphics[width=0.75\textwidth]{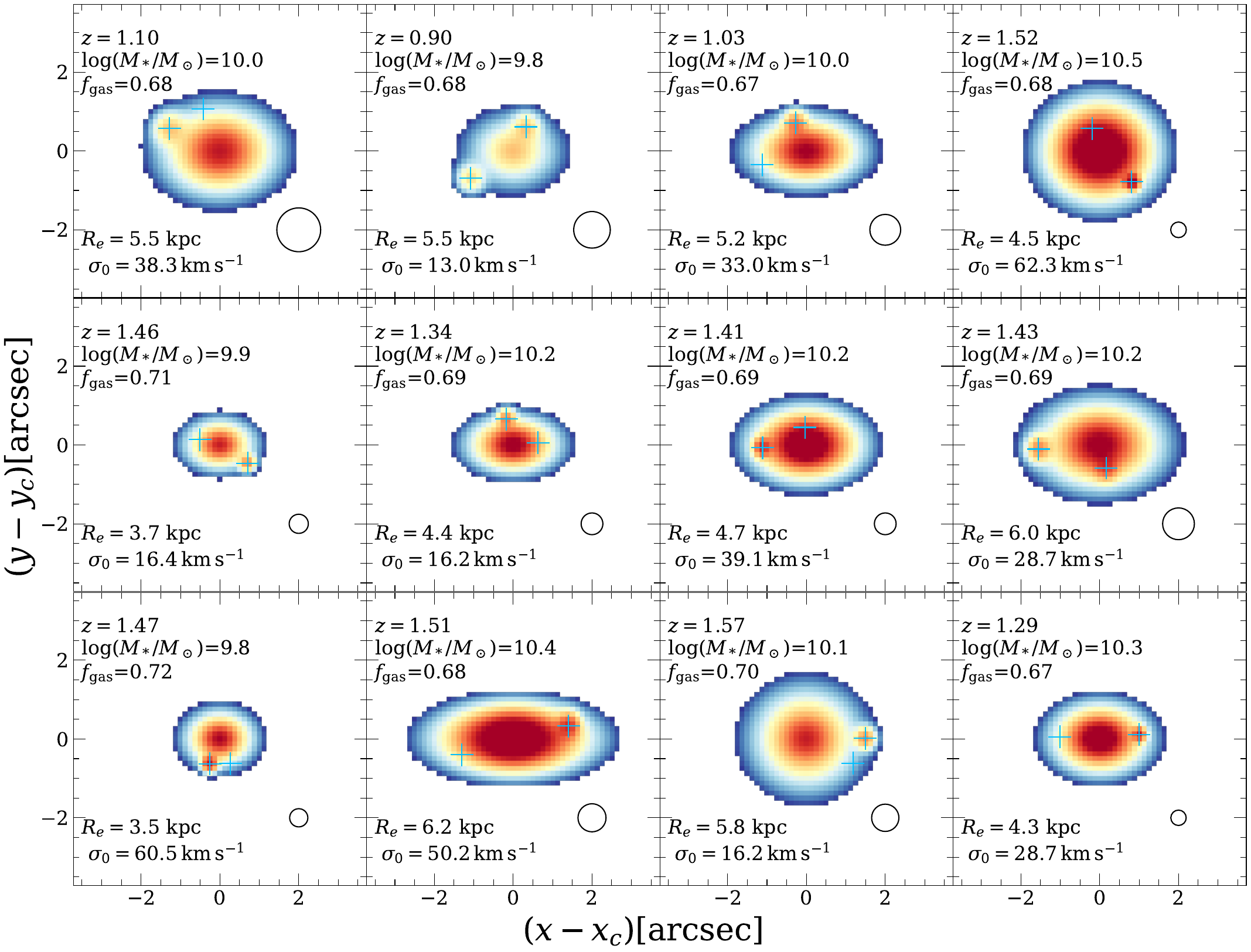}
    \caption{A gallery showing the (noiseless) zeroth moment maps of 12 selected clumpy galaxies out of a total of 500, created using \dy.
    The color map represents the light intensity from blue to red.
    Cyan crosses mark the locations of the clumps. The beam size is shown at the bottom right corner of each panel. The properties of the smooth galaxy component are also listed on the left.}
    \label{fig:clumpgal}
\end{figure*}

To supplement Fig.~\ref{fig:running_med_disp}, Fig.~\ref{fig:running_med_disp_highSN} shows the same median trends of $\sigma$ recovery but focuses only on the S/N $\geq11$ recovery.

\section{Spearman's and MIC's correlation matrices}\label{app:corr_matrix}
\begin{figure}
\centering
        \includegraphics[width=0.4\textwidth]{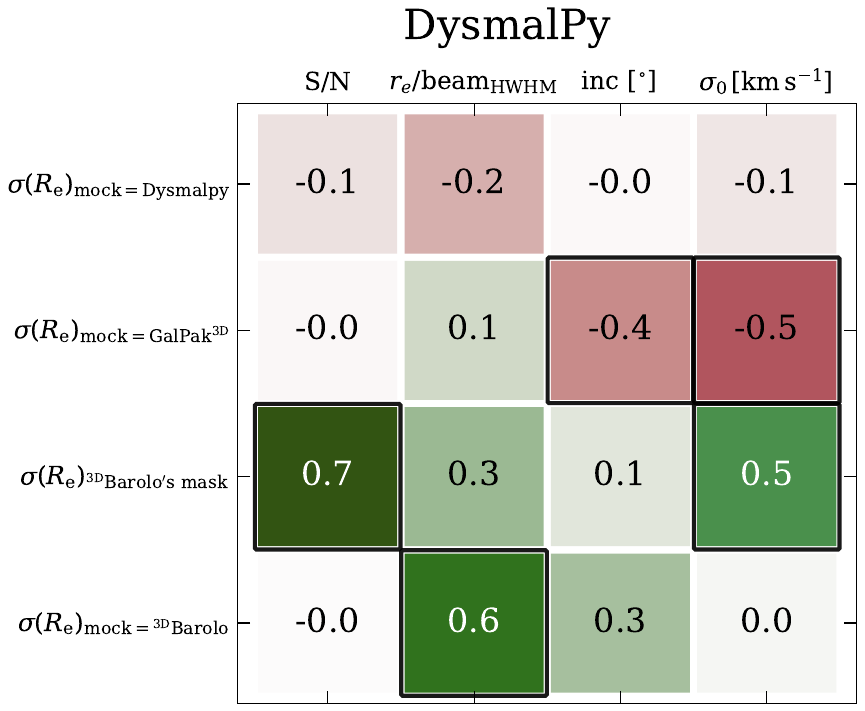}
        \includegraphics[width=0.4\textwidth]{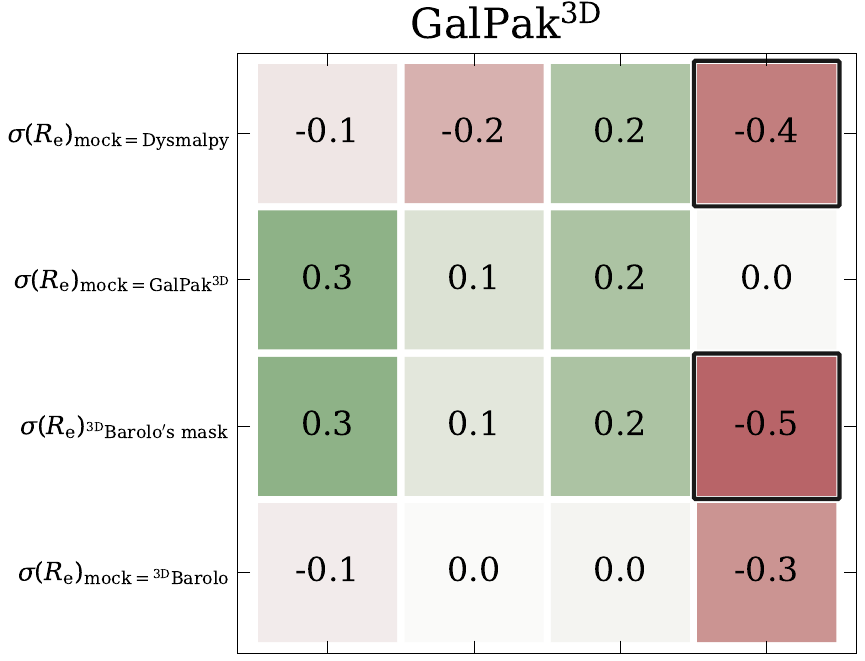}
        \includegraphics[width=0.4\textwidth]{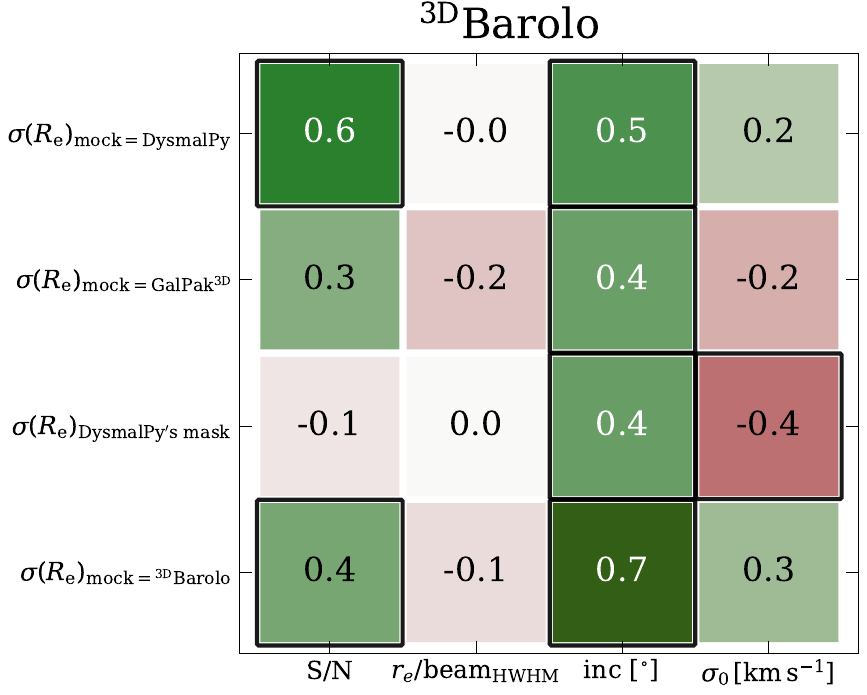}

    \caption{Spearman rank correlation matrix showing the strength of correlations between quality of $\sigma$ recovery and the signal-to-noise (S/N), number of beams in an effective radius (\ReRbeam), inclination angle ($i$), and intrinsic velocity dispersion ($\sigma_\mathrm{intrinsic}$).
    Four cases are shown here: 
    (i) baseline \dy-generated mock models ($\sigma(R_{\rm e})_\text{mock=\dy}$), 
    (ii) \gp\ mock models ($\sigma(R_{\rm e})_{\rm mock=\gp}$), 
    (iii) \bb's masking applied in \dy\ and \gp\ ($\sigma(R_{\rm e})_\text{\bb\'s mask}$) and vice versa for \dy's masking applied in \bb\ ($\sigma(R_{\rm e})_\text{ \dy’s mask}$), and
    (iv) \bb\ mock models($\sigma(R_{\rm e})_{\rm mock=\bb}$).
    The quality of $\sigma$ recovery is represented by the ratio
    $\sigma(R_{\rm e})_\mathrm{model}/\sigma(R_{\rm e})_\mathrm{intrinsic}$ as in the main text, but here for simplicity, the $y$-axis label shows only the numerator.
    The black boxes highlight moderate or stronger correlations with absolute Spearman score $\geq$\,$0.4$.
    Unless ``mock $=$ \gp" or ``mock $=$ \bb" is indicated, the intrinsic dispersion value
    $\sigma(R_{\rm e})_{\rm intrinsic}$ is always that of the baseline
    mock models generated by \dy.}
    \label{fig:cm_spearman_dy}
\end{figure}

Fig.~\ref{fig:cm_spearman_dy} shows the Spearman's $\rho$ between the ratio
\sigrecRe\ and the S/N, \incl, \ReRbeam, and intrinsic $\sigma$ for the
\dy-, \gp- and \bb-generated model set.

\section{Ratios between $V_{\lowercase{\rm rot}}$ and $\sigma$}\label{app:vsig_sig}
\begin{figure*}
    \centering
    \includegraphics[width=0.95\textwidth]{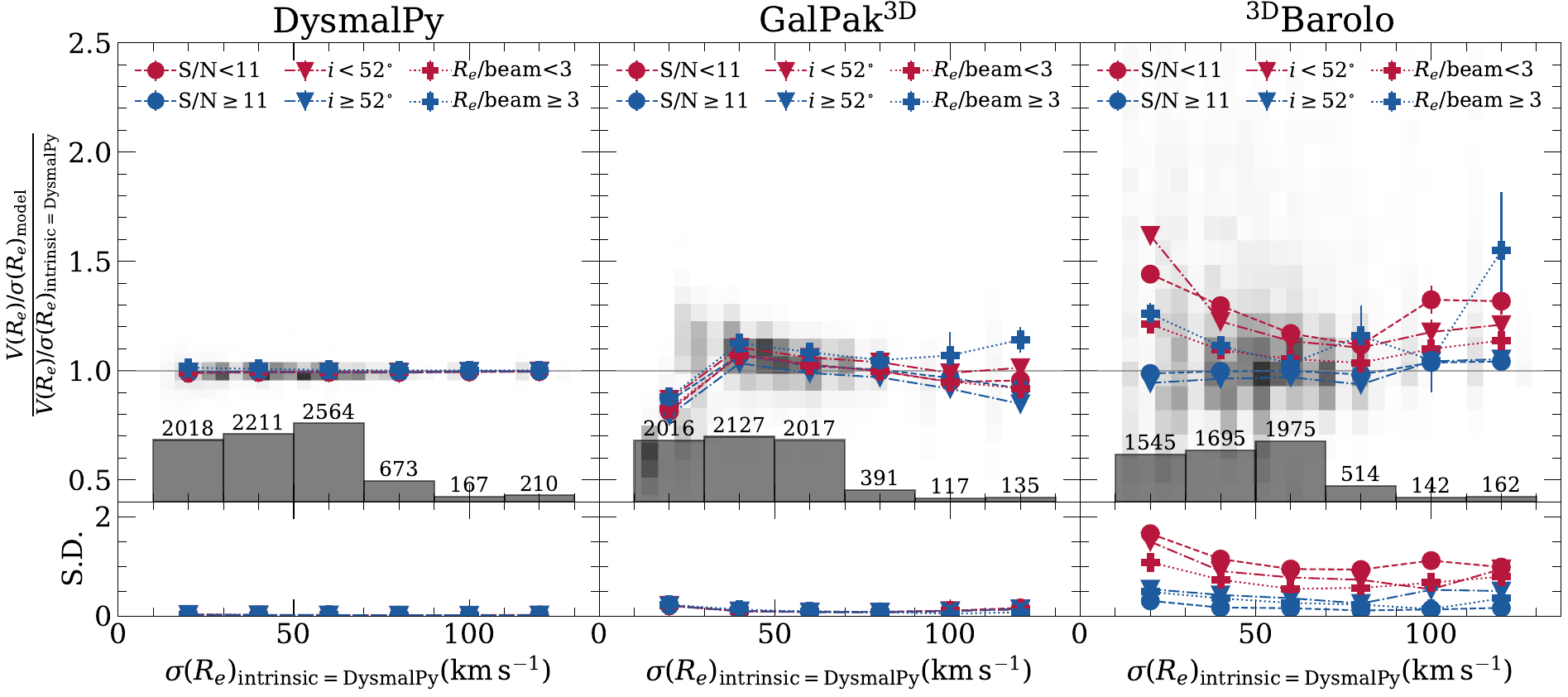}
    \caption{Comparisons between the recovered and intrinsic ratios of 
    rotation velocity over velocity dispersion, evaluated at the effective radius ($V(R_{\rm e})/\sigma(R_{\rm e})$), for the baseline \dy-generated mocks.
    The 2D histograms in the background represent the distribution of the parameters. 
    The choices of line styles and markers for the running median curves are identical to Figure\,\ref{fig:velratio_vel} and Figure\,\ref{fig:dispratio_disp}. Given that $V(R_e)$ is satisfactorily recovered (Section\,\ref{sec:vel_results}) by all three codes, the primary factor responsible for most of the systematic biases in $V(R_{\rm e})/\sigma(R_{\rm e})$ lies in the recovered $\sigma(R_{\rm e})$. 
    In the case of {\gp} (\textit{middle}), the predominant overestimation of $\sigma$ occurs when intrinsic $\sigma < 30{\rm,km,s^{-1}}$, where the self-gravity term becomes dominant (see Figure\,\ref{fig:gp_sigprof_gradient}), 
    leading to the underestimation of $V(R_e)/\sigma$ in the same regime. As for {\bb} (\textit{right}), $V(R_{\rm e})/\sigma(R_{\rm e})$ is most overestimated when the signal-to-noise (S/N) is lower than 11, a result of the underestimated $\sigma(R_{\rm e})$.
    }
    \label{fig:velsig_runmed}
\end{figure*}

Fig.~\ref{fig:velsig_runmed} is the same as Fig.~\ref{fig:velratio_vel} and Fig.~\ref{fig:dispratio_disp} but showing the ratios of recovered and intrinsic $V(R_{\rm e})/\sigma(R_{\rm e})$ for the \dy-generated baseline models.

\section{PV diagrams and channel maps}\label{app:pv_chan}
\begin{sidewaysfigure}
\centering
\vspace{-10cm}
\gridline{\fig{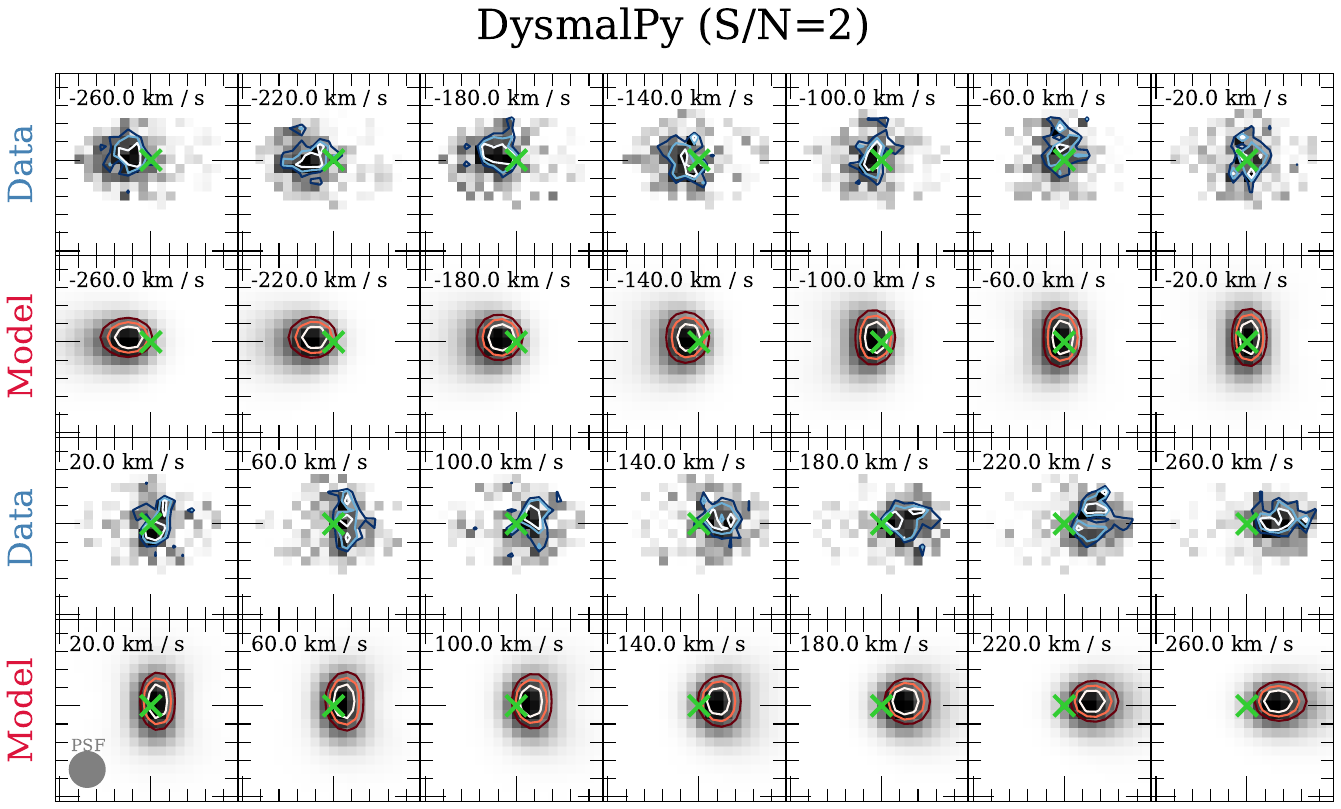}{0.33\textwidth}{}
          \fig{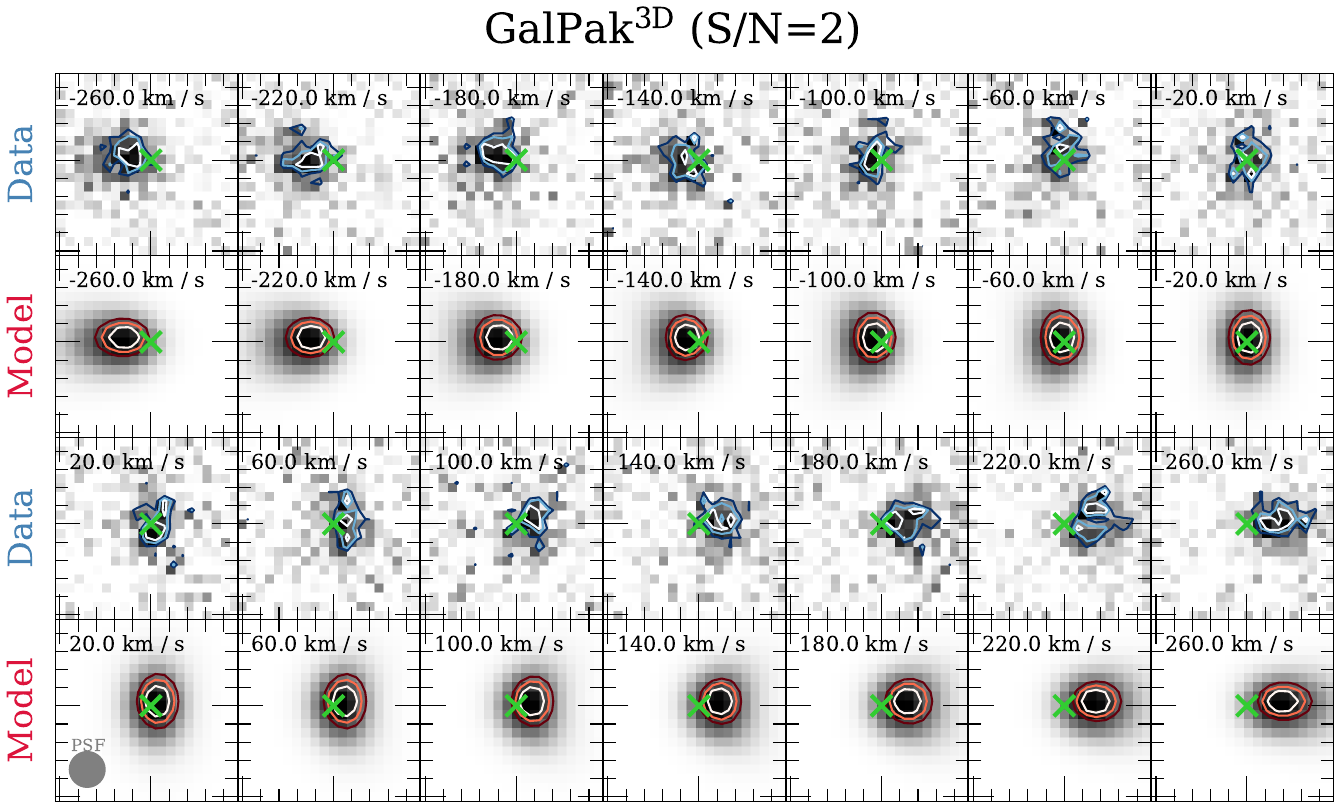}{0.33\textwidth}{}
          \fig{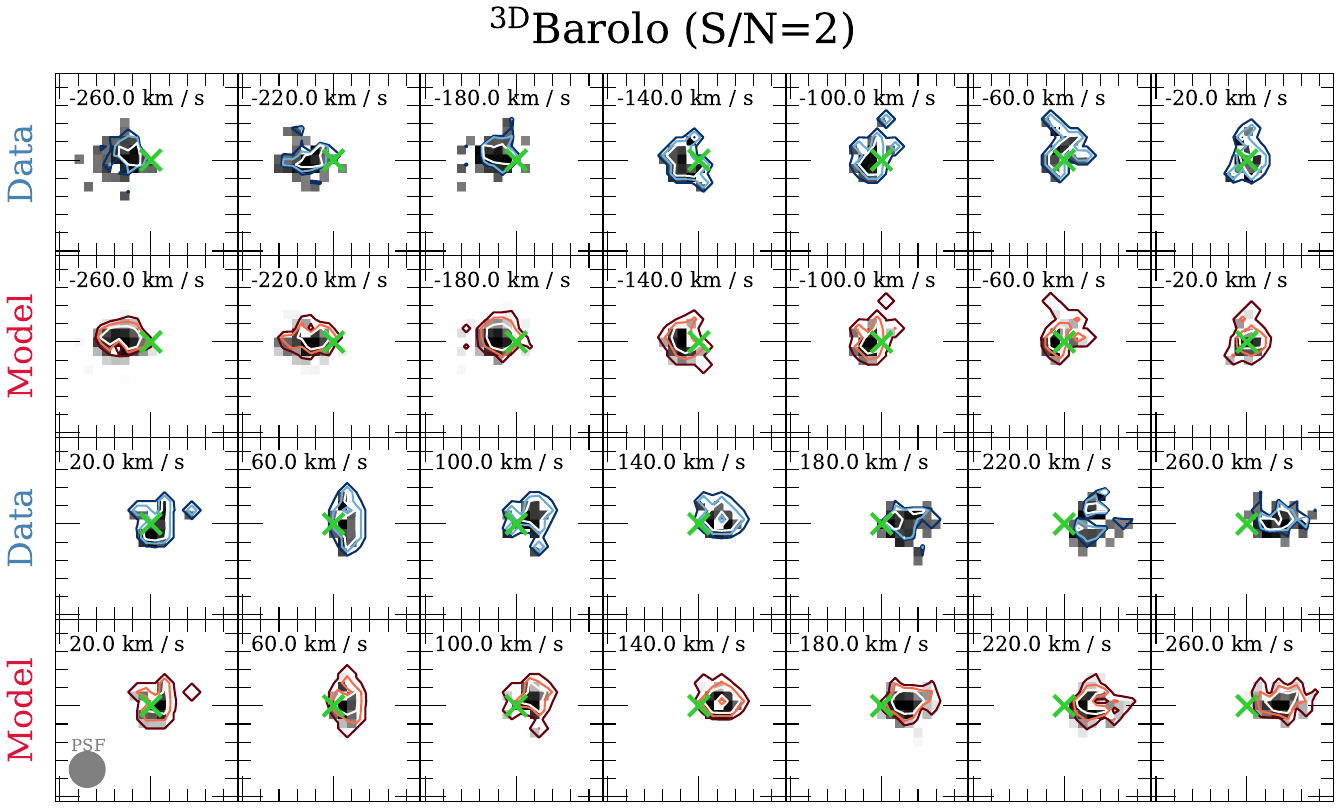}{0.33\textwidth}{}
          }
\vspace{-0.6cm}
\gridline{\fig{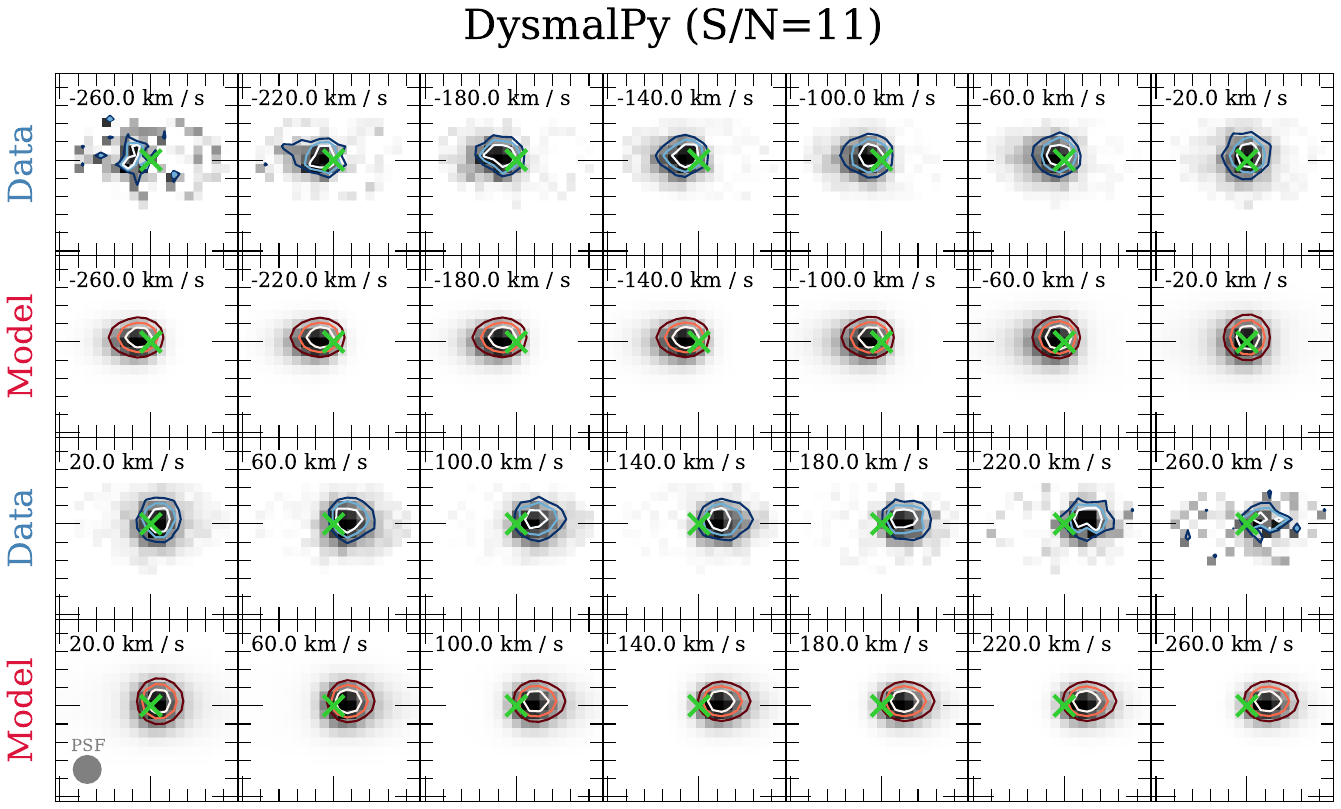}{0.33\textwidth}{}
          \fig{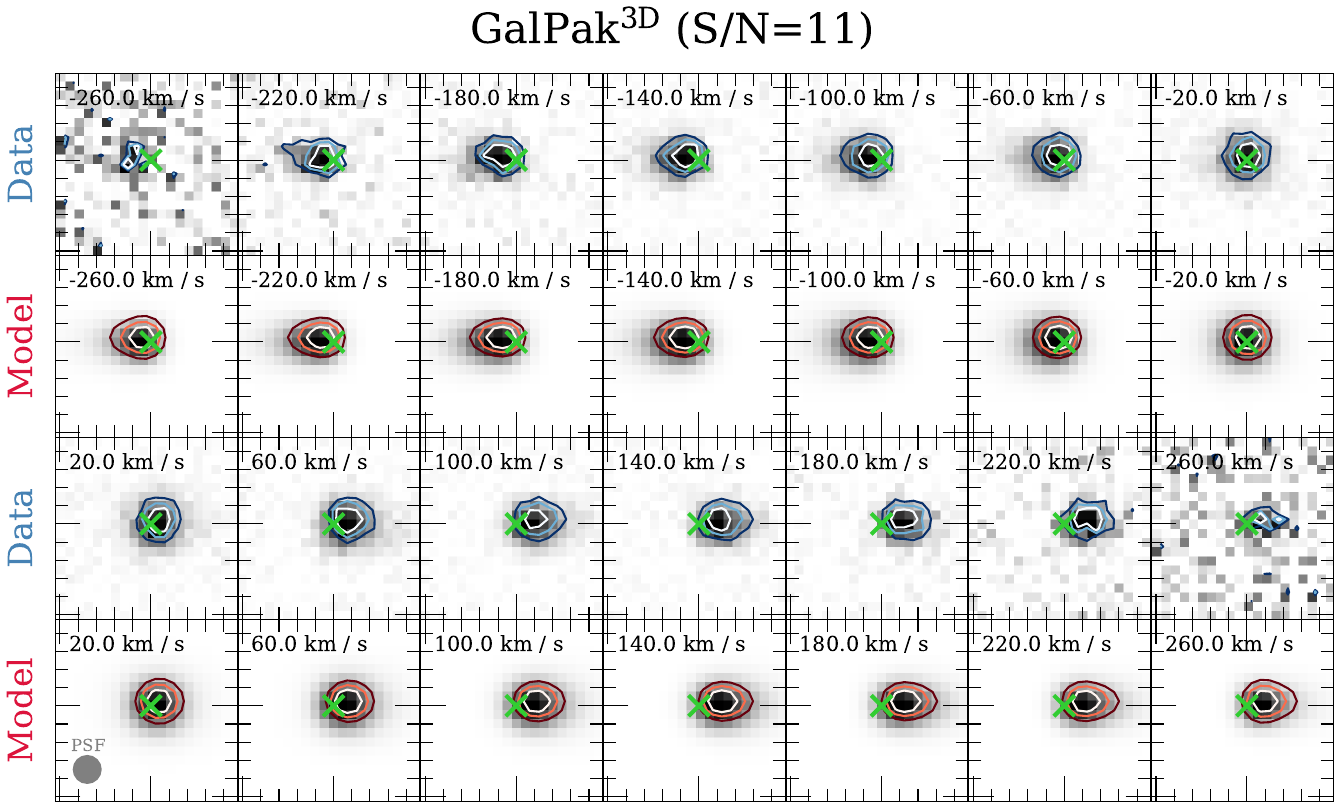}{0.33\textwidth}{}
          \fig{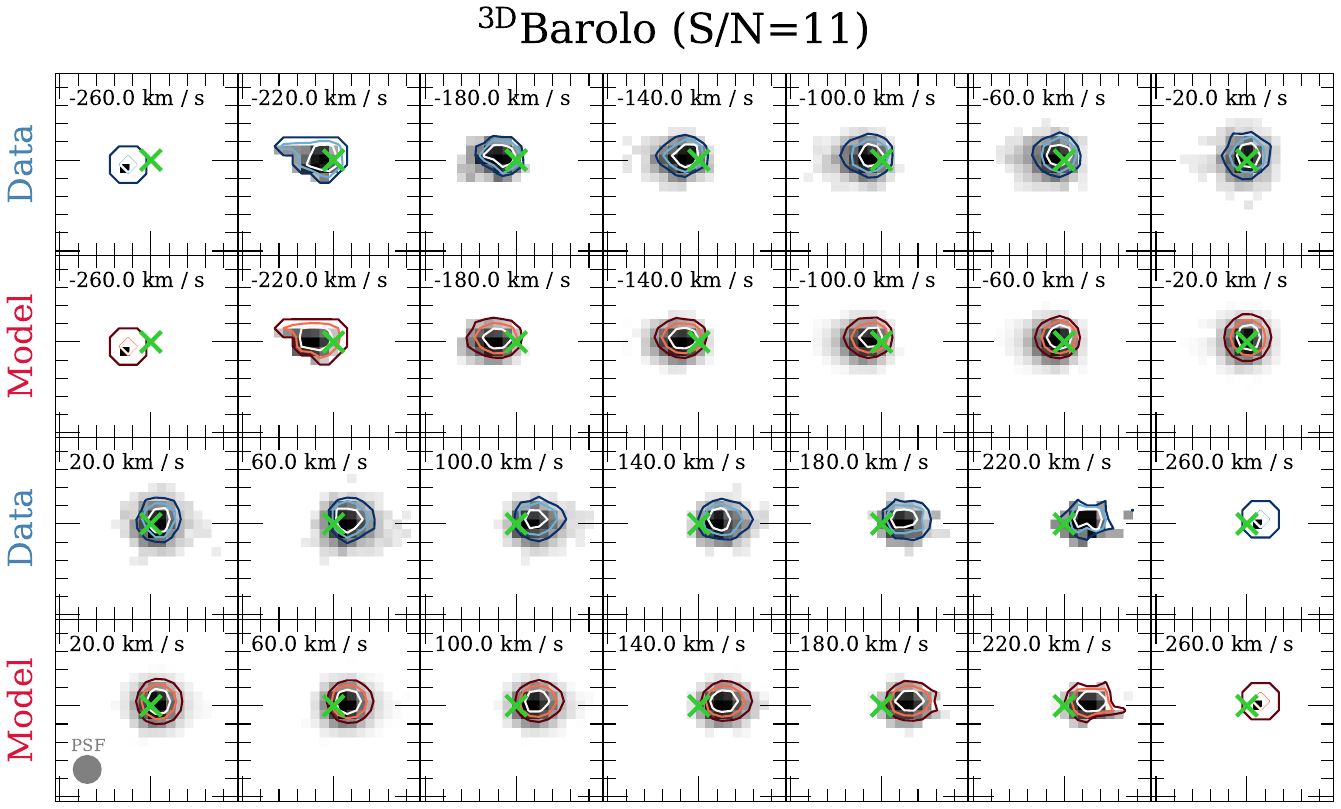}{0.33\textwidth}{}
          }
\vspace{-0.6cm}
\gridline{\fig{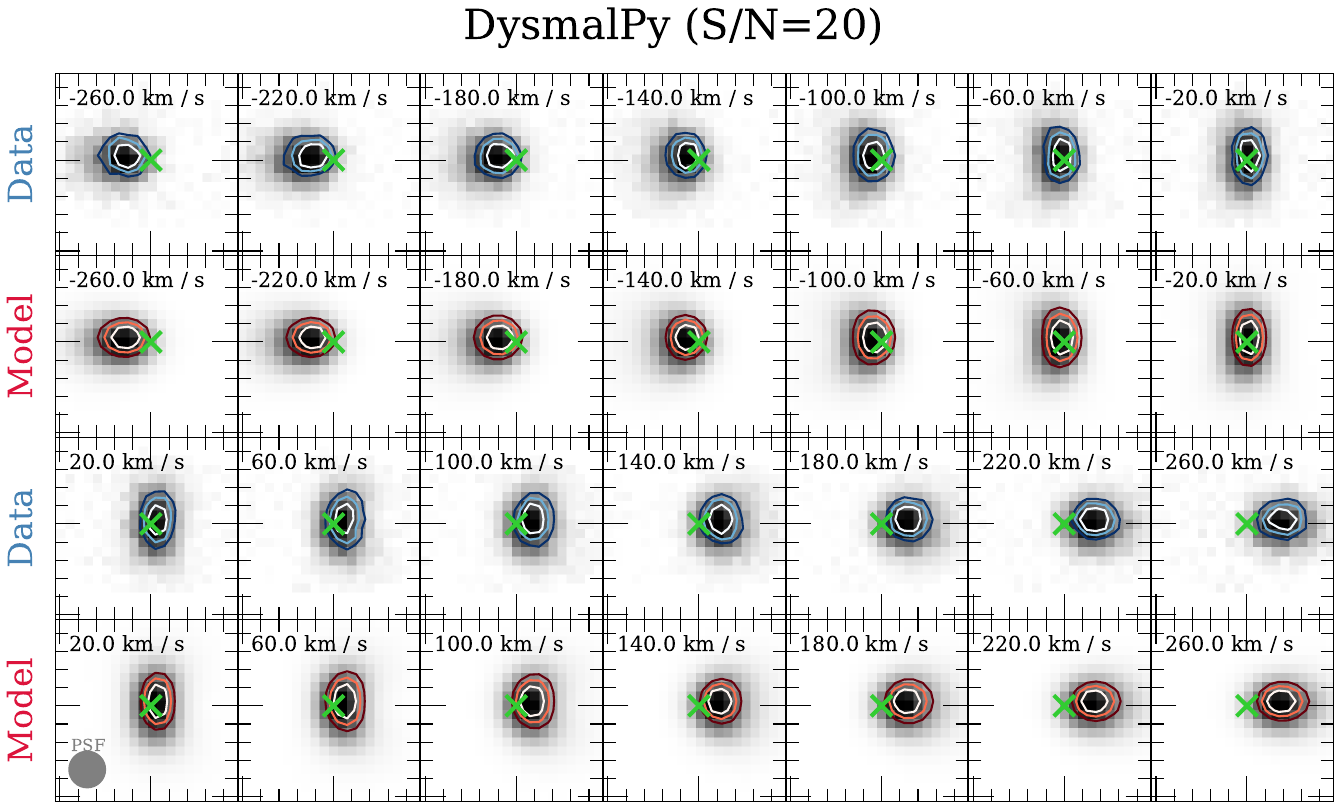}{0.33\textwidth}{}
          \fig{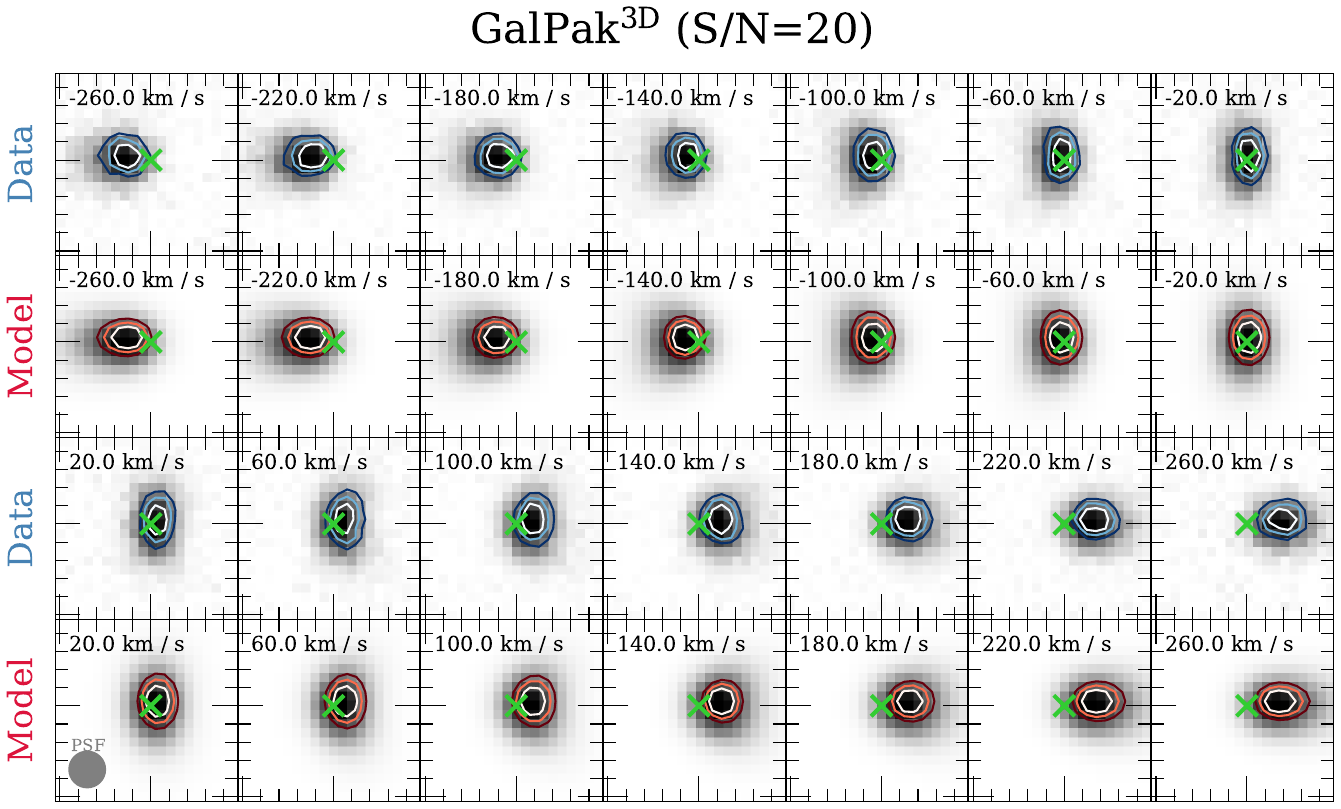}{0.33\textwidth}{}
          \fig{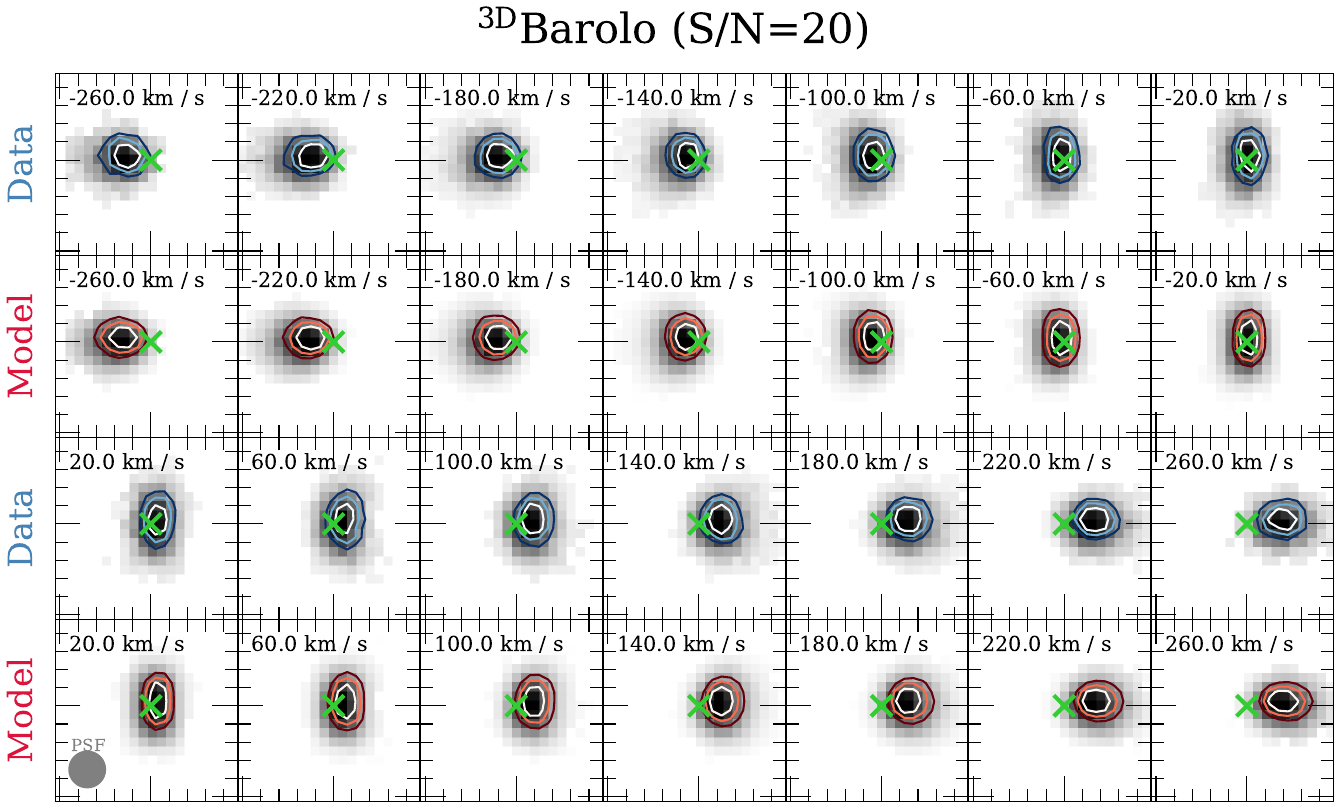}{0.33\textwidth}{}
          }
\caption{Channel maps in step of $40\,{\rm km\,s^{-1}}$ of the three example galaxies in Figure~\ref{fig:bbmasked_spectrum}, arranged in order of increasing signal-to-noise (S/N) ratio from top to bottom, similar to Figure~\ref{fig:bbmasked_spectrum}. The \textit{second} and \textit{fourth rows} of each panel display the model channel maps, overlaid with red contours, from the respective codes as labeled. Note that the channel width of the mocks is $10\,{\rm km\,s^{-1}}$, so the maps shown here represent every 4 channels.}
 \label{fig:channelmaps}
\end{sidewaysfigure}

\begin{figure*}
    \centering
    \includegraphics[width=0.95\textwidth]{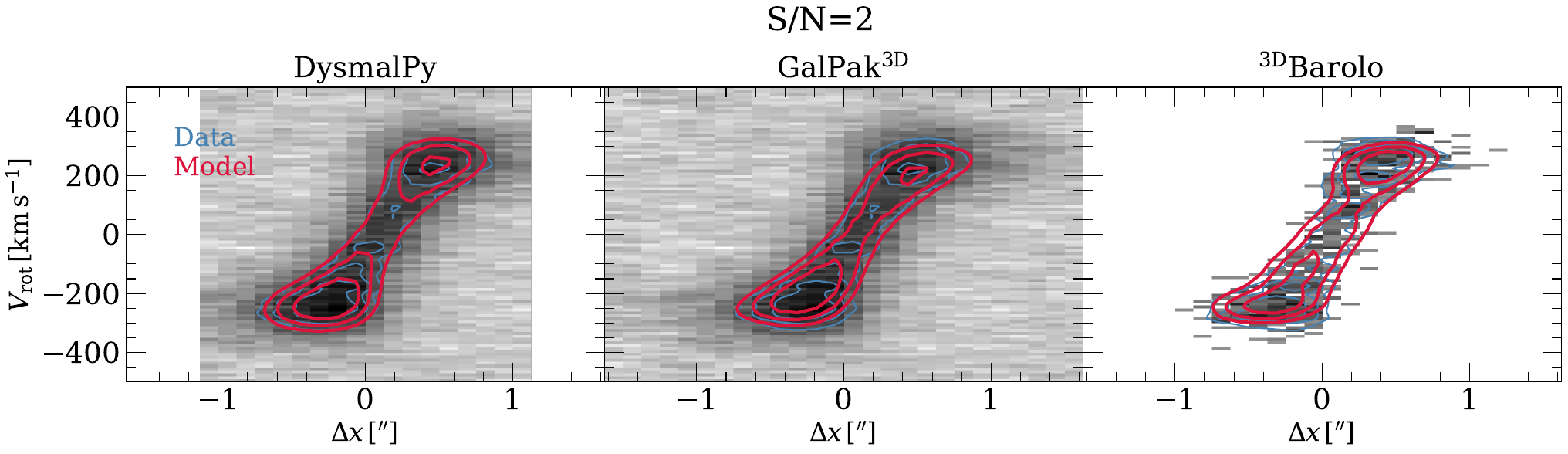}
    \includegraphics[width=0.95\textwidth]{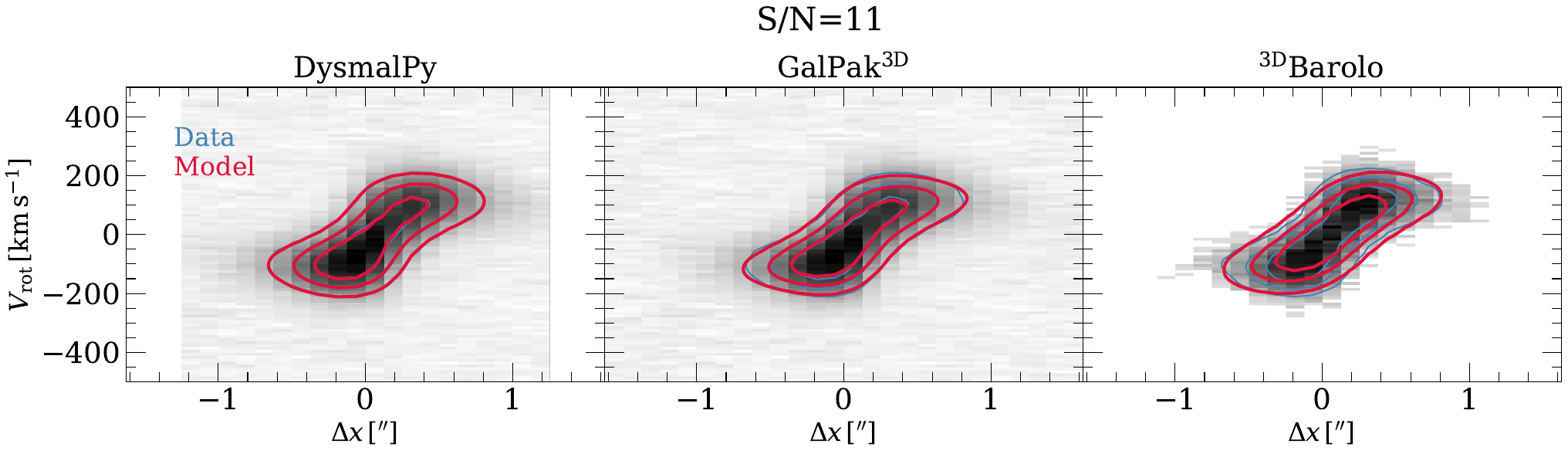}
    \includegraphics[width=0.95\textwidth]{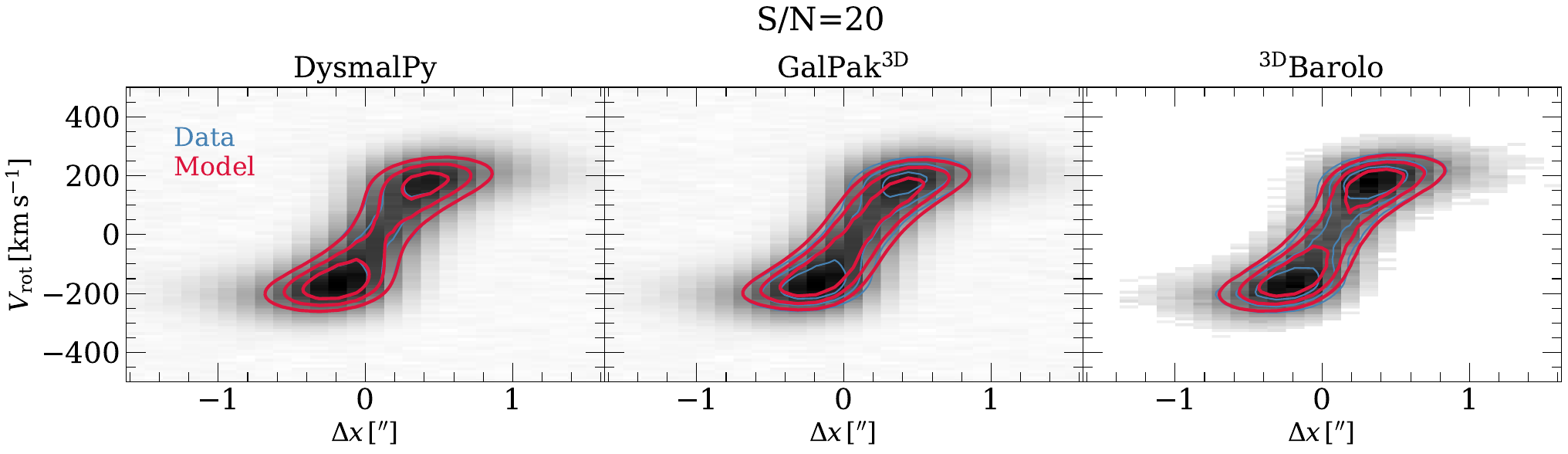}
    \caption{Position-velocity (PV) diagrams for the three example galaxies in Figure~\ref{fig:bbmasked_spectrum} arranged in increasing signal-to-noise (S/N) from top to bottom. The background image shows the PV diagram extracted from the data, masked by the respective codes for \dy\ and \bb, with the blue contours overlaid. The red contours are those of the PV diagrams of the model from the annotated codes.
    }
    \label{fig:pv_diagrams}
\end{figure*}

We show in Figures~\ref{fig:channelmaps} and \ref{fig:pv_diagrams} the channel maps at every 4 channels and PV diagrams of the three example galaxies in Fig.~\ref{fig:bbmasked_spectrum}, respectively. The PV diagrams are extracted using \texttt{PVSLICE} task in \bb, and are derived from the data cube masked by each of the respective codes, namely \dy\ and \bb.

\begin{longrotatetable}
\begin{deluxetable*}{llclc}
\centering
\tabletypesize{\small}
\tablewidth{0pt}
\tablecaption{Parameters and Priors used in this study.}
\tablehead{
\colhead{Software [method, $N_{\rm free}$]} & \colhead{Parameters [unit]} & \colhead{\texttt{Parameters Names}} & \colhead{Priors/Values} & \colhead{\texttt{Input Priors/Options}} }
\startdata 
\texttt{DysmalPy} [mpfit, 3] & PA [deg] & \texttt{pa} &[90] & \texttt{pa=90} \\
         &&&&\texttt{pa\char`_fixed=True}\\
         & Inclination [deg] & \texttt{inc} &[True value]& \texttt{inc\char`_fixed=True}  \\
         & $\log{(M_{\rm bar}/ M_\odot)}$ & \texttt{total\char`_mass}\tablenotemark{$^{\dag}$}& [8.0,13.0]&\texttt{total\char`_mass=10.5}\\
         &&&&\texttt{total\char`_mass\char`_bounds=8.0  13.0}\\
         &&&&\texttt{total\char`_mass\char`_fixed=False}\\
         &&&&\texttt{pressure\char`_support=True}\\
         &&&&\texttt{noord\char`_flat=True}\\
         &$R_{\rm e, d}/(h_z\sqrt{2\ln2})$  &\texttt{invq\char`_disk}&[True value]&\texttt{*}\\
        & $R_{\rm e, b}/(h_z\sqrt{2\ln2})$ &\texttt{invq\char`_bulge}&[True value]&\texttt{*}\\
         & $R_{\rm e, d}$ [kpc]& \texttt{r\char`_eff\char`_disk} & [True value]&\texttt{r\char`_eff\char`_disk\char`_fixed=True} \\
         & $R_{\rm e, b}$ [kpc]&\texttt{r\char`_eff\char`_bulge} & [True value]&\texttt{r\char`_eff\char`_bulge\char`_fixed=True}\\
         & B/T & \texttt{bt}& [True value]& \texttt{bt\char`_fixed=True}\\
         & $h_z$ [kpc]& \texttt{sigmaz} & Tied to $R_{\rm e, d}$ &\texttt{sigmaz\char`_fixed=False}\\
         &&&& \texttt{zheight\char`_tied=True} \\
         & $f_{\rm DM}$ & \texttt{fdm} &[0,1]&\texttt{fdm=0.5} \\
         &&&&\texttt{fdm\char`_fixed=False}\\
         &&&&\texttt{fdm\char`_bounds=0.0 1.0}\\
         & $\sigma_0$ [${\rm km\,s^{-1}}$]& \texttt{sigma0} &[0,150]&\texttt{sigma0=40}\\
         &&&&\texttt{sigma0\char`_fixed=False}\\
         &&&&\texttt{sigma0\char`_bounds=0.0 150.0}\\
         & $\log{(M_{\rm vir}/M_\odot)}$ & \texttt{mvirial} & Tied to $f_{\rm DM}$ &\texttt{mvirial\char`_tied=True}\\
         &&&&\texttt{mvirial\char`_fixed=True}\\
         &&&&\texttt{adiabatic\char`_contract=False}\\
         &&&&\texttt{halo\char`_profile\char`_type=NFW}\\
         &&&&\texttt{include\char`_halo=True}\\
         & $c_{\rm halo}$ & \texttt{halo\char`_conc} & [True value]  &\texttt{halo\char`_conc\char`_fixed=True}  \\
         \cline{2-5}
        & PSF [\arcsec] & \texttt{psf\char`_fwhm} & [True value] &\texttt{psf\char`_type=Gaussian}\\
        & LSF [${\rm km\,s^{-1}}$]& \texttt{sig\char`_inst\char`_res} & 40 &\texttt{use\char`_lsf=True}\\
        &&&&\texttt{data\char`_inst\char`_corr=True}\\
         & Masking & \texttt{auto\char`_gen\char`_3D\char`_mask}&&\texttt{True} \\ &Integrated S/N threshold&\texttt{auto\char`_gen\char`_mask\char`_snr\char`_int\char`_flux\char`_thresh} & 3 & \texttt{3} \\
         \cline{2-5}
         & Light distribution: S\'ersic & \texttt{fitflux}&True&\texttt{True} \\
         & Disk S\'ersic index ($n_{\rm d}$) & \texttt{n\char`_disk}& Exponential&\texttt{1}\\
         & Bulge S\'ersic index ($n_{\rm b}$)& \texttt{n\char`_bulge}& de Vaucouleurs&\texttt{4} \\
         \hline
{\gp} [mcmc, 2]   & PA [deg] & \texttt{fixed.pa\tablenotemark{\scriptsize a}} & [90] & \texttt{90}\\
         & Inclination [deg] & \texttt{fixed.inclination} & [True value] & *  \\
         & $\sigma_0$ [${\rm km\,s^{-1}}$] & \texttt{velocity\char`_dispersion\tablenotemark{\scriptsize b}} & [0,150] &\texttt{0,150}\\
         &&&&\texttt{dispersion\char`_profile='thick'}\tablenotemark{\scriptsize c}\\
         & $V_{\rm max}$ [${\rm km\,s^{-1}}$]& \texttt{maximum\char`_velocity\tablenotemark{\scriptsize b}} & [50,400]&\texttt{50,400}\\         &&&&\texttt{rotation\char`_curve='arctan'}\tablenotemark{\scriptsize c}\\
         & $R_{\rm e}$ [\arcsec]   & \texttt{fixed.radius} &[True value] &*\\
         & $R_{\rm turn}$ [\arcsec]  & \texttt{fixed.turnover\char`_radius} &[$0.25R_{\rm e}$] & \texttt{radius/4}\\
         \cline{2-5}
          & PSF [\arcsec]& \texttt{fwhm}\tablenotemark{\scriptsize d} & [True value] & * \\
          & LSF [${\rm km\,s^{-1}}$] & \texttt{fwhm}\tablenotemark{\scriptsize d} & $40\times2.3548$ &\texttt{94.192}\\
          \cline{2-5}
          & Light distribution & \texttt{type}&S\`ersic& \texttt{\texttt{ModelSersic}}\\
         &   & \texttt{flux\char`_profile}& Exponential&\texttt{exponential}\\
         &  & \texttt{thickness\char`_profile}& Gaussian &\texttt{gaussian} \\
          & $(h_z\sqrt{2\ln2})/R_{\rm e}$ & \texttt{aspect}& [True value] &\texttt{*} \\
         \hline
{\bb} [Nelder-Mead, 2\tablenotemark{$^\ddagger$}]   & PA [deg] & \texttt{PA}&[90] & \texttt{DELTAPA=0} \\
         & Inclination [deg] & \texttt{INC} &[True value]&\texttt{DELTAINC=0} \\
         & Gaussian disk scale height $h_z$ [\arcsec] & \texttt{Z0=*} &[True value]&\texttt{LTYPE=1}\\
         & $V_{\rm rot}(r)$ [${\rm km\,s^{-1}}$] & \texttt{VROT=200}& [0,400] &\texttt{DELTAVROT=200}   \\
         &&&&\texttt{VSYS=0}\\
         &&&&\texttt{ADRIFT=False}\\
         & $\sigma(r)$ [${\rm km\,s^{-1}}$] &\texttt{VDISP=40} &[0,150]&\texttt{MINVDISP=0} \\
         &&&&\texttt{MAXVDISP=150}\\
         &Free parameters &\texttt{FREE} &$V_{\rm rot}(R)$, $\sigma(R)$&\texttt{FREE=VROT VDISP} \\
         \cline{2-5}
         & PSF [arcsec] & \texttt{BEAMFWHM} & [True value]& * \\
         & LSF [channels] & \texttt{LINEAR}\tablenotemark{\scriptsize e} & $40/10$ & \texttt{4} \\
         \cline{2-5}
         & Masking/source finding &\texttt{MASK}&&\texttt{SEARCH} \\ 
         &         &\texttt{SNRCUT} & 3 & \texttt{3} \\
         &         &\texttt{SEARCH}&True&\texttt{True}\\
         \cline{2-5}
         & Light distribution (normalization) & \texttt{NORM}&&\texttt{LOCAL} \\
         & Scale height profile & \texttt{LTYPE} & Gaussian & \texttt{1}\\
         \cline{2-5}
         & Weighting function & \texttt{WFUNC} & Uniform & \texttt{0}\\
         & Minimization function & \texttt{FTYPE} & chi-square & \texttt{1} 
\enddata
\tablecomments{The $x, y, z$ centroid positions are fixed to the known values for all packages. The parameters not listed here are either not relevant for the fitting, e.g. plotting styles, or we use the default values.} Conversion between arcsecond and kpc is based on the specified cosmology.
\tablenotetext{a}{\texttt{fixed =  GalaxyParameters().copy()}, which is then assigned to the argument \texttt{known\char`_parameters} in
\texttt{GalPaK3D.run\char`_mcmc()}.}
\tablenotetext{b}{The lower and upper values are assigned to \texttt{min\char`_boundaries} and \texttt{max\char`_boundaries}, respectively, as a \texttt{GalaxyParameters} object. These boundaries are then passed as arguments with matching names to \texttt{GalPaK3D.run\char`_mcmc()}.}
\tablenotetext{c}{As arguments of \texttt{DefaultModel()} and serves as an argument in \texttt{GalPaK3D.run\char`_mcmc()}.}
\tablenotetext{d}{As the argument of \texttt{GaussianPointSpreadFunction()} and \texttt{GaussianLineSpreadFunction()}, respectively, and are passed to \texttt{Instrument()}, which eventually assigned to object \texttt{GalPaK3D}.}
\tablenotetext{e}{In unit of number of channels.}
\tablenotetext{$$^{\dag}$$}{Total baryonic mass.}
\tablenotetext{$$^{\ddagger}$$}{For each ring.}
\label{tab:prior_table}
\end{deluxetable*}
\end{longrotatetable} 
\bibliography{main}{}

\begin{thebibliography}{}
\expandafter\ifx\csname natexlab\endcsname\relax\def\natexlab#1{#1}\fi
\providecommand{\url}[1]{\href{#1}{#1}}
\providecommand{\dodoi}[1]{doi:~\href{http://doi.org/#1}{\nolinkurl{#1}}}
\providecommand{\doeprint}[1]{\href{http://ascl.net/#1}{\nolinkurl{http://ascl.net/#1}}}
\providecommand{\doarXiv}[1]{\href{https://arxiv.org/abs/#1}{\nolinkurl{https://arxiv.org/abs/#1}}}

\bibitem[{Albanese {et~al.}(2012)Albanese, Filosi, Visintainer, Riccadonna,
  Jurman, \& Furlanello}]{Albanese2012}
Albanese, D., Filosi, M., Visintainer, R., {et~al.} 2012, Bioinformatics, 29,
  407, \dodoi{10.1093/bioinformatics/bts707}

\bibitem[{{Astropy Collaboration} {et~al.}(2013){Astropy Collaboration},
  {Robitaille}, {Tollerud}, {Greenfield}, {Droettboom}, {Bray}, {Aldcroft},
  {Davis}, {Ginsburg}, {Price-Whelan}, {Kerzendorf}, {Conley}, {Crighton},
  {Barbary}, {Muna}, {Ferguson}, {Grollier}, {Parikh}, {Nair}, {Unther},
  {Deil}, {Woillez}, {Conseil}, {Kramer}, {Turner}, {Singer}, {Fox}, {Weaver},
  {Zabalza}, {Edwards}, {Azalee Bostroem}, {Burke}, {Casey}, {Crawford},
  {Dencheva}, {Ely}, {Jenness}, {Labrie}, {Lim}, {Pierfederici}, {Pontzen},
  {Ptak}, {Refsdal}, {Servillat}, \& {Streicher}}]{Astropy2013}
{Astropy Collaboration}, {Robitaille}, T.~P., {Tollerud}, E.~J., {et~al.} 2013,
  \aap, 558, A33, \dodoi{10.1051/0004-6361/201322068}

\bibitem[{Bacchini(2020)}]{Bacchini2020}
Bacchini, C. 2020, PhD thesis, University of Groningen,
  \dodoi{10.33612/diss.133157780}

\bibitem[{{Bacon} {et~al.}(1995){Bacon}, {Adam}, {Baranne}, {Courtes}, {Dubet},
  {Dubois}, {Emsellem}, {Ferruit}, {Georgelin}, {Monnet}, {Pecontal},
  {Rousset}, \& {Say}}]{Bacon1995}
{Bacon}, R., {Adam}, G., {Baranne}, A., {et~al.} 1995, \aaps, 113, 347

\bibitem[{{Bacon} {et~al.}(2015){Bacon}, {Brinchmann}, {Richard}, {Contini},
  {Drake}, {Franx}, {Tacchella}, {Vernet}, {Wisotzki}, {Blaizot}, {Bouch{\'e}},
  {Bouwens}, {Cantalupo}, {Carollo}, {Carton}, {Caruana}, {Cl{\'e}ment},
  {Dreizler}, {Epinat}, {Guiderdoni}, {Herenz}, {Husser}, {Kamann}, {Kerutt},
  {Kollatschny}, {Krajnovic}, {Lilly}, {Martinsson}, {Michel-Dansac},
  {Patricio}, {Schaye}, {Shirazi}, {Soto}, {Soucail}, {Steinmetz}, {Urrutia},
  {Weilbacher}, \& {de Zeeuw}}]{Bacon2015}
{Bacon}, R., {Brinchmann}, J., {Richard}, J., {et~al.} 2015, \aap, 575, A75,
  \dodoi{10.1051/0004-6361/201425419}

\bibitem[{{Begeman}(1989)}]{Begeman1989}
{Begeman}, K.~G. 1989, \aap, 223, 47

\bibitem[{{Bekiaris} {et~al.}(2016){Bekiaris}, {Glazebrook}, {Fluke}, \&
  {Abraham}}]{Bekiaris2016}
{Bekiaris}, G., {Glazebrook}, K., {Fluke}, C.~J., \& {Abraham}, R. 2016,
  \mnras, 455, 754, \dodoi{10.1093/mnras/stv2292}

\bibitem[{{Bewketu Belete} {et~al.}(2021){Bewketu Belete}, {Andreani},
  {Fern{\'a}ndez-Ontiveros}, {Hatziminaoglou}, {Combes}, {Sirressi}, {Slater},
  {Ricci}, {Dasyra}, {Cicone}, {Aalto}, {Spinoglio}, {Imanishi}, \& {De
  Medeiros}}]{BewketuBelete2021}
{Bewketu Belete}, A., {Andreani}, P., {Fern{\'a}ndez-Ontiveros}, J.~A.,
  {et~al.} 2021, \aap, 654, A24, \dodoi{10.1051/0004-6361/202140492}

\bibitem[{{Bischetti} {et~al.}(2021){Bischetti}, {Feruglio}, {Piconcelli},
  {Duras}, {P{\'e}rez-Torres}, {Herrero}, {Venturi}, {Carniani}, {Bruni},
  {Gavignaud}, {Testa}, {Bongiorno}, {Brusa}, {Circosta}, {Cresci},
  {D'Odorico}, {Maiolino}, {Marconi}, {Mingozzi}, {Pappalardo}, {Perna},
  {Traianou}, {Travascio}, {Vietri}, {Zappacosta}, \& {Fiore}}]{Bischetti2021}
{Bischetti}, M., {Feruglio}, C., {Piconcelli}, E., {et~al.} 2021, \aap, 645,
  A33, \dodoi{10.1051/0004-6361/202039057}

\bibitem[{{Biswas} {et~al.}(2023){Biswas}, {Kalinova}, {Roy}, {Patra}, \&
  {Tyulneva}}]{Biswas2023}
{Biswas}, P., {Kalinova}, V., {Roy}, N., {Patra}, N.~N., \& {Tyulneva}, N.
  2023, \mnras, 524, 6213, \dodoi{10.1093/mnras/stad2285}

\bibitem[{{Boomsma} {et~al.}(2008){Boomsma}, {Oosterloo}, {Fraternali}, {van
  der Hulst}, \& {Sancisi}}]{Boomsma2008}
{Boomsma}, R., {Oosterloo}, T.~A., {Fraternali}, F., {van der Hulst}, J.~M., \&
  {Sancisi}, R. 2008, \aap, 490, 555, \dodoi{10.1051/0004-6361:200810120}

\bibitem[{{Bouch{\'e}} {et~al.}(2015){Bouch{\'e}}, {Carfantan}, {Schroetter},
  {Michel-Dansac}, \& {Contini}}]{Bouche2015}
{Bouch{\'e}}, N., {Carfantan}, H., {Schroetter}, I., {Michel-Dansac}, L., \&
  {Contini}, T. 2015, \aj, 150, 92, \dodoi{10.1088/0004-6256/150/3/92}

\bibitem[{{Bouch{\'e}} {et~al.}(2021){Bouch{\'e}}, {Genel}, {Pellissier},
  {Dubois}, {Contini}, {Epinat}, {Pillepich}, {Krajnovi{\'c}}, {Nelson},
  {Abril-Melgarejo}, {Richard}, {Boogaard}, {Maseda}, {Mercier}, {Bacon},
  {Steinmetz}, \& {Vogelsberger}}]{Bouche2021}
{Bouch{\'e}}, N.~F., {Genel}, S., {Pellissier}, A., {et~al.} 2021, \aap, 654,
  A49, \dodoi{10.1051/0004-6361/202040225}

\bibitem[{{Bouch{\'e}} {et~al.}(2022){Bouch{\'e}}, {Bera}, {Krajnovi{\'c}},
  {Emsellem}, {Mercier}, {Schaye}, {Epinat}, {Richard}, {Zoutendijk},
  {Abril-Melgarejo}, {Brinchmann}, {Bacon}, {Contini}, {Boogaard}, {Wisotzki},
  {Maseda}, \& {Steinmetz}}]{Bouche2022}
{Bouch{\'e}}, N.~F., {Bera}, S., {Krajnovi{\'c}}, D., {et~al.} 2022, \aap, 658,
  A76, \dodoi{10.1051/0004-6361/202141762}

\bibitem[{{Bournaud} {et~al.}(2009){Bournaud}, {Elmegreen}, \&
  {Martig}}]{Bournaud2009}
{Bournaud}, F., {Elmegreen}, B.~G., \& {Martig}, M. 2009, \apjl, 707, L1,
  \dodoi{10.1088/0004-637X/707/1/L1}

\bibitem[{{Burkert} {et~al.}(2010){Burkert}, {Genzel}, {Bouch{\'e}}, {Cresci},
  {Khochfar}, {Sommer-Larsen}, {Sternberg}, {Naab}, {F{\"o}rster Schreiber},
  {Tacconi}, {Shapiro}, {Hicks}, {Lutz}, {Davies}, {Buschkamp}, \&
  {Genel}}]{Burkert2010}
{Burkert}, A., {Genzel}, R., {Bouch{\'e}}, N., {et~al.} 2010, \apj, 725, 2324,
  \dodoi{10.1088/0004-637X/725/2/2324}

\bibitem[{{Burkert} {et~al.}(2016){Burkert}, {F{\"o}rster Schreiber}, {Genzel},
  {Lang}, {Tacconi}, {Wisnioski}, {Wuyts}, {Bandara}, {Beifiori}, {Bender},
  {Brammer}, {Chan}, {Davies}, {Dekel}, {Fabricius}, {Fossati}, {Kulkarni},
  {Lutz}, {Mendel}, {Momcheva}, {Nelson}, {Naab}, {Renzini}, {Saglia},
  {Sharples}, {Sternberg}, {Wilman}, \& {Wuyts}}]{Burkert2016}
{Burkert}, A., {F{\"o}rster Schreiber}, N.~M., {Genzel}, R., {et~al.} 2016,
  \apj, 826, 214, \dodoi{10.3847/0004-637X/826/2/214}

\bibitem[{{Cao} {et~al.}(2023){Cao}, {Wong}, {Bolatto}, {Leroy}, {Rosolowsky},
  {Utomo}, {S{\'a}nchez}, {Barrera-Ballesteros}, {Levy}, {Colombo}, {Blitz},
  {Vogel}, {Puschnig}, {Villanueva}, \& {Rubio}}]{Cao2023}
{Cao}, Y., {Wong}, T., {Bolatto}, A.~D., {et~al.} 2023, \apjs, 268, 3,
  \dodoi{10.3847/1538-4365/acd840}

\bibitem[{{Contini} {et~al.}(2016){Contini}, {Epinat}, {Bouch{\'e}},
  {Brinchmann}, {Boogaard}, {Ventou}, {Bacon}, {Richard}, {Weilbacher},
  {Wisotzki}, {Krajnovi{\'c}}, {Vielfaure}, {Emsellem}, {Finley}, {Inami},
  {Schaye}, {Swinbank}, {Gu{\'e}rou}, {Martinsson}, {Michel-Dansac},
  {Schroetter}, {Shirazi}, \& {Soucail}}]{Contini2016}
{Contini}, T., {Epinat}, B., {Bouch{\'e}}, N., {et~al.} 2016, \aap, 591, A49,
  \dodoi{10.1051/0004-6361/201527866}

\bibitem[{{Courteau}(1997)}]{Courteau1997}
{Courteau}, S. 1997, \aj, 114, 2402, \dodoi{10.1086/118656}

\bibitem[{{Cresci} {et~al.}(2009){Cresci}, {Hicks}, {Genzel}, {F{\"o}rster
  Schreiber}, {Davies}, {Bouch{\'e}}, {Buschkamp}, {Genel}, {Shapiro},
  {Tacconi}, {Sommer-Larsen}, {Burkert}, {Eisenhauer}, {Gerhard}, {Lutz},
  {Naab}, {Sternberg}, {Cimatti}, {Daddi}, {Erb}, {Kurk}, {Lilly}, {Renzini},
  {Shapley}, {Steidel}, \& {Caputi}}]{Cresci2009}
{Cresci}, G., {Hicks}, E.~K.~S., {Genzel}, R., {et~al.} 2009, \apj, 697, 115,
  \dodoi{10.1088/0004-637X/697/1/115}

\bibitem[{{Davies} {et~al.}(2011){Davies}, {F{\"o}rster Schreiber}, {Cresci},
  {Genzel}, {Bouch{\'e}}, {Burkert}, {Buschkamp}, {Genel}, {Hicks}, {Kurk},
  {Lutz}, {Newman}, {Shapiro}, {Sternberg}, {Tacconi}, \& {Wuyts}}]{Davies2011}
{Davies}, R., {F{\"o}rster Schreiber}, N.~M., {Cresci}, G., {et~al.} 2011,
  \apj, 741, 69, \dodoi{10.1088/0004-637X/741/2/69}

\bibitem[{{Davies} {et~al.}(2009){Davies}, {Maciejewski}, {Hicks}, {Tacconi},
  {Genzel}, \& {Engel}}]{Davies2009}
{Davies}, R.~I., {Maciejewski}, W., {Hicks}, E.~K.~S., {et~al.} 2009, \apj,
  702, 114, \dodoi{10.1088/0004-637X/702/1/114}

\bibitem[{{Davies} {et~al.}(2004{\natexlab{a}}){Davies}, {Tacconi}, \&
  {Genzel}}]{Davies2004a}
{Davies}, R.~I., {Tacconi}, L.~J., \& {Genzel}, R. 2004{\natexlab{a}}, \apj,
  602, 148, \dodoi{10.1086/380995}

\bibitem[{{Davies} {et~al.}(2004{\natexlab{b}}){Davies}, {Tacconi}, \&
  {Genzel}}]{Davies2004b}
---. 2004{\natexlab{b}}, \apj, 613, 781, \dodoi{10.1086/423315}

\bibitem[{{Davies} {et~al.}(2014){Davies}, {Maciejewski}, {Hicks}, {Emsellem},
  {Erwin}, {Burtscher}, {Dumas}, {Lin}, {Malkan}, {M{\"u}ller-S{\'a}nchez},
  {Orban de Xivry}, {Rosario}, {Schnorr-M{\"u}ller}, \& {Tran}}]{Davies2014}
{Davies}, R.~I., {Maciejewski}, W., {Hicks}, E.~K.~S., {et~al.} 2014, \apj,
  792, 101, \dodoi{10.1088/0004-637X/792/2/101}

\bibitem[{{Davis} {et~al.}(2017){Davis}, {Bureau}, {Onishi}, {Cappellari},
  {Iguchi}, \& {Sarzi}}]{Davis2017}
{Davis}, T.~A., {Bureau}, M., {Onishi}, K., {et~al.} 2017, \mnras, 468, 4675,
  \dodoi{10.1093/mnras/stw3217}

\bibitem[{{Davis} {et~al.}(2013){Davis}, {Alatalo}, {Bureau}, {Cappellari},
  {Scott}, {Young}, {Blitz}, {Crocker}, {Bayet}, {Bois}, {Bournaud}, {Davies},
  {de Zeeuw}, {Duc}, {Emsellem}, {Khochfar}, {Krajnovi{\'c}}, {Kuntschner},
  {Lablanche}, {McDermid}, {Morganti}, {Naab}, {Oosterloo}, {Sarzi}, {Serra},
  \& {Weijmans}}]{Davis2013}
{Davis}, T.~A., {Alatalo}, K., {Bureau}, M., {et~al.} 2013, \mnras, 429, 534,
  \dodoi{10.1093/mnras/sts353}

\bibitem[{{de Blok} {et~al.}(2024){de Blok}, {Healy}, {Maccagni}, {Pisano},
  {Bosma}, {English}, {Jarrett}, {Marasco}, {Meurer}, {Veronese}, {Bigiel},
  {Chemin}, {Fraternali}, {Holwerda}, {Kamphuis}, {Kl{\"o}ckner}, {Kleiner},
  {Leroy}, {Mogotsi}, {Oman}, {Schinnerer}, {Verdes-Montenegro}, {Westmeier},
  {Wong}, {Zabel}, {Amram}, {Carignan}, {Combes}, {Brinks}, {Dettmar},
  {Gibson}, {Jozsa}, {Koribalski}, {McGaugh}, {Oosterloo}, {Spekkens},
  {Schr{\"o}der}, {Adams}, {Athanassoula}, {Bershady}, {Beswick}, {Blyth},
  {Elson}, {Frank}, {Heald}, {Henning}, {Kurapati}, {Loubser}, {Lucero},
  {Meyer}, {Namumba}, {Oh}, {Sardone}, {Sheth}, {Smith}, {Sorgho}, {Walter},
  {Williams}, {Woudt}, \& {Zijlstra}}]{deBlok2024}
{de Blok}, W.~J.~G., {Healy}, J., {Maccagni}, F.~M., {et~al.} 2024, \aap, 688,
  A109, \dodoi{10.1051/0004-6361/202348297}

\bibitem[{{Deg} {et~al.}(2022){Deg}, {Spekkens}, {Westmeier}, {Reynolds},
  {Venkataraman}, {Goliath}, {Shen}, {Halloran}, {Bosma}, {Catinella}, {de
  Blok}, {D{\'e}nes}, {DiTeodoro}, {Elagali}, {For}, {Howlett}, {J{\'o}zsa},
  {Kamphuis}, {Kleiner}, {Koribalski}, {Lee-Waddell}, {Lelli}, {Lin},
  {Murugeshan}, {Oh}, {Rhee}, {Scott}, {Staveley-Smith}, {van der Hulst},
  {Verdes-Montenegro}, {Wang}, \& {Wong}}]{Deg2022}
{Deg}, N., {Spekkens}, K., {Westmeier}, T., {et~al.} 2022, \pasa, 39, e059,
  \dodoi{10.1017/pasa.2022.43}

\bibitem[{{Dekel} {et~al.}(2022){Dekel}, {Mandelker}, {Bournaud}, {Ceverino},
  {Guo}, \& {Primack}}]{Dekel2022}
{Dekel}, A., {Mandelker}, N., {Bournaud}, F., {et~al.} 2022, \mnras, 511, 316,
  \dodoi{10.1093/mnras/stab3810}

\bibitem[{{Dekel} {et~al.}(2009){Dekel}, {Birnboim}, {Engel}, {Freundlich},
  {Goerdt}, {Mumcuoglu}, {Neistein}, {Pichon}, {Teyssier}, \&
  {Zinger}}]{Dekel2009}
{Dekel}, A., {Birnboim}, Y., {Engel}, G., {et~al.} 2009, \nat, 457, 451,
  \dodoi{10.1038/nature07648}

\bibitem[{{Di Teodoro} \& {Fraternali}(2015)}]{DiTeodoro2015}
{Di Teodoro}, E.~M., \& {Fraternali}, F. 2015, \mnras, 451, 3021,
  \dodoi{10.1093/mnras/stv1213}

\bibitem[{{Di Teodoro} {et~al.}(2016){Di Teodoro}, {Fraternali}, \&
  {Miller}}]{DiTeodoro2016}
{Di Teodoro}, E.~M., {Fraternali}, F., \& {Miller}, S.~H. 2016, \aap, 594, A77,
  \dodoi{10.1051/0004-6361/201628315}

\bibitem[{{Dutton} \& {Macci{\`o}}(2014)}]{Dutton2014}
{Dutton}, A.~A., \& {Macci{\`o}}, A.~V. 2014, \mnras, 441, 3359,
  \dodoi{10.1093/mnras/stu742}

\bibitem[{{Eisenhauer} {et~al.}(2003){Eisenhauer}, {Tecza}, {Thatte}, {Genzel},
  {Abuter}, {Iserlohe}, {Schreiber}, {Huber}, {Roehrle}, {Horrobin},
  {Schegerer}, {Baker}, {Bender}, {Davies}, {Lehnert}, {Lutz}, {Nesvadba},
  {Ott}, {Seitz}, {Schoedel}, {Tacconi}, {Bonnet}, {Castillo}, {Conzelmann},
  {Donaldson}, {Finger}, {Gillet}, {Hubin}, {Kissler-Patig}, {Lizon}, {Monnet},
  \& {Stroebele}}]{Eisenhauer2003}
{Eisenhauer}, F., {Tecza}, M., {Thatte}, N., {et~al.} 2003, The Messenger, 113,
  17

\bibitem[{{Elmegreen} \& {Elmegreen}(2006)}]{ElmegreenElmegreen2006}
{Elmegreen}, B.~G., \& {Elmegreen}, D.~M. 2006, \apj, 650, 644,
  \dodoi{10.1086/507578}

\bibitem[{{Elmegreen} {et~al.}(2005){Elmegreen}, {Elmegreen}, {Vollbach},
  {Foster}, \& {Ferguson}}]{Elmegreen2005}
{Elmegreen}, B.~G., {Elmegreen}, D.~M., {Vollbach}, D.~R., {Foster}, E.~R., \&
  {Ferguson}, T.~E. 2005, \apj, 634, 101, \dodoi{10.1086/496952}

\bibitem[{{Erwin}(2015)}]{erwin2015}
{Erwin}, P. 2015, \apj, 799, 226, \dodoi{10.1088/0004-637X/799/2/226}

\bibitem[{{Fan} {et~al.}(2019){Fan}, {Knudsen}, {Han}, \& {Tan}}]{Fan2019}
{Fan}, L., {Knudsen}, K.~K., {Han}, Y., \& {Tan}, Q.-h. 2019, \apj, 887, 74,
  \dodoi{10.3847/1538-4357/ab5059}

\bibitem[{{Foreman-Mackey}(2016)}]{Foreman-Mackey2016}
{Foreman-Mackey}, D. 2016, The Journal of Open Source Software, 1, 24,
  \dodoi{10.21105/joss.00024}

\bibitem[{{F{\"o}rster Schreiber} \& {Wuyts}(2020)}]{fs2020}
{F{\"o}rster Schreiber}, N.~M., \& {Wuyts}, S. 2020, \araa, 58, 661,
  \dodoi{10.1146/annurev-astro-032620-021910}

\bibitem[{{F{\"o}rster Schreiber} {et~al.}(2009){F{\"o}rster Schreiber},
  {Genzel}, {Bouch{\'e}}, {Cresci}, {Davies}, {Buschkamp}, {Shapiro},
  {Tacconi}, {Hicks}, {Genel}, {Shapley}, {Erb}, {Steidel}, {Lutz},
  {Eisenhauer}, {Gillessen}, {Sternberg}, {Renzini}, {Cimatti}, {Daddi},
  {Kurk}, {Lilly}, {Kong}, {Lehnert}, {Nesvadba}, {Verma}, {McCracken},
  {Arimoto}, {Mignoli}, \& {Onodera}}]{nmfs2009}
{F{\"o}rster Schreiber}, N.~M., {Genzel}, R., {Bouch{\'e}}, N., {et~al.} 2009,
  \apj, 706, 1364, \dodoi{10.1088/0004-637X/706/2/1364}

\bibitem[{{F{\"o}rster Schreiber} {et~al.}(2011){F{\"o}rster Schreiber},
  {Shapley}, {Genzel}, {Bouch{\'e}}, {Cresci}, {Davies}, {Erb}, {Genel},
  {Lutz}, {Newman}, {Shapiro}, {Steidel}, {Sternberg}, \&
  {Tacconi}}]{nmfs2011b}
{F{\"o}rster Schreiber}, N.~M., {Shapley}, A.~E., {Genzel}, R., {et~al.} 2011,
  \apj, 739, 45, \dodoi{10.1088/0004-637X/739/1/45}

\bibitem[{{F{\"o}rster Schreiber} {et~al.}(2018){F{\"o}rster Schreiber},
  {Renzini}, {Mancini}, {Genzel}, {Bouch{\'e}}, {Cresci}, {Hicks}, {Lilly},
  {Peng}, {Burkert}, {Carollo}, {Cimatti}, {Daddi}, {Davies}, {Genel}, {Kurk},
  {Lang}, {Lutz}, {Mainieri}, {McCracken}, {Mignoli}, {Naab}, {Oesch},
  {Pozzetti}, {Scodeggio}, {Shapiro Griffin}, {Shapley}, {Sternberg},
  {Tacchella}, {Tacconi}, {Wuyts}, \& {Zamorani}}]{nmfs2018}
{F{\"o}rster Schreiber}, N.~M., {Renzini}, A., {Mancini}, C., {et~al.} 2018,
  \apjs, 238, 21, \dodoi{10.3847/1538-4365/aadd49}

\bibitem[{{Fraternali} {et~al.}(2021){Fraternali}, {Karim}, {Magnelli},
  {G{\'o}mez-Guijarro}, {Jim{\'e}nez-Andrade}, \& {Posses}}]{Fraternali2021}
{Fraternali}, F., {Karim}, A., {Magnelli}, B., {et~al.} 2021, \aap, 647, A194,
  \dodoi{10.1051/0004-6361/202039807}

\bibitem[{{Freundlich} {et~al.}(2019){Freundlich}, {Combes}, {Tacconi},
  {Genzel}, {Garcia-Burillo}, {Neri}, {Contini}, {Bolatto}, {Lilly},
  {Salom{\'e}}, {Bicalho}, {Boissier}, {Boone}, {Bouch{\'e}}, {Bournaud},
  {Burkert}, {Carollo}, {Cooper}, {Cox}, {Feruglio}, {F{\"o}rster Schreiber},
  {Juneau}, {Lippa}, {Lutz}, {Naab}, {Renzini}, {Saintonge}, {Sternberg},
  {Walter}, {Weiner}, {Wei{\ss}}, \& {Wuyts}}]{Freundlich2019}
{Freundlich}, J., {Combes}, F., {Tacconi}, L.~J., {et~al.} 2019, \aap, 622,
  A105, \dodoi{10.1051/0004-6361/201732223}

\bibitem[{{Fujimoto} {et~al.}(2021){Fujimoto}, {Oguri}, {Brammer}, {Yoshimura},
  {Laporte}, {Gonz{\'a}lez-L{\'o}pez}, {Caminha}, {Kohno}, {Zitrin}, {Richard},
  {Ouchi}, {Bauer}, {Smail}, {Hatsukade}, {Ono}, {Kokorev}, {Umehata},
  {Schaerer}, {Knudsen}, {Sun}, {Magdis}, {Valentino}, {Ao}, {Toft},
  {Dessauges-Zavadsky}, {Shimasaku}, {Caputi}, {Kusakabe}, {Morokuma-Matsui},
  {Shotaro}, {Egami}, {Lee}, {Rawle}, \& {Espada}}]{Fujimoto2021}
{Fujimoto}, S., {Oguri}, M., {Brammer}, G., {et~al.} 2021, \apj, 911, 99,
  \dodoi{10.3847/1538-4357/abd7ec}

\bibitem[{{Fujimoto} {et~al.}(2024){Fujimoto}, {Ouchi}, {Kohno}, {Valentino},
  {Gim{\'e}nez-Arteaga}, {Brammer}, {Furtak}, {Kohandel}, {Oguri},
  {Pallottini}, {Richard}, {Zitrin}, {Bauer}, {Boylan-Kolchin},
  {Dessauges-Zavadsky}, {Egami}, {Finkelstein}, {Ma}, {Smail}, {Watson},
  {Hutchison}, {Rigby}, {Welch}, {Ao}, {Bradley}, {Caminha}, {Caputi},
  {Espada}, {Endsley}, {Fudamoto}, {Gonz{\'a}lez-L{\'o}pez}, {Hatsukade},
  {Koekemoer}, {Kokorev}, {Laporte}, {Lee}, {Magdis}, {Ono}, {Rizzo},
  {Shibuya}, {Shimasaku}, {Sun}, {Toft}, {Umehata}, {Wang}, \&
  {Yajima}}]{Fujimoto2024}
{Fujimoto}, S., {Ouchi}, M., {Kohno}, K., {et~al.} 2024, arXiv e-prints,
  arXiv:2402.18543, \dodoi{10.48550/arXiv.2402.18543}

\bibitem[{{Genzel} {et~al.}(2006){Genzel}, {Tacconi}, {Eisenhauer},
  {F{\"o}rster Schreiber}, {Cimatti}, {Daddi}, {Bouch{\'e}}, {Davies},
  {Lehnert}, {Lutz}, {Nesvadba}, {Verma}, {Abuter}, {Shapiro}, {Sternberg},
  {Renzini}, {Kong}, {Arimoto}, \& {Mignoli}}]{Genzel2006}
{Genzel}, R., {Tacconi}, L.~J., {Eisenhauer}, F., {et~al.} 2006, \nat, 442,
  786, \dodoi{10.1038/nature05052}

\bibitem[{{Genzel} {et~al.}(2008){Genzel}, {Burkert}, {Bouch{\'e}}, {Cresci},
  {F{\"o}rster Schreiber}, {Shapley}, {Shapiro}, {Tacconi}, {Buschkamp},
  {Cimatti}, {Daddi}, {Davies}, {Eisenhauer}, {Erb}, {Genel}, {Gerhard},
  {Hicks}, {Lutz}, {Naab}, {Ott}, {Rabien}, {Renzini}, {Steidel}, {Sternberg},
  \& {Lilly}}]{Genzel2008}
{Genzel}, R., {Burkert}, A., {Bouch{\'e}}, N., {et~al.} 2008, \apj, 687, 59,
  \dodoi{10.1086/591840}

\bibitem[{{Genzel} {et~al.}(2011){Genzel}, {Newman}, {Jones}, {F{\"o}rster
  Schreiber}, {Shapiro}, {Genel}, {Lilly}, {Renzini}, {Tacconi}, {Bouch{\'e}},
  {Burkert}, {Cresci}, {Buschkamp}, {Carollo}, {Ceverino}, {Davies}, {Dekel},
  {Eisenhauer}, {Hicks}, {Kurk}, {Lutz}, {Mancini}, {Naab}, {Peng},
  {Sternberg}, {Vergani}, \& {Zamorani}}]{Genzel2011}
{Genzel}, R., {Newman}, S., {Jones}, T., {et~al.} 2011, \apj, 733, 101,
  \dodoi{10.1088/0004-637X/733/2/101}

\bibitem[{{Genzel} {et~al.}(2014){Genzel}, {F{\"o}rster Schreiber}, {Rosario},
  {Lang}, {Lutz}, {Wisnioski}, {Wuyts}, {Wuyts}, {Bandara}, {Bender}, {Berta},
  {Kurk}, {Mendel}, {Tacconi}, {Wilman}, {Beifiori}, {Brammer}, {Burkert},
  {Buschkamp}, {Chan}, {Carollo}, {Davies}, {Eisenhauer}, {Fabricius},
  {Fossati}, {Kriek}, {Kulkarni}, {Lilly}, {Mancini}, {Momcheva}, {Naab},
  {Nelson}, {Renzini}, {Saglia}, {Sharples}, {Sternberg}, {Tacchella}, \& {van
  Dokkum}}]{Genzel2014b}
{Genzel}, R., {F{\"o}rster Schreiber}, N.~M., {Rosario}, D., {et~al.} 2014,
  \apj, 796, 7, \dodoi{10.1088/0004-637X/796/1/7}

\bibitem[{{Genzel} {et~al.}(2017){Genzel}, {F{\"o}rster Schreiber},
  {{\"U}bler}, {Lang}, {Naab}, {Bender}, {Tacconi}, {Wisnioski}, {Wuyts},
  {Alexander}, {Beifiori}, {Belli}, {Brammer}, {Burkert}, {Carollo}, {Chan},
  {Davies}, {Fossati}, {Galametz}, {Genel}, {Gerhard}, {Lutz}, {Mendel},
  {Momcheva}, {Nelson}, {Renzini}, {Saglia}, {Sternberg}, {Tacchella},
  {Tadaki}, \& {Wilman}}]{Genzel2017}
{Genzel}, R., {F{\"o}rster Schreiber}, N.~M., {{\"U}bler}, H., {et~al.} 2017,
  \nat, 543, 397, \dodoi{10.1038/nature21685}

\bibitem[{{Genzel} {et~al.}(2020){Genzel}, {Price}, {{\"U}bler}, {F{\"o}rster
  Schreiber}, {Shimizu}, {Tacconi}, {Bender}, {Burkert}, {Contursi}, {Coogan},
  {Davies}, {Davies}, {Dekel}, {Herrera-Camus}, {Lee}, {Lutz}, {Naab}, {Neri},
  {Nestor}, {Renzini}, {Saglia}, {Schuster}, {Sternberg}, {Wisnioski}, \&
  {Wuyts}}]{Genzel2020}
{Genzel}, R., {Price}, S.~H., {{\"U}bler}, H., {et~al.} 2020, \apj, 902, 98,
  \dodoi{10.3847/1538-4357/abb0ea}

\bibitem[{{Genzel} {et~al.}(2023){Genzel}, {Jolly}, {Liu}, {Price}, {Lee},
  {F{\"o}rster Schreiber}, {Tacconi}, {Herrera-Camus}, {Barfety}, {Burkert},
  {Cao}, {Davies}, {Dekel}, {Lee}, {Lutz}, {Naab}, {Neri}, {Nestor Shachar},
  {Pastras}, {Pulsoni}, {Renzini}, {Schuster}, {Shimizu}, {Stanley},
  {Sternberg}, \& {{\"U}bler}}]{Genzel2023}
{Genzel}, R., {Jolly}, J.~B., {Liu}, D., {et~al.} 2023, \apj, 957, 48,
  \dodoi{10.3847/1538-4357/acef1a}

\bibitem[{{Ginzburg} {et~al.}(2022){Ginzburg}, {Dekel}, {Mandelker}, \&
  {Krumholz}}]{Ginzburg2022}
{Ginzburg}, O., {Dekel}, A., {Mandelker}, N., \& {Krumholz}, M.~R. 2022,
  \mnras, 513, 6177, \dodoi{10.1093/mnras/stac1324}

\bibitem[{{Girard} {et~al.}(2018){Girard}, {Dessauges-Zavadsky}, {Schaerer},
  {Cirasuolo}, {Turner}, {Cava}, {Rodr{\'\i}guez-Mu{\~n}oz}, {Richard}, \&
  {P{\'e}rez-Gonz{\'a}lez}}]{Girard2018}
{Girard}, M., {Dessauges-Zavadsky}, M., {Schaerer}, D., {et~al.} 2018, \aap,
  613, A72, \dodoi{10.1051/0004-6361/201731988}

\bibitem[{{Glazebrook}(2013)}]{Glazebrook2013}
{Glazebrook}, K. 2013, \pasa, 30, e056, \dodoi{10.1017/pasa.2013.34}

\bibitem[{{Guo} {et~al.}(2015){Guo}, {Ferguson}, {Bell}, {Koo}, {Conselice},
  {Giavalisco}, {Kassin}, {Lu}, {Lucas}, {Mandelker}, {McIntosh}, {Primack},
  {Ravindranath}, {Barro}, {Ceverino}, {Dekel}, {Faber}, {Fang}, {Koekemoer},
  {Noeske}, {Rafelski}, \& {Straughn}}]{Guo2015}
{Guo}, Y., {Ferguson}, H.~C., {Bell}, E.~F., {et~al.} 2015, \apj, 800, 39,
  \dodoi{10.1088/0004-637X/800/1/39}

\bibitem[{Harris {et~al.}(2020)Harris, Millman, van~der Walt, Gommers,
  Virtanen, Cournapeau, Wieser, Taylor, Berg, Smith, Kern, Picus, Hoyer, van
  Kerkwijk, Brett, Haldane, del R{\'{i}}o, Wiebe, Peterson,
  G{\'{e}}rard-Marchant, Sheppard, Reddy, Weckesser, Abbasi, Gohlke, \&
  Oliphant}]{harris2020}
Harris, C.~R., Millman, K.~J., van~der Walt, S.~J., {et~al.} 2020, Nature, 585,
  357, \dodoi{10.1038/s41586-020-2649-2}

\bibitem[{{Herrera-Camus} {et~al.}(2022){Herrera-Camus}, {F{\"o}rster
  Schreiber}, {Price}, {{\"U}bler}, {Bolatto}, {Davies}, {Fisher}, {Genzel},
  {Lutz}, {Naab}, {Nestor}, {Shimizu}, {Sternberg}, {Tacconi}, \&
  {Tadaki}}]{Herrera-Camus2022}
{Herrera-Camus}, R., {F{\"o}rster Schreiber}, N.~M., {Price}, S.~H., {et~al.}
  2022, \aap, 665, L8, \dodoi{10.1051/0004-6361/202142562}

\bibitem[{{Hodge} {et~al.}(2012){Hodge}, {Carilli}, {Walter}, {de Blok},
  {Riechers}, {Daddi}, \& {Lentati}}]{Hodge2012}
{Hodge}, J.~A., {Carilli}, C.~L., {Walter}, F., {et~al.} 2012, \apj, 760, 11,
  \dodoi{10.1088/0004-637X/760/1/11}

\bibitem[{{Hogan} {et~al.}(2021){Hogan}, {Rigopoulou}, {Magdis},
  {Pereira-Santaella}, {Garc{\'\i}a-Bernete}, {Thatte}, {Grisdale}, \&
  {Huang}}]{Hogan2021}
{Hogan}, L., {Rigopoulou}, D., {Magdis}, G.~E., {et~al.} 2021, \mnras, 503,
  5329, \dodoi{10.1093/mnras/stab527}

\bibitem[{{Hogan} {et~al.}(2022){Hogan}, {Rigopoulou}, {Garc{\'\i}a-Burillo},
  {Alonso-Herrero}, {Barrufet}, {Combes}, {Garc{\'\i}a-Bernete}, {Magdis},
  {Pereira-Santaella}, {Thatte}, \& {Wei{\ss}}}]{Hogan2022}
{Hogan}, L., {Rigopoulou}, D., {Garc{\'\i}a-Burillo}, S., {et~al.} 2022,
  \mnras, 512, 2371, \dodoi{10.1093/mnras/stac520}

\bibitem[{{Huang} {et~al.}(2023){Huang}, {Kawabe}, {Kohno}, {Saito},
  {Mizukoshi}, {Iono}, {Michiyama}, {Tamura}, {Hayward}, \&
  {Umehata}}]{Huang2023}
{Huang}, S., {Kawabe}, R., {Kohno}, K., {et~al.} 2023, \apjl, 958, L26,
  \dodoi{10.3847/2041-8213/acff63}

\bibitem[{{Hung} {et~al.}(2019){Hung}, {Hayward}, {Yuan}, {Boylan-Kolchin},
  {Faucher-Gigu{\`e}re}, {Hopkins}, {Kere{\v{s}}}, {Murray}, \&
  {Wetzel}}]{Hung2019}
{Hung}, C.-L., {Hayward}, C.~C., {Yuan}, T., {et~al.} 2019, \mnras, 482, 5125,
  \dodoi{10.1093/mnras/sty2970}

\bibitem[{{Hunter}(2007)}]{Hunter2007}
{Hunter}, J.~D. 2007, Computing in Science and Engineering, 9, 90,
  \dodoi{10.1109/MCSE.2007.55}

\bibitem[{{Iorio} {et~al.}(2017){Iorio}, {Fraternali}, {Nipoti}, {Di Teodoro},
  {Read}, \& {Battaglia}}]{Iorio2017}
{Iorio}, G., {Fraternali}, F., {Nipoti}, C., {et~al.} 2017, \mnras, 466, 4159,
  \dodoi{10.1093/mnras/stw3285}

\bibitem[{{Jim{\'e}nez} {et~al.}(2023){Jim{\'e}nez}, {Lagos}, {Ludlow}, \&
  {Wisnioski}}]{Jimenez2023}
{Jim{\'e}nez}, E., {Lagos}, C. d.~P., {Ludlow}, A.~D., \& {Wisnioski}, E. 2023,
  \mnras, 524, 4346, \dodoi{10.1093/mnras/stad2119}

\bibitem[{{Johnson} {et~al.}(2018){Johnson}, {Harrison}, {Swinbank}, {Tiley},
  {Stott}, {Bower}, {Smail}, {Bunker}, {Sobral}, {Turner}, {Best}, {Bureau},
  {Cirasuolo}, {Jarvis}, {Magdis}, {Sharples}, {Bland-Hawthorn}, {Catinella},
  {Cortese}, {Croom}, {Federrath}, {Glazebrook}, {Sweet}, {Bryant}, {Goodwin},
  {Konstantopoulos}, {Lawrence}, {Medling}, {Owers}, \&
  {Richards}}]{Johnson2018}
{Johnson}, H.~L., {Harrison}, C.~M., {Swinbank}, A.~M., {et~al.} 2018, \mnras,
  474, 5076, \dodoi{10.1093/mnras/stx3016}

\bibitem[{{Jones} {et~al.}(2021){Jones}, {Vergani}, {Romano}, {Ginolfi},
  {Fudamoto}, {B{\'e}thermin}, {Fujimoto}, {Lemaux}, {Morselli}, {Capak},
  {Cassata}, {Faisst}, {Le F{\`e}vre}, {Schaerer}, {Silverman}, {Yan},
  {Boquien}, {Cimatti}, {Dessauges-Zavadsky}, {Ibar}, {Maiolino}, {Rizzo},
  {Talia}, \& {Zamorani}}]{Jones2021}
{Jones}, G.~C., {Vergani}, D., {Romano}, M., {et~al.} 2021, \mnras, 507, 3540,
  \dodoi{10.1093/mnras/stab2226}

\bibitem[{{J{\'o}zsa} {et~al.}(2007){J{\'o}zsa}, {Kenn}, {Klein}, \&
  {Oosterloo}}]{Jozsa2007}
{J{\'o}zsa}, G.~I.~G., {Kenn}, F., {Klein}, U., \& {Oosterloo}, T.~A. 2007,
  \aap, 468, 731, \dodoi{10.1051/0004-6361:20066164}

\bibitem[{{Kamphuis} {et~al.}(2015){Kamphuis}, {J{\'o}zsa}, {Oh}, {Spekkens},
  {Urbancic}, {Serra}, {Koribalski}, \& {Dettmar}}]{Kamphuis2015}
{Kamphuis}, P., {J{\'o}zsa}, G.~I.~G., {Oh}, S. .~H., {et~al.} 2015, \mnras,
  452, 3139, \dodoi{10.1093/mnras/stv1480}

\bibitem[{{Kassin} {et~al.}(2012){Kassin}, {Weiner}, {Faber}, {Gardner},
  {Willmer}, {Coil}, {Cooper}, {Devriendt}, {Dutton}, {Guhathakurta}, {Koo},
  {Metevier}, {Noeske}, \& {Primack}}]{Kassin2012}
{Kassin}, S.~A., {Weiner}, B.~J., {Faber}, S.~M., {et~al.} 2012, \apj, 758,
  106, \dodoi{10.1088/0004-637X/758/2/106}

\bibitem[{Koposov {et~al.}(2023)Koposov, Speagle, Barbary, Ashton, Bennett,
  Buchner, Scheffler, Cook, Talbot, Guillochon, Cubillos, Ramos, Johnson, Lang,
  Ilya, Dartiailh, Nitz, McCluskey, \& Archibald}]{koposov2023}
Koposov, S., Speagle, J., Barbary, K., {et~al.} 2023, joshspeagle/dynesty:
  v2.1.3, v2.1.3,  Zenodo, \dodoi{10.5281/zenodo.8408702}

\bibitem[{{Kretschmer} {et~al.}(2021){Kretschmer}, {Dekel}, {Freundlich},
  {Lapiner}, {Ceverino}, \& {Primack}}]{Kretschmer2021}
{Kretschmer}, M., {Dekel}, A., {Freundlich}, J., {et~al.} 2021, \mnras, 503,
  5238, \dodoi{10.1093/mnras/stab833}

\bibitem[{{Krumholz} {et~al.}(2018){Krumholz}, {Burkhart}, {Forbes}, \&
  {Crocker}}]{Krumholz2018}
{Krumholz}, M.~R., {Burkhart}, B., {Forbes}, J.~C., \& {Crocker}, R.~M. 2018,
  \mnras, 477, 2716, \dodoi{10.1093/mnras/sty852}

\bibitem[{{Lang} {et~al.}(2014){Lang}, {Wuyts}, {Somerville}, {F{\"o}rster
  Schreiber}, {Genzel}, {Bell}, {Brammer}, {Dekel}, {Faber}, {Ferguson},
  {Grogin}, {Kocevski}, {Koekemoer}, {Lutz}, {McGrath}, {Momcheva}, {Nelson},
  {Primack}, {Rosario}, {Skelton}, {Tacconi}, {van Dokkum}, \&
  {Whitaker}}]{Lang2014}
{Lang}, P., {Wuyts}, S., {Somerville}, R.~S., {et~al.} 2014, \apj, 788, 11,
  \dodoi{10.1088/0004-637X/788/1/11}

\bibitem[{{Lang} {et~al.}(2017){Lang}, {F{\"o}rster Schreiber}, {Genzel},
  {Wuyts}, {Wisnioski}, {Beifiori}, {Belli}, {Bender}, {Brammer}, {Burkert},
  {Chan}, {Davies}, {Fossati}, {Galametz}, {Kulkarni}, {Lutz}, {Mendel},
  {Momcheva}, {Naab}, {Nelson}, {Saglia}, {Seitz}, {Tacchella}, {Tacconi},
  {Tadaki}, {{\"U}bler}, {van Dokkum}, \& {Wilman}}]{Lang2017}
{Lang}, P., {F{\"o}rster Schreiber}, N.~M., {Genzel}, R., {et~al.} 2017, \apj,
  840, 92, \dodoi{10.3847/1538-4357/aa6d82}

\bibitem[{{Larkin} {et~al.}(2006){Larkin}, {Barczys}, {Krabbe}, {Adkins},
  {Aliado}, {Amico}, {Brims}, {Campbell}, {Canfield}, {Gasaway}, {Honey},
  {Iserlohe}, {Johnson}, {Kress}, {LaFreniere}, {Lyke}, {Magnone}, {Magnone},
  {McElwain}, {Moon}, {Quirrenbach}, {Skulason}, {Song}, {Spencer}, {Weiss}, \&
  {Wright}}]{Larkin2006}
{Larkin}, J., {Barczys}, M., {Krabbe}, A., {et~al.} 2006, in Society of
  Photo-Optical Instrumentation Engineers (SPIE) Conference Series, Vol. 6269,
  Ground-based and Airborne Instrumentation for Astronomy, ed. I.~S. {McLean}
  \& M.~{Iye}, 62691A, \dodoi{10.1117/12.672061}

\bibitem[{{Lelli} {et~al.}(2021){Lelli}, {Di Teodoro}, {Fraternali}, {Man},
  {Zhang}, {De Breuck}, {Davis}, \& {Maiolino}}]{Lelli2021}
{Lelli}, F., {Di Teodoro}, E.~M., {Fraternali}, F., {et~al.} 2021, Science,
  371, 713, \dodoi{10.1126/science.abc1893}

\bibitem[{{Lelli} {et~al.}(2023){Lelli}, {Zhang}, {Bisbas}, {Lin},
  {Papadopoulos}, {Schombert}, {Di Teodoro}, {Marasco}, \&
  {McGaugh}}]{Lelli2023}
{Lelli}, F., {Zhang}, Z.-Y., {Bisbas}, T.~G., {et~al.} 2023, \aap, 672, A106,
  \dodoi{10.1051/0004-6361/202245105}

\bibitem[{{Lin} {et~al.}(2016){Lin}, {Davies}, {Burtscher}, {Contursi},
  {Genzel}, {Gonz{\'a}lez-Alfonso}, {Graci{\'a}-Carpio}, {Janssen}, {Lutz},
  {Orban de Xivry}, {Rosario}, {Schnorr-M{\"u}ller}, {Sternberg}, {Sturm}, \&
  {Tacconi}}]{Lin2016}
{Lin}, M.-Y., {Davies}, R.~I., {Burtscher}, L., {et~al.} 2016, \mnras, 458,
  1375, \dodoi{10.1093/mnras/stw401}

\bibitem[{{Liu} {et~al.}(2023){Liu}, {F{\"o}rster Schreiber}, {Genzel}, {Lutz},
  {Price}, {Lee}, {Baker}, {Burkert}, {Coogan}, {Davies}, {Davies},
  {Herrera-Camus}, {Kodama}, {Lee}, {Nestor}, {Pulsoni}, {Renzini}, {Sharon},
  {Shimizu}, {Tacconi}, {Tadaki}, \& {{\"U}bler}}]{Liu2023}
{Liu}, D., {F{\"o}rster Schreiber}, N.~M., {Genzel}, R., {et~al.} 2023, \apj,
  942, 98, \dodoi{10.3847/1538-4357/aca46b}

\bibitem[{{Loiacono} {et~al.}(2019){Loiacono}, {Talia}, {Fraternali},
  {Cimatti}, {Di Teodoro}, \& {Caminha}}]{Loiacono2019}
{Loiacono}, F., {Talia}, M., {Fraternali}, F., {et~al.} 2019, \mnras, 489, 681,
  \dodoi{10.1093/mnras/stz2170}

\bibitem[{{Madau} \& {Dickinson}(2014)}]{Madau2014}
{Madau}, P., \& {Dickinson}, M. 2014, \araa, 52, 415,
  \dodoi{10.1146/annurev-astro-081811-125615}

\bibitem[{{Mancera Pi{\~n}a} {et~al.}(2019){Mancera Pi{\~n}a}, {Fraternali},
  {Adams}, {Marasco}, {Oosterloo}, {Oman}, {Leisman}, {di Teodoro}, {Posti},
  {Battipaglia}, {Cannon}, {Gault}, {Haynes}, {Janowiecki}, {McAllan}, {Pagel},
  {Reiter}, {Rhode}, {Salzer}, \& {Smith}}]{MancenaPina2019}
{Mancera Pi{\~n}a}, P.~E., {Fraternali}, F., {Adams}, E. A.~K., {et~al.} 2019,
  \apjl, 883, L33, \dodoi{10.3847/2041-8213/ab40c7}

\bibitem[{{Mancera Pi{\~n}a} {et~al.}(2020){Mancera Pi{\~n}a}, {Fraternali},
  {Oman}, {Adams}, {Bacchini}, {Marasco}, {Oosterloo}, {Pezzulli}, {Posti},
  {Leisman}, {Cannon}, {di Teodoro}, {Gault}, {Haynes}, {Reiter}, {Rhode},
  {Salzer}, \& {Smith}}]{ManceraPina2020}
{Mancera Pi{\~n}a}, P.~E., {Fraternali}, F., {Oman}, K.~A., {et~al.} 2020,
  \mnras, 495, 3636, \dodoi{10.1093/mnras/staa1256}

\bibitem[{{Mancini} {et~al.}(2011){Mancini}, {F{\"o}rster Schreiber},
  {Renzini}, {Cresci}, {Hicks}, {Peng}, {Vergani}, {Lilly}, {Carollo},
  {Pozzetti}, {Zamorani}, {Daddi}, {Genzel}, {Maraston}, {McCracken},
  {Tacconi}, {Bouch{\'e}}, {Davies}, {Oesch}, {Shapiro}, {Mainieri}, {Lutz},
  {Mignoli}, \& {Sternberg}}]{Mancini2011}
{Mancini}, C., {F{\"o}rster Schreiber}, N.~M., {Renzini}, A., {et~al.} 2011,
  \apj, 743, 86, \dodoi{10.1088/0004-637X/743/1/86}

\bibitem[{{Markwardt}(2009)}]{Markwardt2009}
{Markwardt}, C.~B. 2009, in Astronomical Society of the Pacific Conference
  Series, Vol. 411, Astronomical Data Analysis Software and Systems XVIII, ed.
  D.~A. {Bohlender}, D.~{Durand}, \& P.~{Dowler}, 251,
  \dodoi{10.48550/arXiv.0902.2850}

\bibitem[{{Mason} {et~al.}(2017){Mason}, {Treu}, {Fontana}, {Jones},
  {Morishita}, {Amorin}, {Brada{\v{c}}}, {Quinn Finney}, {Grillo}, {Henry},
  {Hoag}, {Huang}, {Schmidt}, {Trenti}, \& {Vulcani}}]{Mason2017}
{Mason}, C.~A., {Treu}, T., {Fontana}, A., {et~al.} 2017, \apj, 838, 14,
  \dodoi{10.3847/1538-4357/aa60c4}

\bibitem[{{Mogotsi} {et~al.}(2016){Mogotsi}, {de Blok}, {Cald{\'u}-Primo},
  {Walter}, {Ianjamasimanana}, \& {Leroy}}]{Mogotsi2016}
{Mogotsi}, K.~M., {de Blok}, W.~J.~G., {Cald{\'u}-Primo}, A., {et~al.} 2016,
  \aj, 151, 15, \dodoi{10.3847/0004-6256/151/1/15}

\bibitem[{{Moster} {et~al.}(2018){Moster}, {Naab}, \& {White}}]{Moster2018}
{Moster}, B.~P., {Naab}, T., \& {White}, S. D.~M. 2018, \mnras, 477, 1822,
  \dodoi{10.1093/mnras/sty655}

\bibitem[{{M{\"u}ller-S{\'a}nchez} {et~al.}(2013){M{\"u}ller-S{\'a}nchez},
  {Prieto}, {Mezcua}, {Davies}, {Malkan}, \& {Elitzur}}]{MuellerSanchez2013}
{M{\"u}ller-S{\'a}nchez}, F., {Prieto}, M.~A., {Mezcua}, M., {et~al.} 2013,
  \apjl, 763, L1, \dodoi{10.1088/2041-8205/763/1/L1}

\bibitem[{{Navarro} {et~al.}(1996){Navarro}, {Frenk}, \& {White}}]{Navarro1996}
{Navarro}, J.~F., {Frenk}, C.~S., \& {White}, S. D.~M. 1996, \apj, 462, 563,
  \dodoi{10.1086/177173}

\bibitem[{{Neeleman} {et~al.}(2023){Neeleman}, {Walter}, {Decarli}, {Drake},
  {Eilers}, {Meyer}, \& {Venemans}}]{Neeleman2023}
{Neeleman}, M., {Walter}, F., {Decarli}, R., {et~al.} 2023, \apj, 958, 132,
  \dodoi{10.3847/1538-4357/ad05d2}

\bibitem[{{Neeleman} {et~al.}(2021){Neeleman}, {Novak}, {Venemans}, {Walter},
  {Decarli}, {Kaasinen}, {Schindler}, {Ba{\~n}ados}, {Carilli}, {Drake}, {Fan},
  \& {Rix}}]{Neeleman2021}
{Neeleman}, M., {Novak}, M., {Venemans}, B.~P., {et~al.} 2021, \apj, 911, 141,
  \dodoi{10.3847/1538-4357/abe70f}

\bibitem[{Nelder \& Mead(1965)}]{NelderMead1965}
Nelder, J.~A., \& Mead, R. 1965, The Computer Journal, 7, 308,
  \dodoi{10.1093/comjnl/7.4.308}

\bibitem[{{Nestor Shachar} {et~al.}(2023){Nestor Shachar}, {Price},
  {F{\"o}rster Schreiber}, {Genzel}, {Shimizu}, {Tacconi}, {{\"U}bler},
  {Burkert}, {Davies}, {Dekel}, {Herrera-Camus}, {Lee}, {Liu}, {Lutz}, {Naab},
  {Neri}, {Renzini}, {Saglia}, {Schuster}, {Sternberg}, {Wisnioski}, \&
  {Wuyts}}]{Nestor2023}
{Nestor Shachar}, A., {Price}, S.~H., {F{\"o}rster Schreiber}, N.~M., {et~al.}
  2023, \apj, 944, 78, \dodoi{10.3847/1538-4357/aca9cf}

\bibitem[{{Noordermeer}(2008)}]{Noordermeer2008}
{Noordermeer}, E. 2008, \mnras, 385, 1359,
  \dodoi{10.1111/j.1365-2966.2008.12837.x}

\bibitem[{{Oh} {et~al.}(2015){Oh}, {Hunter}, {Brinks}, {Elmegreen}, {Schruba},
  {Walter}, {Rupen}, {Young}, {Simpson}, {Johnson}, {Herrmann}, {Ficut-Vicas},
  {Cigan}, {Heesen}, {Ashley}, \& {Zhang}}]{Oh2015}
{Oh}, S.-H., {Hunter}, D.~A., {Brinks}, E., {et~al.} 2015, \aj, 149, 180,
  \dodoi{10.1088/0004-6256/149/6/180}

\bibitem[{{Parlanti} {et~al.}(2023){Parlanti}, {Carniani}, {Pallottini},
  {Cignoni}, {Cresci}, {Kohandel}, {Mannucci}, \& {Marconi}}]{Parlanti2023}
{Parlanti}, E., {Carniani}, S., {Pallottini}, A., {et~al.} 2023, \aap, 673,
  A153, \dodoi{10.1051/0004-6361/202245603}

\bibitem[{{Parlanti} {et~al.}(2024){Parlanti}, {Carniani}, {{\"U}bler},
  {Venturi}, {Circosta}, {D'Eugenio}, {Arribas}, {Bunker}, {Charlot},
  {L{\"u}tzgendorf}, {Maiolino}, {Perna}, {Rodr{\'\i}guez Del Pino}, {Willott},
  {B{\"o}ker}, {Cameron}, {Chevallard}, {Cresci}, {Jones}, {Kumari},
  {Lamperti}, \& {Scholtz}}]{Parlanti2024}
{Parlanti}, E., {Carniani}, S., {{\"U}bler}, H., {et~al.} 2024, \aap, 684, A24,
  \dodoi{10.1051/0004-6361/202347914}

\bibitem[{{Perna} {et~al.}(2022){Perna}, {Arribas}, {Colina}, {Pereira
  Santaella}, {Lamperti}, {Di Teodoro}, {{\"U}bler}, {Costantin}, {Maiolino},
  {Cresci}, {Bellocchi}, {Catal{\'a}n-Torrecilla}, {Cazzoli}, \& {Piqueras
  L{\'o}pez}}]{Perna2022}
{Perna}, M., {Arribas}, S., {Colina}, L., {et~al.} 2022, \aap, 662, A94,
  \dodoi{10.1051/0004-6361/202142659}

\bibitem[{{P{\'e}roux} {et~al.}(2013){P{\'e}roux}, {Bouch{\'e}}, {Kulkarni}, \&
  {York}}]{Peroux2013}
{P{\'e}roux}, C., {Bouch{\'e}}, N., {Kulkarni}, V.~P., \& {York}, D.~G. 2013,
  \mnras, 436, 2650, \dodoi{10.1093/mnras/stt1760}

\bibitem[{{Pope} {et~al.}(2023){Pope}, {McKinney}, {Kamieneski}, {Battisti},
  {Aretxaga}, {Brammer}, {Diego}, {Hughes}, {Keller}, {Marchesini}, {Mizener},
  {Monta{\~n}a}, {Murphy}, {Whitaker}, {Wilson}, \& {Yun}}]{Pope2023}
{Pope}, A., {McKinney}, J., {Kamieneski}, P., {et~al.} 2023, \apjl, 951, L46,
  \dodoi{10.3847/2041-8213/acdf5a}

\bibitem[{{Posses} {et~al.}(2023){Posses}, {Aravena}, {Gonz{\'a}lez-L{\'o}pez},
  {Assef}, {Lambert}, {Jones}, {Bouwens}, {Brisbin}, {D{\'\i}az-Santos},
  {Herrera-Camus}, {Ricci}, \& {Smit}}]{Posses2023}
{Posses}, A.~C., {Aravena}, M., {Gonz{\'a}lez-L{\'o}pez}, J., {et~al.} 2023,
  \aap, 669, A46, \dodoi{10.1051/0004-6361/202243399}

\bibitem[{{Price} {et~al.}(2021){Price}, {Shimizu}, {Genzel}, {{\"U}bler},
  {F{\"o}rster Schreiber}, {Tacconi}, {Davies}, {Coogan}, {Lutz}, {Wuyts},
  {Wisnioski}, {Nestor}, {Sternberg}, {Burkert}, {Bender}, {Contursi},
  {Davies}, {Herrera-Camus}, {Lee}, {Naab}, {Neri}, {Renzini}, {Saglia},
  {Schruba}, \& {Schuster}}]{Price2021}
{Price}, S.~H., {Shimizu}, T.~T., {Genzel}, R., {et~al.} 2021, \apj, 922, 143,
  \dodoi{10.3847/1538-4357/ac22ad}

\bibitem[{{Puglisi} {et~al.}(2023){Puglisi}, {Dudzevi{\v{c}}i{\={u}}t{\.{e}}},
  {Swinbank}, {Gillman}, {Tiley}, {Bower}, {Cirasuolo}, {Cortese},
  {Glazebrook}, {Harrison}, {Ibar}, {Molina}, {Obreschkow}, {Oman}, {Schaller},
  {Shankar}, \& {Sharples}}]{Puglisi2023}
{Puglisi}, A., {Dudzevi{\v{c}}i{\={u}}t{\.{e}}}, U., {Swinbank}, M., {et~al.}
  2023, \mnras, 524, 2814, \dodoi{10.1093/mnras/stad1966}

\bibitem[{Reshef {et~al.}(2011)Reshef, Reshef, Finucane, Grossman, McVean,
  Turnbaugh, Lander, Mitzenmacher, \& Sabeti}]{Reshef2011}
Reshef, D.~N., Reshef, Y.~A., Finucane, H.~K., {et~al.} 2011, Science, 334,
  1518, \dodoi{10.1126/science.1205438}

\bibitem[{{Rizzo} {et~al.}(2022){Rizzo}, {Kohandel}, {Pallottini}, {Zanella},
  {Ferrara}, {Vallini}, \& {Toft}}]{Rizzo2022}
{Rizzo}, F., {Kohandel}, M., {Pallottini}, A., {et~al.} 2022, \aap, 667, A5,
  \dodoi{10.1051/0004-6361/202243582}

\bibitem[{{Rizzo} {et~al.}(2021){Rizzo}, {Vegetti}, {Fraternali}, {Stacey}, \&
  {Powell}}]{Rizzo2021}
{Rizzo}, F., {Vegetti}, S., {Fraternali}, F., {Stacey}, H.~R., \& {Powell}, D.
  2021, \mnras, 507, 3952, \dodoi{10.1093/mnras/stab2295}

\bibitem[{{Rizzo} {et~al.}(2020){Rizzo}, {Vegetti}, {Powell}, {Fraternali},
  {McKean}, {Stacey}, \& {White}}]{Rizzo2020}
{Rizzo}, F., {Vegetti}, S., {Powell}, D., {et~al.} 2020, \nat, 584, 201,
  \dodoi{10.1038/s41586-020-2572-6}

\bibitem[{{Rizzo} {et~al.}(2023){Rizzo}, {Roman-Oliveira}, {Fraternali},
  {Frickmann}, {Valentino}, {Brammer}, {Zanella}, {Kokorev}, {Popping},
  {Whitaker}, {Kohandel}, {Magdis}, {Di Mascolo}, {Ikeda}, {Jin}, \&
  {Toft}}]{Rizzo2023}
{Rizzo}, F., {Roman-Oliveira}, F., {Fraternali}, F., {et~al.} 2023, \aap, 679,
  A129, \dodoi{10.1051/0004-6361/202346444}

\bibitem[{{Rodighiero} {et~al.}(2011){Rodighiero}, {Daddi}, {Baronchelli},
  {Cimatti}, {Renzini}, {Aussel}, {Popesso}, {Lutz}, {Andreani}, {Berta},
  {Cava}, {Elbaz}, {Feltre}, {Fontana}, {F{\"o}rster Schreiber},
  {Franceschini}, {Genzel}, {Grazian}, {Gruppioni}, {Ilbert}, {Le Floch},
  {Magdis}, {Magliocchetti}, {Magnelli}, {Maiolino}, {McCracken}, {Nordon},
  {Poglitsch}, {Santini}, {Pozzi}, {Riguccini}, {Tacconi}, {Wuyts}, \&
  {Zamorani}}]{Rodighiero2011}
{Rodighiero}, G., {Daddi}, E., {Baronchelli}, I., {et~al.} 2011, \apjl, 739,
  L40, \dodoi{10.1088/2041-8205/739/2/L40}

\bibitem[{{Rogstad} {et~al.}(1974){Rogstad}, {Lockhart}, \&
  {Wright}}]{Rogstad1974}
{Rogstad}, D.~H., {Lockhart}, I.~A., \& {Wright}, M.~C.~H. 1974, \apj, 193,
  309, \dodoi{10.1086/153164}

\bibitem[{{Roman-Oliveira} {et~al.}(2023){Roman-Oliveira}, {Fraternali}, \&
  {Rizzo}}]{Roman-Oliveira2023}
{Roman-Oliveira}, F., {Fraternali}, F., \& {Rizzo}, F. 2023, \mnras, 521, 1045,
  \dodoi{10.1093/mnras/stad530}

\bibitem[{{Roper} {et~al.}(2023){Roper}, {Oman}, {Frenk},
  {Ben{\'\i}tez-Llambay}, {Navarro}, \& {Santos-Santos}}]{Roper2023}
{Roper}, F.~A., {Oman}, K.~A., {Frenk}, C.~S., {et~al.} 2023, \mnras, 521,
  1316, \dodoi{10.1093/mnras/stad549}

\bibitem[{{Rowland} {et~al.}(2024){Rowland}, {Hodge}, {Bouwens}, {Pi{\~n}a},
  {Hygate}, {Algera}, {Aravena}, {Bowler}, {da Cunha}, {Dayal}, {Ferrara},
  {Herard-Demanche}, {Inami}, {van Leeuwen}, {de Looze}, {Oesch}, {Pallottini},
  {Phillips}, {Rybak}, {Schouws}, {Smit}, {Sommovigo}, {Stefanon}, \& {van der
  Werf}}]{Rowland2024}
{Rowland}, L.~E., {Hodge}, J., {Bouwens}, R., {et~al.} 2024, \mnras,
  \dodoi{10.1093/mnras/stae2217}

\bibitem[{{Sancisi}(2004)}]{Sancisi2004}
{Sancisi}, R. 2004, in Dark Matter in Galaxies, ed. S.~{Ryder}, D.~{Pisano},
  M.~{Walker}, \& K.~{Freeman}, Vol. 220, 233,
  \dodoi{10.48550/arXiv.astro-ph/0311348}

\bibitem[{{Sani} {et~al.}(2012){Sani}, {Davies}, {Sternberg},
  {Graci{\'a}-Carpio}, {Hicks}, {Krips}, {Tacconi}, {Genzel}, {Vollmer},
  {Schinnerer}, {Garc{\'\i}a-Burillo}, {Usero}, \& {Orban de Xivry}}]{Sani2012}
{Sani}, E., {Davies}, R.~I., {Sternberg}, A., {et~al.} 2012, \mnras, 424, 1963,
  \dodoi{10.1111/j.1365-2966.2012.21333.x}

\bibitem[{{Sargent} {et~al.}(2012){Sargent}, {B{\'e}thermin}, {Daddi}, \&
  {Elbaz}}]{Sargent2012}
{Sargent}, M.~T., {B{\'e}thermin}, M., {Daddi}, E., \& {Elbaz}, D. 2012, \apjl,
  747, L31, \dodoi{10.1088/2041-8205/747/2/L31}

\bibitem[{{Sharda} {et~al.}(2019){Sharda}, {da Cunha}, {Federrath},
  {Wisnioski}, {Di Teodoro}, {Tadaki}, {Yun}, {Aretxaga}, \&
  {Kawabe}}]{Sharda2019}
{Sharda}, P., {da Cunha}, E., {Federrath}, C., {et~al.} 2019, \mnras, 487,
  4305, \dodoi{10.1093/mnras/stz1543}

\bibitem[{{Sharma} {et~al.}(2023){Sharma}, {Freundlich}, {van de Ven},
  {Famaey}, {Salucci}, {Martorano}, \& {Renaud}}]{Sharma2023}
{Sharma}, G., {Freundlich}, J., {van de Ven}, G., {et~al.} 2023, arXiv
  e-prints, arXiv:2309.04541, \dodoi{10.48550/arXiv.2309.04541}

\bibitem[{{Sharma} {et~al.}(2021){Sharma}, {Salucci}, {Harrison}, {van de Ven},
  \& {Lapi}}]{Sharma2021}
{Sharma}, G., {Salucci}, P., {Harrison}, C.~M., {van de Ven}, G., \& {Lapi}, A.
  2021, \mnras, 503, 1753, \dodoi{10.1093/mnras/stab249}

\bibitem[{{Sharma} {et~al.}(2022){Sharma}, {Salucci}, \& {van de
  Ven}}]{Sharma2022}
{Sharma}, G., {Salucci}, P., \& {van de Ven}, G. 2022, \aap, 659, A40,
  \dodoi{10.1051/0004-6361/202141822}

\bibitem[{{Sharon} {et~al.}(2019){Sharon}, {Tagore}, {Baker}, {Rivera},
  {Keeton}, {Lutz}, {Genzel}, {Wilner}, {Hicks}, {Allam}, \&
  {Tucker}}]{Sharon2019}
{Sharon}, C.~E., {Tagore}, A.~S., {Baker}, A.~J., {et~al.} 2019, \apj, 879, 52,
  \dodoi{10.3847/1538-4357/ab22b9}

\bibitem[{{Sharples} {et~al.}(2013){Sharples}, {Bender}, {Agudo Berbel},
  {Bezawada}, {Castillo}, {Cirasuolo}, {Davidson}, {Davies}, {Dubbeldam},
  {Fairley}, {Finger}, {F{\"o}rster Schreiber}, {Gonte}, {Hess}, {Jung},
  {Lewis}, {Lizon}, {Muschielok}, {Pasquini}, {Pirard}, {Popovic}, {Ramsay},
  {Rees}, {Richter}, {Riquelme}, {Rodrigues}, {Saviane}, {Schlichter},
  {Schmidtobreick}, {Segovia}, {Smette}, {Szeifert}, {van Kesteren}, {Wegner},
  \& {Wiezorrek}}]{Sharples2013}
{Sharples}, R., {Bender}, R., {Agudo Berbel}, A., {et~al.} 2013, The Messenger,
  151, 21

\bibitem[{Sicking(1997)}]{Sicking1997}
Sicking, F. 1997, PhD thesis

\bibitem[{{Simons} {et~al.}(2018){Simons}, {Kassin}, {Weiner}, {Heckman},
  {Trump}, \& {SIGMA}}]{Simons2018}
{Simons}, R.~C., {Kassin}, S.~A., {Weiner}, B., {et~al.} 2018, in American
  Astronomical Society Meeting Abstracts, Vol. 231, American Astronomical
  Society Meeting Abstracts \#231, 309.02

\bibitem[{{Speagle} {et~al.}(2014){Speagle}, {Steinhardt}, {Capak}, \&
  {Silverman}}]{Speagle2014}
{Speagle}, J.~S., {Steinhardt}, C.~L., {Capak}, P.~L., \& {Silverman}, J.~D.
  2014, \apjs, 214, 15, \dodoi{10.1088/0067-0049/214/2/15}

\bibitem[{Spearman(1904)}]{Spearman1904}
Spearman, C. 1904, The American Journal of Psychology, 15, 72.
\newblock \url{http://www.jstor.org/stable/1412159}

\bibitem[{{Straatman} {et~al.}(2022){Straatman}, {van der Wel}, {van Houdt},
  {Bezanson}, {Bell}, {van Dokkum}, {D'Eugenio}, {Franx}, {Gallazzi}, {de
  Graaff}, {Maseda}, {Meidt}, {Muzzin}, {Sobral}, \& {Wu}}]{Straatman2022}
{Straatman}, C. M.~S., {van der Wel}, A., {van Houdt}, J., {et~al.} 2022, \apj,
  928, 126, \dodoi{10.3847/1538-4357/ac4e18}

\bibitem[{{Su} {et~al.}(2022){Su}, {Lin}, {Pan}, {L{\'o}pez Cob{\'a}}, {Hsieh},
  {S{\'a}nchez}, {Thorp}, {Bureau}, \& {Ellison}}]{Su2022}
{Su}, Y.-C., {Lin}, L., {Pan}, H.-A., {et~al.} 2022, \apj, 934, 173,
  \dodoi{10.3847/1538-4357/ac77fd}

\bibitem[{Tacconi {et~al.}(2020)Tacconi, Genzel, \& Sternberg}]{Tacconi2020}
Tacconi, L.~J., Genzel, R., \& Sternberg, A. 2020, Annual Review of Astronomy
  and Astrophysics, 58, 157, \dodoi{10.1146/annurev-astro-082812-141034}

\bibitem[{{Tacconi} {et~al.}(2013){Tacconi}, {Neri}, {Genzel}, {Combes},
  {Bolatto}, {Cooper}, {Wuyts}, {Bournaud}, {Burkert}, {Comerford}, {Cox},
  {Davis}, {F{\"o}rster Schreiber}, {Garc{\'\i}a-Burillo}, {Gracia-Carpio},
  {Lutz}, {Naab}, {Newman}, {Omont}, {Saintonge}, {Shapiro Griffin}, {Shapley},
  {Sternberg}, \& {Weiner}}]{Tacconi2013}
{Tacconi}, L.~J., {Neri}, R., {Genzel}, R., {et~al.} 2013, \apj, 768, 74,
  \dodoi{10.1088/0004-637X/768/1/74}

\bibitem[{{Tacconi} {et~al.}(2018){Tacconi}, {Genzel}, {Saintonge}, {Combes},
  {Garc{\'\i}a-Burillo}, {Neri}, {Bolatto}, {Contini}, {F{\"o}rster Schreiber},
  {Lilly}, {Lutz}, {Wuyts}, {Accurso}, {Boissier}, {Boone}, {Bouch{\'e}},
  {Bournaud}, {Burkert}, {Carollo}, {Cooper}, {Cox}, {Feruglio}, {Freundlich},
  {Herrera-Camus}, {Juneau}, {Lippa}, {Naab}, {Renzini}, {Salome}, {Sternberg},
  {Tadaki}, {{\"U}bler}, {Walter}, {Weiner}, \& {Weiss}}]{Tacconi2018}
{Tacconi}, L.~J., {Genzel}, R., {Saintonge}, A., {et~al.} 2018, \apj, 853, 179,
  \dodoi{10.3847/1538-4357/aaa4b4}

\bibitem[{{Tadaki} {et~al.}(2018){Tadaki}, {Iono}, {Yun}, {Aretxaga},
  {Hatsukade}, {Hughes}, {Ikarashi}, {Izumi}, {Kawabe}, {Kohno}, {Lee},
  {Matsuda}, {Nakanishi}, {Saito}, {Tamura}, {Ueda}, {Umehata}, {Wilson},
  {Michiyama}, {Ando}, \& {Kamieneski}}]{Tadaki2018}
{Tadaki}, K., {Iono}, D., {Yun}, M.~S., {et~al.} 2018, \nat, 560, 613,
  \dodoi{10.1038/s41586-018-0443-1}

\bibitem[{{Tadaki} {et~al.}(2017){Tadaki}, {Kodama}, {Nelson}, {Belli},
  {F{\"o}rster Schreiber}, {Genzel}, {Hayashi}, {Herrera-Camus}, {Koyama},
  {Lang}, {Lutz}, {Shimakawa}, {Tacconi}, {{\"U}bler}, {Wisnioski}, {Wuyts},
  {Hatsukade}, {Lippa}, {Nakanishi}, {Ikarashi}, {Kohno}, {Suzuki}, {Tamura},
  \& {Tanaka}}]{Tadaki2017}
{Tadaki}, K.-i., {Kodama}, T., {Nelson}, E.~J., {et~al.} 2017, \apjl, 841, L25,
  \dodoi{10.3847/2041-8213/aa7338}

\bibitem[{{Tadaki} {et~al.}(2020){Tadaki}, {Iono}, {Yun}, {Aretxaga},
  {Hatsukade}, {Lee}, {Michiyama}, {Nakanishi}, {Saito}, {Ueda}, \&
  {Umehata}}]{Tadaki2020}
{Tadaki}, K.-i., {Iono}, D., {Yun}, M.~S., {et~al.} 2020, \apj, 889, 141,
  \dodoi{10.3847/1538-4357/ab64f4}

\bibitem[{{Tamburro} {et~al.}(2009){Tamburro}, {Rix}, {Leroy}, {Mac Low},
  {Walter}, {Kennicutt}, {Brinks}, \& {de Blok}}]{Tamburro2009}
{Tamburro}, D., {Rix}, H.~W., {Leroy}, A.~K., {et~al.} 2009, \aj, 137, 4424,
  \dodoi{10.1088/0004-6256/137/5/4424}

\bibitem[{{Tsukui} \& {Iguchi}(2021)}]{Tsukui2021}
{Tsukui}, T., \& {Iguchi}, S. 2021, Science, 372, 1201,
  \dodoi{10.1126/science.abe9680}

\bibitem[{{{\"U}bler} {et~al.}(2018){{\"U}bler}, {Genzel}, {Tacconi},
  {F{\"o}rster Schreiber}, {Neri}, {Contursi}, {Belli}, {Nelson}, {Lang},
  {Shimizu}, {Davies}, {Herrera-Camus}, {Lutz}, {Plewa}, {Price}, {Schuster},
  {Sternberg}, {Tadaki}, {Wisnioski}, \& {Wuyts}}]{Uebler2018}
{{\"U}bler}, H., {Genzel}, R., {Tacconi}, L.~J., {et~al.} 2018, \apjl, 854,
  L24, \dodoi{10.3847/2041-8213/aaacfa}

\bibitem[{{{\"U}bler} {et~al.}(2019){{\"U}bler}, {Genzel}, {Wisnioski},
  {F{\"o}rster Schreiber}, {Shimizu}, {Price}, {Tacconi}, {Belli}, {Wilman},
  {Fossati}, {Mendel}, {Davies}, {Beifiori}, {Bender}, {Brammer}, {Burkert},
  {Chan}, {Davies}, {Fabricius}, {Galametz}, {Herrera-Camus}, {Lang}, {Lutz},
  {Momcheva}, {Naab}, {Nelson}, {Saglia}, {Tadaki}, {van Dokkum}, \&
  {Wuyts}}]{Uebler2019}
{{\"U}bler}, H., {Genzel}, R., {Wisnioski}, E., {et~al.} 2019, \apj, 880, 48,
  \dodoi{10.3847/1538-4357/ab27cc}

\bibitem[{{{\"U}bler} {et~al.}(2021){{\"U}bler}, {Genel}, {Sternberg},
  {Genzel}, {Price}, {F{\"o}rster Schreiber}, {Shimizu}, {Pillepich}, {Nelson},
  {Burkert}, {Davies}, {Hernquist}, {Lang}, {Lutz}, {Pakmor}, \&
  {Tacconi}}]{Uebler2021}
{{\"U}bler}, H., {Genel}, S., {Sternberg}, A., {et~al.} 2021, \mnras, 500,
  4597, \dodoi{10.1093/mnras/staa3464}

\bibitem[{{{\"U}bler} {et~al.}(2024{\natexlab{a}}){{\"U}bler}, {D'Eugenio},
  {Perna}, {Arribas}, {Jones}, {Bunker}, {Carniani}, {Charlot}, {Maiolino},
  {Rodr{\'\i}guez del Pino}, {Willott}, {B{\"o}ker}, {Cresci}, {Kumari},
  {Lamperti}, {Parlanti}, {Scholtz}, \& {Venturi}}]{uebler2024b}
{{\"U}bler}, H., {D'Eugenio}, F., {Perna}, M., {et~al.} 2024{\natexlab{a}},
  \mnras, 533, 4287, \dodoi{10.1093/mnras/stae1993}

\bibitem[{{{\"U}bler} {et~al.}(2024{\natexlab{b}}){{\"U}bler}, {F{\"o}rster
  Schreiber}, {van der Wel}, {Bezanson}, {Price}, {D'Eugenio}, {Wisnioski},
  {Genzel}, {Tacconi}, {Wuyts}, {Naab}, {Lutz}, {Straatman}, {Shimizu},
  {Davies}, {Liu}, \& {Mendel}}]{uebler2024}
{{\"U}bler}, H., {F{\"o}rster Schreiber}, N.~M., {van der Wel}, A., {et~al.}
  2024{\natexlab{b}}, \mnras, 527, 9206, \dodoi{10.1093/mnras/stad3826}

\bibitem[{{van Albada} {et~al.}(1985){van Albada}, {Bahcall}, {Begeman}, \&
  {Sancisi}}]{vanAlbada1985}
{van Albada}, T.~S., {Bahcall}, J.~N., {Begeman}, K., \& {Sancisi}, R. 1985,
  \apj, 295, 305, \dodoi{10.1086/163375}

\bibitem[{{van der Hulst} {et~al.}(1992){van der Hulst}, {Terlouw}, {Begeman},
  {Zwitser}, \& {Roelfsema}}]{vdH1992}
{van der Hulst}, J.~M., {Terlouw}, J.~P., {Begeman}, K.~G., {Zwitser}, W., \&
  {Roelfsema}, P.~R. 1992, in Astronomical Society of the Pacific Conference
  Series, Vol.~25, Astronomical Data Analysis Software and Systems I, ed. D.~M.
  {Worrall}, C.~{Biemesderfer}, \& J.~{Barnes}, 131

\bibitem[{{van der Wel} {et~al.}(2014){van der Wel}, {Franx}, {van Dokkum},
  {Skelton}, {Momcheva}, {Whitaker}, {Brammer}, {Bell}, {Rix}, {Wuyts},
  {Ferguson}, {Holden}, {Barro}, {Koekemoer}, {Chang}, {McGrath},
  {H{\"a}ussler}, {Dekel}, {Behroozi}, {Fumagalli}, {Leja}, {Lundgren},
  {Maseda}, {Nelson}, {Wake}, {Patel}, {Labb{\'e}}, {Faber}, {Grogin}, \&
  {Kocevski}}]{vanderWel2014}
{van der Wel}, A., {Franx}, M., {van Dokkum}, P.~G., {et~al.} 2014, \apj, 788,
  28, \dodoi{10.1088/0004-637X/788/1/28}

\bibitem[{{van Houdt} {et~al.}(2021){van Houdt}, {van der Wel}, {Bezanson},
  {Franx}, {d'Eugenio}, {Barisic}, {Bell}, {Gallazzi}, {de Graaff}, {Maseda},
  {Pacifici}, {van de Sande}, {Sobral}, {Straatman}, \& {Wu}}]{vanHoudt2021}
{van Houdt}, J., {van der Wel}, A., {Bezanson}, R., {et~al.} 2021, \apj, 923,
  11, \dodoi{10.3847/1538-4357/ac1f29}

\bibitem[{{Varidel} \& {Croom}(2023)}]{Varidel2023}
{Varidel}, M., \& {Croom}, S. 2023, {Blobby3D: Bayesian inference for gas
  kinematics}, Astrophysics Source Code Library, record ascl:2303.005.
\newblock \doeprint{2303.005}

\bibitem[{{Varidel} {et~al.}(2019){Varidel}, {Croom}, {Lewis}, {Brewer}, {Di
  Teodoro}, {Bland-Hawthorn}, {Bryant}, {Federrath}, {Foster}, {Glazebrook},
  {Goodwin}, {Groves}, {Hopkins}, {Lawrence}, {L{\'o}pez-S{\'a}nchez},
  {Medling}, {Owers}, {Richards}, {Scalzo}, {Scott}, {Sweet}, {Taranu}, \& {van
  de Sande}}]{Varidel2019}
{Varidel}, M.~R., {Croom}, S.~M., {Lewis}, G.~F., {et~al.} 2019, \mnras, 485,
  4024, \dodoi{10.1093/mnras/stz670}

\bibitem[{{Virtanen} {et~al.}(2020){Virtanen}, {Gommers}, {Oliphant},
  {Haberland}, {Reddy}, {Cournapeau}, {Burovski}, {Peterson}, {Weckesser},
  {Bright}, {van der Walt}, {Brett}, {Wilson}, {Millman}, {Mayorov}, {Nelson},
  {Jones}, {Kern}, {Larson}, {Carey}, {Polat}, {Feng}, {Moore}, {VanderPlas},
  {Laxalde}, {Perktold}, {Cimrman}, {Henriksen}, {Quintero}, {Harris},
  {Archibald}, {Ribeiro}, {Pedregosa}, {van Mulbregt}, \& {SciPy 1. 0
  Contributors}}]{Virtanen2020}
{Virtanen}, P., {Gommers}, R., {Oliphant}, T.~E., {et~al.} 2020, Nature
  Methods, 17, 261, \dodoi{10.1038/s41592-019-0686-2}

\bibitem[{{Weitzel} {et~al.}(1996){Weitzel}, {Krabbe}, {Kroker}, {Thatte},
  {Tacconi-Garman}, {Cameron}, \& {Genzel}}]{Weitzel1996}
{Weitzel}, L., {Krabbe}, A., {Kroker}, H., {et~al.} 1996, \aaps, 119, 531

\bibitem[{{Wellons} {et~al.}(2020){Wellons}, {Faucher-Gigu{\`e}re},
  {Angl{\'e}s-Alc{\'a}zar}, {Hayward}, {Feldmann}, {Hopkins}, \&
  {Kere{\v{s}}}}]{Wellons2020}
{Wellons}, S., {Faucher-Gigu{\`e}re}, C.-A., {Angl{\'e}s-Alc{\'a}zar}, D.,
  {et~al.} 2020, \mnras, 497, 4051, \dodoi{10.1093/mnras/staa2229}

\bibitem[{{Whiting}(2012)}]{Whiting2012}
{Whiting}, M.~T. 2012, \mnras, 421, 3242,
  \dodoi{10.1111/j.1365-2966.2012.20548.x}

\bibitem[{{Wilson} {et~al.}(2011){Wilson}, {Warren}, {Irwin}, {Knapen},
  {Israel}, {Serjeant}, {Attewell}, {Bendo}, {Brinks}, {Butner}, {Clements},
  {Leech}, {Matthews}, {M{\"u}hle}, {Mortier}, {Parkin}, {Petitpas}, {Tan},
  {Tilanus}, {Usero}, {Vaccari}, {van der Werf}, {Wiegert}, \&
  {Zhu}}]{Wilson2011}
{Wilson}, C.~D., {Warren}, B.~E., {Irwin}, J., {et~al.} 2011, \mnras, 410,
  1409, \dodoi{10.1111/j.1365-2966.2010.17646.x}

\bibitem[{{Wisnioski} {et~al.}(2015){Wisnioski}, {F{\"o}rster Schreiber},
  {Wuyts}, {Wuyts}, {Bandara}, {Wilman}, {Genzel}, {Bender}, {Davies},
  {Fossati}, {Lang}, {Mendel}, {Beifiori}, {Brammer}, {Chan}, {Fabricius},
  {Fudamoto}, {Kulkarni}, {Kurk}, {Lutz}, {Nelson}, {Momcheva}, {Rosario},
  {Saglia}, {Seitz}, {Tacconi}, \& {van Dokkum}}]{Wisnioski2015}
{Wisnioski}, E., {F{\"o}rster Schreiber}, N.~M., {Wuyts}, S., {et~al.} 2015,
  \apj, 799, 209, \dodoi{10.1088/0004-637X/799/2/209}

\bibitem[{{Wisnioski} {et~al.}(2019){Wisnioski}, {F{\"o}rster Schreiber},
  {Fossati}, {Mendel}, {Wilman}, {Genzel}, {Bender}, {Wuyts}, {Davies},
  {{\"U}bler}, {Bandara}, {Beifiori}, {Belli}, {Brammer}, {Chan}, {Davies},
  {Fabricius}, {Galametz}, {Lang}, {Lutz}, {Nelson}, {Momcheva}, {Price},
  {Rosario}, {Saglia}, {Seitz}, {Shimizu}, {Tacconi}, {Tadaki}, {van Dokkum},
  \& {Wuyts}}]{Wisnioski2019}
{Wisnioski}, E., {F{\"o}rster Schreiber}, N.~M., {Fossati}, M., {et~al.} 2019,
  \apj, 886, 124, \dodoi{10.3847/1538-4357/ab4db8}

\bibitem[{{Wuyts} {et~al.}(2012){Wuyts}, {F{\"o}rster Schreiber}, {Genzel},
  {Guo}, {Barro}, {Bell}, {Dekel}, {Faber}, {Ferguson}, {Giavalisco}, {Grogin},
  {Hathi}, {Huang}, {Kocevski}, {Koekemoer}, {Koo}, {Lotz}, {Lutz}, {McGrath},
  {Newman}, {Rosario}, {Saintonge}, {Tacconi}, {Weiner}, \& {van der
  Wel}}]{Wuyts2012}
{Wuyts}, S., {F{\"o}rster Schreiber}, N.~M., {Genzel}, R., {et~al.} 2012, \apj,
  753, 114, \dodoi{10.1088/0004-637X/753/2/114}

\bibitem[{{Wuyts} {et~al.}(2013){Wuyts}, {F{\"o}rster Schreiber}, {Nelson},
  {van Dokkum}, {Brammer}, {Chang}, {Faber}, {Ferguson}, {Franx}, {Fumagalli},
  {Genzel}, {Grogin}, {Kocevski}, {Koekemoer}, {Lundgren}, {Lutz}, {McGrath},
  {Momcheva}, {Rosario}, {Skelton}, {Tacconi}, {van der Wel}, \&
  {Whitaker}}]{Wuyts2013}
{Wuyts}, S., {F{\"o}rster Schreiber}, N.~M., {Nelson}, E.~J., {et~al.} 2013,
  \apj, 779, 135, \dodoi{10.1088/0004-637X/779/2/135}

\bibitem[{{Wuyts} {et~al.}(2016){Wuyts}, {F{\"o}rster Schreiber}, {Wisnioski},
  {Genzel}, {Burkert}, {Bandara}, {Beifiori}, {Belli}, {Bender}, {Brammer},
  {Chan}, {Davies}, {Fossati}, {Galametz}, {Kulkarni}, {Lang}, {Lutz},
  {Mendel}, {Momcheva}, {Naab}, {Nelson}, {Saglia}, {Seitz}, {Tacconi},
  {Tadaki}, {{\"U}bler}, {van Dokkum}, {Wilman}, \& {Wuyts}}]{Wuyts2016}
{Wuyts}, S., {F{\"o}rster Schreiber}, N.~M., {Wisnioski}, E., {et~al.} 2016,
  \apj, 831, 149, \dodoi{10.3847/0004-637X/831/2/149}

\bibitem[{{Zabl} {et~al.}(2019){Zabl}, {Bouch{\'e}}, {Schroetter}, {Wendt},
  {Finley}, {Schaye}, {Conseil}, {Contini}, {Marino}, {Mitchell}, {Muzahid},
  {Pezzulli}, \& {Wisotzki}}]{Zabl2019}
{Zabl}, J., {Bouch{\'e}}, N.~F., {Schroetter}, I., {et~al.} 2019, \mnras, 485,
  1961, \dodoi{10.1093/mnras/stz392}

\bibitem[{{Zabl} {et~al.}(2020){Zabl}, {Bouch{\'e}}, {Schroetter}, {Wendt},
  {Contini}, {Schaye}, {Marino}, {Muzahid}, {Pezzulli}, {Verhamme}, \&
  {Wisotzki}}]{Zabl2020}
---. 2020, \mnras, 492, 4576, \dodoi{10.1093/mnras/stz3607}

\bibitem[{{Zabl} {et~al.}(2021){Zabl}, {Bouch{\'e}}, {Wisotzki}, {Schaye},
  {Leclercq}, {Garel}, {Wendt}, {Schroetter}, {Muzahid}, {Cantalupo},
  {Contini}, {Bacon}, {Brinchmann}, \& {Richard}}]{Zabl2021}
{Zabl}, J., {Bouch{\'e}}, N.~F., {Wisotzki}, L., {et~al.} 2021, \mnras, 507,
  4294, \dodoi{10.1093/mnras/stab2165}

\end{thebibliography}
\bibliographystyle{aasjournal}
\end{document}